\documentclass[journal=jacsat,manuscript=article,layout=twocolumn]{achemso}
\usepackage{graphicx}
\usepackage{amsmath}
\usepackage{subcaption}
\usepackage{float}
\usepackage{textcomp}
\usepackage{textgreek}
\usepackage{xcolor}
\usepackage{tablefootnote}
\usepackage{url}
\usepackage{epstopdf}
\usepackage{xr}
\usepackage{nicefrac}
\usepackage{ulem}
\DeclareUnicodeCharacter{2009}{FIXME}
\usepackage{multirow}
\setkeys{acs}{articletitle = true}
\SectionsOn
\SectionNumbersOn

\makeatletter
\newcommand*{\addFileDependency}[1]{
  \typeout{(#1)}
  \@addtofilelist{#1}
  \IfFileExists{#1}{}{\typeout{No file #1.}}
}
\makeatother

\newcommand*{\myexternaldocument}[1]{%
    \externaldocument{#1}%
    \addFileDependency{#1.tex}%
    \addFileDependency{#1.aux}%
}

\myexternaldocument{SI_finalversion}

\title
  {Influence of Dopamine Methacrylamide \\ on swelling behaviour and nanomechanical properties \\of PNIPAM microgels }

\author{Sandra Forg}
\affiliation{Soft Matter at Interfaces (SMI), Institute for physics of condensed matter, Technical University of Darmstadt, Darmstadt, Germany}
\author{Xuhong Guo}
\affiliation{School of Chemical Engineering, East China University of Science and Technology, Shanghai, China}
\author{Regine von Klitzing}
\affiliation{Soft Matter at Interfaces (SMI), Institute for physics of condensed matter, Technical University of Darmstadt, Darmstadt, Germany}
\email{klitzing@smi.tu-darmstadt.de}

\affiliation[Soft Matter at Interfaces (SMI), Institute for physics of condensed matter, Technical University of Darmstadt, Darmstadt, Germany]
{Soft Matter at Interfaces (SMI), Institute for physics of condensed matter, Technical University of Darmstadt, Darmstadt, Germany}
\affiliation[School of Chemical Engineering, East China University of Science and Technology, Shanghai, China]{School of Chemical Engineering, East China University of Science and Technology, Shanghai, China}

\begin{document}
\maketitle

\begin{figure}[H]
\centering
\includegraphics[scale=1]{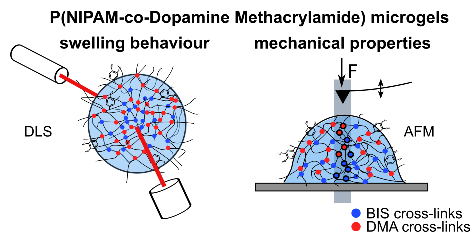}
\caption{for Table of Contents use only}
\end{figure}

\begin{abstract}
\noindent
The combination of the catechol-containing co-monomer dopamine methacrylamide (DMA) with stimuli-responsive microgels such as poly(N-isopropylacrylamide) (PNIPAM) bears a huge potential in research and for applications due to the versatile properties of catechols. This research gives first detailed insights into the influence of DMA on the swelling of PNIPAM microgels and the correlation with their nanomechanical properties. Dynamic light scattering (DLS) was used to analyse the swelling behaviour of microgels in bulk solution. The incorporation of DMA decreases the volume phase transition temperature (VPTT) and completion temperature (VPT CT) due to its higher hydrophobicity when compared to NIPAM, while sharpening the transition. The cross-linking ability of DMA decreases swelling ratios and mesh sizes of the microgels. Microgels adsorbed at the solid surface are characterised by atomic force microscopy (AFM): Scanning provides information about the microgel's shape on the surface and force spectroscopy measurements determines their nanomechanical properties ($E$ modulus). As the DMA content increases, microgels protrude more from the surface, correlating with an increase of $E$ modulus and a stiffening of the microgels - confirming the cross-linking ability of DMA. Force spectroscopy measurements below and above the VPTT display a stiffening of the microgels with the incorporation of DMA and upon heating across it's entire cross-section. The affine network factor $\beta$, derived from the Flory-Rehner theory describing the elasticity and swelling of the microgel network, is linearly correlated with the $E$ moduli of the microgels for both - pure PNIPAM and P(NIPAM-co-DMA) microgels. However, for large amounts of DMA, DMA appears to hinder the microgel shrinking, while still ensuring mechanical stiffness, possibly due to catechol interactions within the microgel network.
\end{abstract}

\section{Introduction}
Hydrogels are defined as polymer networks that can uptake large amounts of water \cite{IUPAC1997}, similar to human body tissues, making them widely applicable especially in the biomedical field \cite{Liu2015, Chai2017, Li2016}. Their micro-network, characterised by the so-called mesh size, facilitates the diffusion and transport of oxygen, nutrients and drugs. They are manufacturable with good biocompatibility and a variety of properties. Due to these versatile properties, hydrogels are used as tissue adhesives \cite{Lee2001, Mi2011, Zhao2022}, in drug delivery \cite{Li2016, Gupta2014, Satarkar2008} or as self-healing materials \cite{Vatankhah2014, Krogsgaard2013, Wang2015}. Nevertheless, hydrogels possess some weaknesses, i.e. a low stiffness, restricting their use for applications demanding mechanical robustness \cite{Chai2017}. For biomedical applications for instance, they need the same stiffness as the tissue they mimic \cite{Liu2015}. To overcome this limitation, stiffness can be introduced and tuned by chemically cross-linking the polymer network. While the cartilage, for example, has a Young's modulus of around 700\,kPa, muscles only possess moduli around $10-15$\,kPa \cite{Liu2015}. Smaller hydrogels, so-called microgels, revealed advantages over larger hydrogels \cite{Li2016, Daly2020, Alzanbaki2021}. Surface coatings can be easier prepared from microgels than from macrogels \cite{Wellert2015, Uhlig2018} and they are easier adaptable to the size of defects, ensuring better filling. \\
The introduction of stimuli-responsiveness by using for example poly(\textit{N}-isopropylacrylamide) (PNIPAM) further promotes the use of such microgels \cite{Gupta2014, Mi2011, Satarkar2008, Metawea2021, Lima2016}, as they respond i.e. faster to outer stimuli than macrogels \cite{Wang2001}. They can be used for cancer therapy \cite{Metawea2021} or as bioadhesives for retinal diseases \cite{Lima2016}. PNIPAM possesses a volume phase transition temperature (VPTT) at around 32$^\circ$C. Introducing hydrophobic or hydrophilic co-monomers allows the VPTT to be tuned to lower or higher temperatures regarding their particular needs. Tuning the sharpness of the transition allows, e.g. drugs incorporated in the microgel network to be released faster or slower, when used as drug delivery systems \cite{Li2022}. \\
Based on marine organisms like mussels, dopamine moieties have gained popularity in science due to the unique properties of catechol groups (contained in amino acids of the mussel's byssus\cite{Waite1981}). They enable the great adhesive properties and mechanical, as well as self-healing properties of mussels, offering great advantages in the biomedical field \cite{Kaushik2015, Balkenende2019, Zhang2017}. The exceptional under-water performance of mussels is attributed to a balance of mussel adhesion and cohesion, where latter one provides the strength of the mussel byssus.  However, extracting these proteins from the mussel byssus provides high costs \cite{Cha2008, Morgan1990, Castillo2017} and complex purifying methods with limited yields \cite{Castillo2017} to lower the risks of negative side-effects when applied to humans, and not to forget, cause harm to the animals. To circumvent these challenges, dopamine methacrylamide (DMA) can be used as a synthetic replacement to mimic the mussel's properties. \\
DMA acts as a cross-linker \cite{Forg2022, Xue2017, Yang2015} and was already included in hydrogels as the sole cross-linker \cite{Xue2017}, theoretically leading to a higher stiffness. It was furthermore included to improve the adhesiveness of materials \cite{Glass2009, Tiu2019, Xiong2018, Forg2022} or to produce self-healing materials \cite{Vatankhah2014}. We recently focused on studying the reaction kinetics of DMA during the polymerisation of P(NIPAM-co-DMA) microgels cross-linked with \textit{N,N'}-methylene-bis(acrylamide) (BIS) \cite{Forg2022}. This research seeked to better control the synthesis process, particularly concerning the scavenging effect of DMA: Phenols are radical scavengers, where ortho-structured phenols, such as catechols, possess the highest scavenging abilities \cite{Thavasi2009}. Therefore, the addition of DMA has to be carefully timed to not prohibit the microgel formation, while ensuring its incorporation \cite{Forg2022}. Together with the work of Xue \textit{et al.}\cite{Xue2017} on the reactivity and cross-linking mechanism of DMA on PNIPAM microgels without additional cross-linker, the synthesis of mussel-inspired P(NIPAM-co-DMA) microgels can be precisely controlled. But so far, there is only limited research \cite{Marcisz2017, Garcia2019, Xue2017, Vatankhah2014, Forg2022} on P(NIPAM-co-DMA) materials. A comprehensive understanding of the influence of DMA onto the swelling behaviour and mechanical properties of PNIPAM microgels is still missing, especially in terms of its cohesive performance. Furthermore, DMA's increased hydrophobicity in comparison to NIPAM, suggests a decrease of the VPTT \cite{Ryo1995, Garcia2019, Vatankhah2014, Feil1993, Marcisz2017}. \\
Therefore, we synthesised P(NIPAM-co-DMA) microgels, varying the DMA and cross-linker BIS concentration. For a full evaluation of their properties, we conducted a multi-faceted characterisation:\\
- \textbf{Swelling behaviour:} Microgels in bulk solution were characterised by dynamic light scattering (DLS). The investigation involves the determination of the VPT, as well as swelling ratios and mesh sizes of the microgels. Additionally, electrophoretic mobility measurements in the swollen and shrunken state of the microgels were carried out to gain information about their surface charge. \\
- \textbf{Nanomechanical properties:} Microgels adsorbed at the solid surface have been examined by atomic force spectroscopy (AFM). The scanning of the microgels allowed the investigation of the microgel's shape under ambient conditions. Force spectroscopy measurements provide information about the elastic modulus $E$ of the microgels. \\
- \textbf{Correlation:} The correlation between the affine network factor $\beta$, derived from the Flory-Rehner theory \cite{Flory1943, Flory1953} describing the elasticity and swelling of the microgels, with the $E$ modulus of the microgels was investigated. This examination aimed to test the validity of the theory for the microgels and detect differences between pure PNIPAM microgels and their P(NIPAM-co-DMA) counterparts.

\section{Experimental details} \label{Experimental Details}
\subsection{Materials}
\textbf{For dopamine methacrylamide (DMA) synthesis:} 3,4-dihydroxyphenethylamine hydrochloride (dopamine-HCl) (purity\,=\,99\,\%) and sodium hydrogen carbonate (purity\,=\,99\,\%) were purchased from Alfa Aesar (Haverhill, Massachusetts, USA). Sodium tetraborate (purity\,=\,99\,\%) was purchased from Sigma Aldrich (St. Louis, Missouri, USA). \textit{n}-Hexane (purity\,$\geq$\,99\,\%) and magnesium sulfate (anhydrous, purity\,$\geq$\,98\,\%) were purchased from TH Geyer (Renningen, Germany). Hydrochloric acid (HCl), isopropanol (purity\,$\geq$\,99.5\,\%), tetrahydrofuran (THF) (purity\,$\geq$\,99.5\,\%) and sodium chloride (NaCl) (purity\,$\geq$\,99.5\,\%) were purchased from Carl Roth (Karlsruhe, Germany). Ethyl acetate (purity\,=\,99.9\,\%) was purchased from VWR chemicals (Radnor, Pennsylvania, USA). Sodium hydroxide (NaOH) solution ($c=1$\,mol/l) was purchased from Merck (Darmstadt, Germany). \\
\textbf{For microgel synthesis:}
\textit{N}-isopropylacrylamide (NIPAM) (purity\,$\geq$\,99\,\%) and \textit{N,N'}-methylene-bis(acrylamide) (BIS) (purity\,$\geq$\,99.5\,\%) were purchased from Sigma Aldrich (St. Louis, Missouri, USA). 2,2'-azobis-2-methyl-propanimidamide dihydrochloride (AAPH) (purity\,$\geq$\,98\,\%) was purchased from Cayman (Ann Arbor, Michigan, USA). Toluhydroquinone (THQ) (purity\,$\geq$\,99\,\%) was purchased from Carl Roth (Karlsruhe, Germany). Ethanol (purity\,$\geq$\,99.8\,\%) was purchased from Fisher Scientific (Waltham, MA, USA).\\
\textbf{NMR solvents:} Dimethyl sulfoxide-d$_6$ (DMSO-d$_6$) (purity\,=\,99\,\%) and methanol-d$_4$ (CD$_3$OD) (purity\,$\geq$\,99.8\,\%) were purchased from Sigma Aldrich (St. Louis, Missouri, USA). \\
\textbf{AFM materials:}
Silicon wafers were purchased from SK Siltron (Gumi City, South Korea). AC160TS-R3 and BL-AC40TS cantilevers were purchased from Olympus Corporation (Tokyo, Japan). \\
\\
All chemicals were used as received without further purification. Purified water was obtained from a Millipore Milli-Q device (Merck, Darmstadt, Germany; resistance 18.2\,M\textOmega\,cm at 25\,$^\circ$C). 

\subsection{Synthesis} 
\subsubsection{DMA synthesis} \label{sec: DMA synthesis}
\begin{figure}[h!]
	\centering
	\includegraphics[width=8.4cm, height=18cm, keepaspectratio]{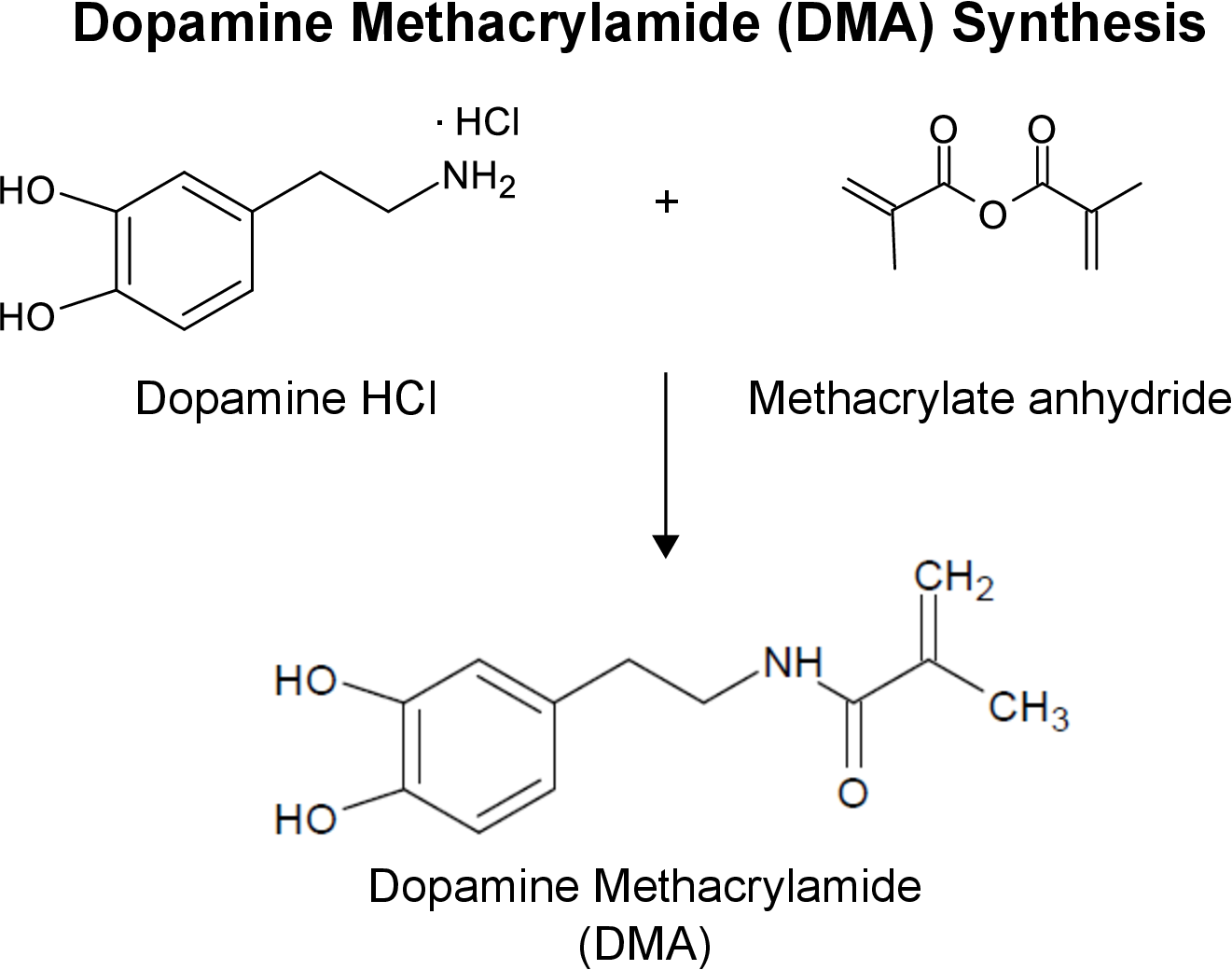}
	\caption{Synthesis scheme of dopamine methacrylamide (DMA).}
	\label{fig:DMA_synthesis_scheme}
\end{figure}

Details on the synthesis of dopamine methacrylamide (DMA) are described in literature \cite{Glass2009}. The synthesis scheme is presented in fig.\,\ref{fig:DMA_synthesis_scheme}. The success of the synthesis was confirmed by $^1$H and $^{13}$C NMR spectra (see exemplary spectra in SI, fig.\,S1). DMA was stored under N$_2$ in a freezer until further use. \\
\\
\\
\textbf{ $^1$H NMR (300\,MHz, DMSO-d$_6$, 301\,K):} $\delta=$\,7.93 (t, 1H, "e"), 7.46-7.95 (m, 3H, "e" and "b"), 6.65-6.58 (m, 2H, "a"), 6.44 (dd, 1H, "a"), 5.62 (s, 1H, "f"), 5.29 (m, 1H, "f"), 3.23 (m, 2H, "d"), 2.55 (t, 2H, "c"), 1.84 (t, 3H, "g"). $^{13}$C NMR (75\,MHz, DMSO-d$_6$): $\delta=$\,167.4 ("h"), 145.1 ("b"), 143.6 ("b"), 140.1 ("j"), 130.4 ("k"), 119.3 ("a"), 118.9 ("f"), 116.0 ("a"), 115.5 ("a"), 41.0 ("d"), 34.7 ("c"), 18.7 ("g"). \\

\subsubsection{Microgel synthesis} \label{sec:PNIPAMsynthesis}
\begin{figure*}[h!]
	\centering
	\includegraphics[width=22cm, height=10cm, keepaspectratio]{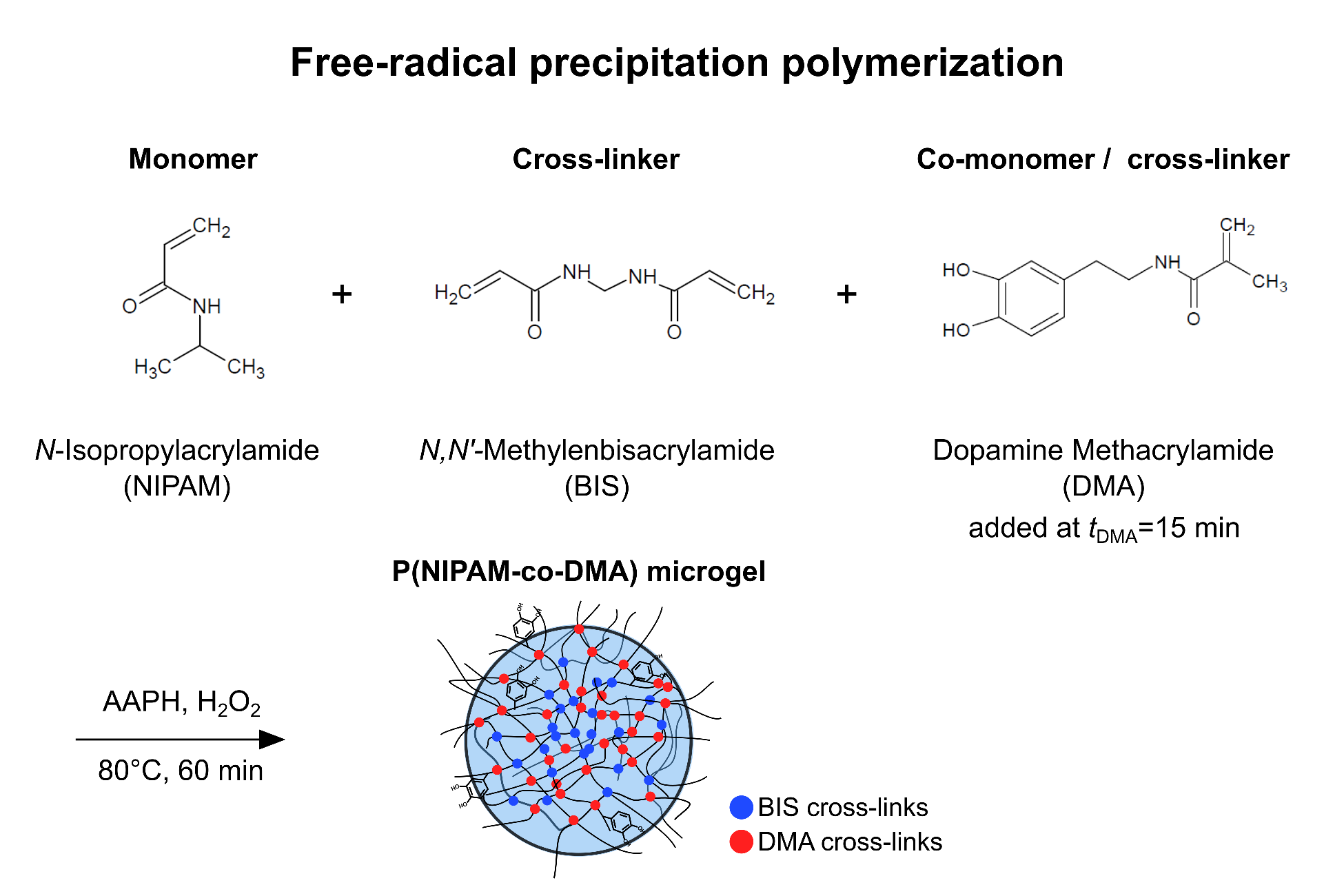}
	\caption{Synthesis scheme of P(NIPAM-co-DMA) synthesis. DMA cross-links the structure additionally, but free catechol moieties are still present \cite{Xue2017}. DMA is added at $t_\text{DMA}=15$\,min. For the pure PNIPAM microgel synthesis, no DMA is added.}
	\label{fig:microgel_synthesis_scheme}
\end{figure*}

\begin{table*}[h!]
\caption{Overview of synthesised microgels and their composition}
\centering
\begin{tabular}{llllll}
\hline
Microgel & $c_\text{NIPAM}$/mol\% & $c_\text{BIS}$/mol\% & $c_\text{DMA, injected}$/mol\% & \multicolumn{2}{l}{$c_\text{DMA, incorporated}$}\\
& & &  & UV-vis\textsuperscript{a}/mol\% & NMR\textsuperscript{a}/mol\%\\
\hline
MG$_\text{C1,D0}$ & 99 & 1 & - & - & -  \\
MG$_\text{C2,D0}$ & 98 & 2 & - & -  & -\\
MG$_\text{C1,D5}$  & 94 &  1 &  5 &  2.2 $\pm$ 0.1 & 2.3  \\
MG$_\text{C2,D5}$  & 93 & 2 & 5 & 2.9 $\pm$ 0 & 2.5 \\
MG$_\text{C2,D10}$  & 88 &  2 & 10 & 6.7 $\pm$ 0.1 & 6.3 \\
MG$_\text{C2,D15}$  & 83 &  2 & 15 & 7.8 $\pm$ 0.3 & 11.0 \\
MG$_\text{C2,D20}$  & 77 &  2 & 20 & 16.5 $\pm$ 0.8  & 21.2 \\
\hline
\multicolumn{6}{l}{\footnotesize{\textsuperscript{a} Values obtained by UV-vis standard addition and NMR spectroscopy as explained in sec.\,\ref{subsec:charactTechniques}}}
\end{tabular}
\label{tab:microgels}
\end{table*}

Microgels were synthesised by surfactant-free precipitation polymerisation \cite{PeltonChibante1986} in a home-built reactor \cite{Witt2019, Forg2022}. The reaction parameters were optimised in a previous publication \cite{Forg2022}. While an early injection of DMA prohibits the PNIPAM polymerisation due to the scavenging ability of DMA, a later injection ensures the PNIPAM polymerisation, but slows down the reaction kinetics of DMA. An injection time of DMA of $t_\text{DMA}=10-15$\,min and an overall reaction time of $t_\text{reac}=60$\,min ensure the full NIPAM monomer conversion, while still leading to large DMA incorporations.  \\
The amount of NIPAM (\textit{N}-isopropylacrylamide) (purity\,$\geq$\,99\,\%) was fixed to $n_\text{NIPAM}=18.6$\,mmol. Either $n_\text{BIS}=0.2$ or $0.4$\,mmol of the cross-linker BIS (\textit{N,N'}-methylene-bis(acrylamide)) were used, which is equivalent to 1\,mol\% and 2\,mol\% in the final microgel structure with respect to the total amount of reactants. Both were dissolved in 120\,ml of purified water, directly poured into the reactor (Batch method \cite{Witt2019, Acciaro2011}), heated to $80$\,$^\circ$C and degassed with N$_2$ for 1\,h under constant stirring. The positively charged, radical initiator AAPH (2,2'-azobis-2-methyl-propanimidamide dihydrochloride; $n_\text{AAPH}=247$\,mmol) was injected to start the reaction. The stirring frequency was maintained at 1000\,rpm. The solution turned opaque within the first 3\,min, confirming a successful nucleation of polymer particles. After $t_\text{DMA}=15$\,min,  $n_\text{DMA}=1, 2.1, 3.3, 5$\,mmol of the co-monomer DMA (dopamine methacrylamide), dissolved in 5-15\,ml of pure ethanol, was injected into the reactor for the synthesis of P(NIPAM-co-DMA) microgels. The amounts are equivalent to $c_\text{DMA, injected}=5, 10, 15, 20$\,mol\% with respect to the total amount of reactants. After $t_\text{reac}=60$\,min, the reaction was stopped by adding $n_\text{THQ}=10$\,mmol of THQ (toluhydroquinone) into the mixture. Analogously, pure PNIPAM microgels were synthesised as a reference system without the addition of DMA. The reaction scheme is presented in fig.\,\ref{fig:microgel_synthesis_scheme}.\\
The microgels were purified by dialysis in purified water for 10\,d, exchanging the water daily. The resulting solution was lyophilised and stored in a freezer until further use. \\
The synthesised microgels and their composition are listed in table\,\ref{tab:microgels}. The notation "C" followed by a number corresponds to the cross-linker BIS amount in mol\% (with respect to the total amount of reactants). The notation "D" followed by a number corresponds to the injected amount of DMA $c_\text{DMA, injected}$ in mol\% (with respect to the total amount of reactants). The incorporated amount of DMA $c_\text{DMA, incorporated}$ was determined by UV-vis standard addition as described in literature \cite{Forg2022} and by NMR spectroscopy as described in Ref.\cite{Peng2019, Vatankhah2014, Forg2022}. 

\subsection{Methods} \label{subsec:charactTechniques}
\subsubsection{Mass Spectrometry (MS)} Time-samples were taken during the microgel synthesis and analysed by mass spectrometry (MS) to study the reactant's consumption and ensure a successful, reproducible polymerisation procedure. For a detailed description see Ref. \cite{Forg2022, Acciaro2011}. Samples were analysed by high performance liquid chromatography electrospray ionization mass spectrometry (HPLC-MS/ESI) on an Impact II mass spectrometer (Bruker Daltonik GmbH, Bremen, Germany).  The consumption of reactants (NIPAM, BIS and DMA) is comparable for all microgels used in this work, confirming a successful, reproducible reaction procedure (see SI; fig.\,S2). The monomer NIPAM and the cross-linker BIS nearly fully react until 15\,min, when the co-monomer DMA was added. 
\subsubsection{UV-vis Absorption Spectroscopy} \label{subsec:UV-Vis } The incorporated amount of DMA $c_\text{DMA, incorporated}$ was determined with UV-vis standard addition as described previously \cite{Forg2022} with a Lambda 650 spectrometer (PerkinElmer, Waltham, USA) at room temperature. The wavelength range was set from 200 to 500\,nm. For each microgel, measurements were repeated twice. The corresponding calibration curves are given in the SI, fig.\,S3. Results are listed in table\,\ref{tab:microgels}.
\subsubsection{Nuclear Magnetic Resonance (NMR)} \label{subsubseq:NMR} NMR spectroscopy was used to confirm the success of the DMA synthesis. $^1$H and $^{13}$C NMR spectra of DMA, solved in DMSO-d$_6$, were measured by a 300\,MHz NMR Avance II spectrometer (Bruker BioSpin GmbH, Rheinstetten, Germany) (see SI, fig.\,S1). \\
NMR spectroscopy was also used to confirm the successfull microgel synthesis.  $^1$H NMR spectra of the synthesised microgels, solved in DMSO-d$_6$ and methanol-d$_4$, were measured with a 500\,MHz DRX Avance 500 (Bruker BioSpin GmbH, Rheinstetten, Germany) (see SI, fig.\,S4 and fig.\,S5). It was used as a second method to determine the incorporated amount of DMA $c_\text{DMA, incorporated}$ based on previous publications \cite{Peng2019, Vatankhah2014, Forg2022}:
\begin{equation}
c_\text{DMA,NMR}=\frac{\frac{\int \text{DMA("h")}}{3\text{H}}}{\frac{\int \text{DMA("h")}}{3\text{H}}+\frac{\int \text{NIPAM("f")}}{1\text{H}}}
\end{equation}\\
The influence of BIS cannot be derived from the spectra, but since only $0.2-0.4$\,mmol of BIS in comparison to the total amount of reactants $20-24$\,mmol has been used, its impact is assumed to be negligible. The deconvolution tool of mestrenova analysis software, which fits a Gaussian function to the peaks of the spectra, was used to minimise the influence of the overlapping NH peaks of NIPAM and BIS with the peaks arising from the catechol peaks of DMA (DMA("h")) (see SI, fig.\,S4). Results can be found in table\,\ref{tab:microgels}.
\subsubsection{Dynamic Light Scattering (DLS)} \label{subsubseq:DLS} 
The hydrodynamic radius $R_\text{H}$ of the microgels dispersed in ultrapure water in dependence of temperature $T$ was measured by dynamic light scattering (DLS). The multi-angle set-up with a solid-state laser (wavelength $\lambda=660$\,nm, power $P=$100\,mW) was purchased from LS Instruments (LS Instruments AG, Fribourg, Switzerland). The measurement angle was varied between 30$^\circ$ and 120$^\circ$ in steps of 5$^\circ$. The temperature was controlled with a thermostat and varied between 15 and 60\,$^\circ$C. For the analysis, the temperature adjusted and measured by the device in the index-matching vat was taken, which can deviate up to 0.15\,$^\circ$C to the set temperature. The measurement time was fixed to 30\,s. The microgel sample concentration was about $c=25$\,\textmu g\,ml$^{-1}$ to exclude particle interaction and multiple scattering events. The LS hardware correlator transfers the signals into correlation functions, which were fitted by a cumulant fit of third order by a home-written script by Marcus Witt. \\
The resulting temperature-dependent swelling curves (fig.\,\ref{fig:VPTT}) of the microgels were fitted with a Boltzmann fit (see SI; fig.\,S6). The derivative $\text{d}R_\text{H}/\text{d}T$ was calculated. While the temperature at the minimum of the derivative function equals the volume phase transition temperature (VPTT), the temperature where $\text{d}R_\text{H}/\text{d}T$ evens out describes the VPT completion temperature (VPT CT) \cite{Garcia2019, Pasparakis2020, Forg2022}. Evening out was defined as a maximum $R_\text{H}$ difference of 2.5\,nm for a difference in temperature of 5$^\circ$C.  \\
The swelling of the microgel network can be described by the Flory-Rehner theory \cite{Flory1943, Flory1953}. Microgel networks possess a permanent structure, due to the incorporation of cross-linker, which is able to recover after a deformation process, i.e. during swelling. Hence, the microgel network is elastic. The change in the elastic free energy $\Delta G_\text{el}$ is mainly of entropic nature and can be calculated by using the affine approximation \cite{Kuhn1946}: the relative deformation of each polymer chain is equal to the deformation of the whole polymer network. It follows:
\begin{equation}
\begin{split}
\Delta G_\text{el} &=-T\Delta S_\text{el} \\\
&=\frac{3k_\text{B}T N_\text{c}}{2}[\lambda^2-1-\ln \lambda]
\end{split}
\end{equation}
$N_\text{c}$ is the number of polymer chains and $\lambda$ defines the volume change due to swelling. In the case of homogeneous and isotropic swelling: 
\begin{equation} 
\lambda=\left(\frac{\Phi_\text{0}}{\Phi}\right)^\frac{1}{3}=\left(\frac{V}{V_\text{0}}\right)^\frac{1}{3}=\left(\frac{R_\text{H}(20^\circ \text{C})}{R_\text{H}(50^\circ \text{C})}\right)
\end{equation}
$\Phi$ is the polymer volume fraction and $V_\text{0}$ is the particle volume. The notation 0 always refers to the reference state, mostly defined as the collapsed state (in the following defined at $50\,^\circ \text{C}$). According to $\lambda$ a swelling ratio $\alpha$ can be derived: 
\begin{equation}
\alpha=\left(\frac{V}{V_\text{0}}\right)=\left(\frac{R_\text{H}(20^\circ \text{C})}{R_\text{H}(50^\circ \text{C})}\right)^3
\label{eq:swellingratio}
\end{equation}
The change in the Gibbs free energy $\Delta G_\text{el}$ results in a change in osmotic pressure:
\begin{equation}
\begin{split}
\Pi_\text{el}&=\frac{k_\text{B}T N_\text{c}}{V_\text{0}}\left[\frac{\Phi}{2\Phi_\text{0}}-\left( \frac{\Phi}{\Phi_\text{0}}\right)^\frac{1}{3}\right] \\\
&=\frac{k_\text{B}T N_\text{c}}{V_\text{0}}\left[\frac{1}{2}\left(\frac{R_\text{H}(50^\circ \text{C})}{R_\text{H}(20\,^\circ \text{C})}\right)^3-\left( \frac{R_\text{H}(50^\circ \text{C})}{R_\text{H}(20^\circ \text{C})}\right)\right]
\end{split}
\end{equation}
The affine network factor is then defined as:
\begin{equation}\label{eq:affine}
\beta=-\left[\frac{1}{2}\left(\frac{R_\text{H}(50^\circ \text{C})}{R_\text{H}(20^\circ \text{C})}\right)^3-\left( \frac{R_\text{H}(50^\circ \text{C})}{R_\text{H}(20^\circ \text{C})}\right)\right]
\end{equation} \\
Mesh sizes of the microgels, which describe the average distance between sequential cross-links, were calculated by \cite{Canal1989}:
\begin{equation} \label{eq:mesh size}
\xi = l \left(\frac{2M_\text{c}}{M}\right)^\frac{1}{2}c_\text{N}^\frac{1}{2}Q^\frac{1}{3} ,
\end{equation}
where $l$ is the length of the C-C bond with $l=0.154$\,nm, $M$ the average molar mass of the monomer NIPAM with $M=113.16$, $c_\text{N}$ the characteristic ratio with $c_\text{N}=6.9$\cite{Peppas1985} and $M_\text{c}$ the molar mass between two cross-links with $M_\text{c}=\frac{n_\text{NIPAM}M_\text{NIPAM}}{n_\text{BIS}}+M_\text{BIS}$\cite{Faenger2006}. $Q$ is the degree of swelling and is assumed to be equal to $\alpha$ during this study. All the results are given in table\,\ref{tab:microgels_characterization_bulk}. Swelling ratios and mesh sizes are shown in fig.\,S7 (see SI) in dependence of the sum of the concentration of BIS $c_\text{BIS}$ and injected DMA $c_\text{DMA, injected}$.

\subsubsection{Electrophoretic Mobility}\label{subsec:elecMob}
The electrophoretic mobility $\mu$ of the microgels at 20$^\circ$C and 50$^\circ$C was measured with a Zetasizer NanoZ (Malvern Panalytical, UK). $\mu$ was measured to gain information about the charge of the microgels. Before starting the measurement, a Malvern zeta potential transfer standard was measured to test the set-up. The microgel sample concentration was 0.5\,mg\,ml$^{-1}$. Each measurement was repeated 10\,times.

\subsubsection{Spin Coating}
Microgels were deposited on piranha-etched silicon wafers. In a pre-drying step, $100-150\,$\textmu l of a $c=0.5$\,mg\,ml$^{-1}$ microgel dispersion was placed on the wafers for 1\,min before spin coating was started with a rotational speed of 1000\,rpm for 1\,min. The procedure results in a loose packing of the microgels at the surface, to study them separately.

\subsubsection{Atomic Force Microscopy (AFM) - Microgel Scanning}\label{subsubseq:Scanning} Microgels, deposited on silicon wafers by spin coating, were scanned in ambient conditions with an AC160TS-R3 cantilever at a MFP-3D AFM (Oxford Instrument, UK) by intermittent contact mode at 20$^\circ$C. Samples were equilibrated for 30\,min. AC160TS-R3 cantilevers have an average spring constant of $k=26$\,N\,m$^{-1}$ and an average resonance frequency of $f_\text{res}=300$\,kHz. Scan sizes were set to 10x10\, \textmu m and a scan rate of 1\,Hz was used. Cross-sections were analysed for 5 individual microgel particles and averaged, as illustrated in fig.\,\ref{fig:dry height microgels}. Corresponding AFM scans can be found in the SI, fig.\,S8. The height $h_\text{amb,AFM}$ and width $w_\text{amb,AFM}$ in ambient conditions are listed in the SI, table\,S1. Results of the $h_\text{amb,AFM}$/$w_\text{amb,AFM}$ ratio are given in table\,\ref{tab:microgels_characterization_surface}.

\subsubsection{Atomic Force Microscopy (AFM) - Force Spectroscopy} \label{subsec:Indentation} 
Atomic force microscopy (AFM) was used to study the mechanical properties of the adsorbed microgel particles by static indentation measurements. Measurements were done with a Cypher AFM (Oxford Instruments, Abingdon, UK) equipped with a CoolerHeater Sample Stage to control the temperature of the samples. Tips were chosen as indenters instead of colloidal probes. They provide information about the mechanical properties (laterally resolved) of the microgels \cite{Schulte2022}. Sharp tips are able to penetrate the hydrogels sufficiently enough, even in the rigid collapsed state \cite{Matzelle2003, Junk2010}, while not damaging the gel \cite{Harmon2003_2, Junk2010}. The BL-AC40TS cantilever was therefore chosen in accordance to previous work of our group \cite{Keenhammer2022} (spring constant $k=0.09$\,N\,m$^{-1}$, resonance frequency $f_\text{res}=110$\,kHz). It's essential to keep the type of cantilever constant when comparing data, as for different cantilevers / tips significant variation in $E$ moduli can be observed \cite{Zemla2020, Sokolov2014, Alcaraz2018}. \\
Before starting the measurement, the spring constant of the cantilever was determined with the thermal noise method \cite{Hutter1993}. Subsequently, both the cantilever and the microgel sample were immersed in ultrapure water and the system was equilibrated for at least 30\,min (at 20$^\circ$C and 50$^\circ$C). To enable the conversion of the cantilever deflection into a force-distance curve, a force measurement on an empty spot of the hard silicon wafer was recorded to determine the inverse optical lever sensitivity (InvOLS). \\
A 2x2\,\textmu m sized area with 30x30 pixel points was measured with the force mapping tool. The size allowed a high resolution of the microgel particles while maintaining reasonable measurement times. The force mapping tool provides information about the topography (topography map), gained from the tip position at the trigger point, which is the point at which the cantilever switches from approaching to retracting. Additionally, it provides information about the lift-off force (adhesion map, lift-off map), defined as the maximum force needed to detach the tip from the surface. Force curves were done with an approach velocity of 800\,nm\,s$^{-1}$. The velocity does not change the results substantially \cite{Hashmi2009}. Even by varying the speed by two orders of magnitudes, the resulting $E$ modulus varied only about 10\%. The influence of the trigger point, was investigated for the pure reference PNIPAM microgel MG$_\text{C2,D0}$ (2\,mol\% cross-linker BIS).

\begin{figure*}[h!]
	\centering
	\includegraphics[width=17cm, height=8cm, keepaspectratio]{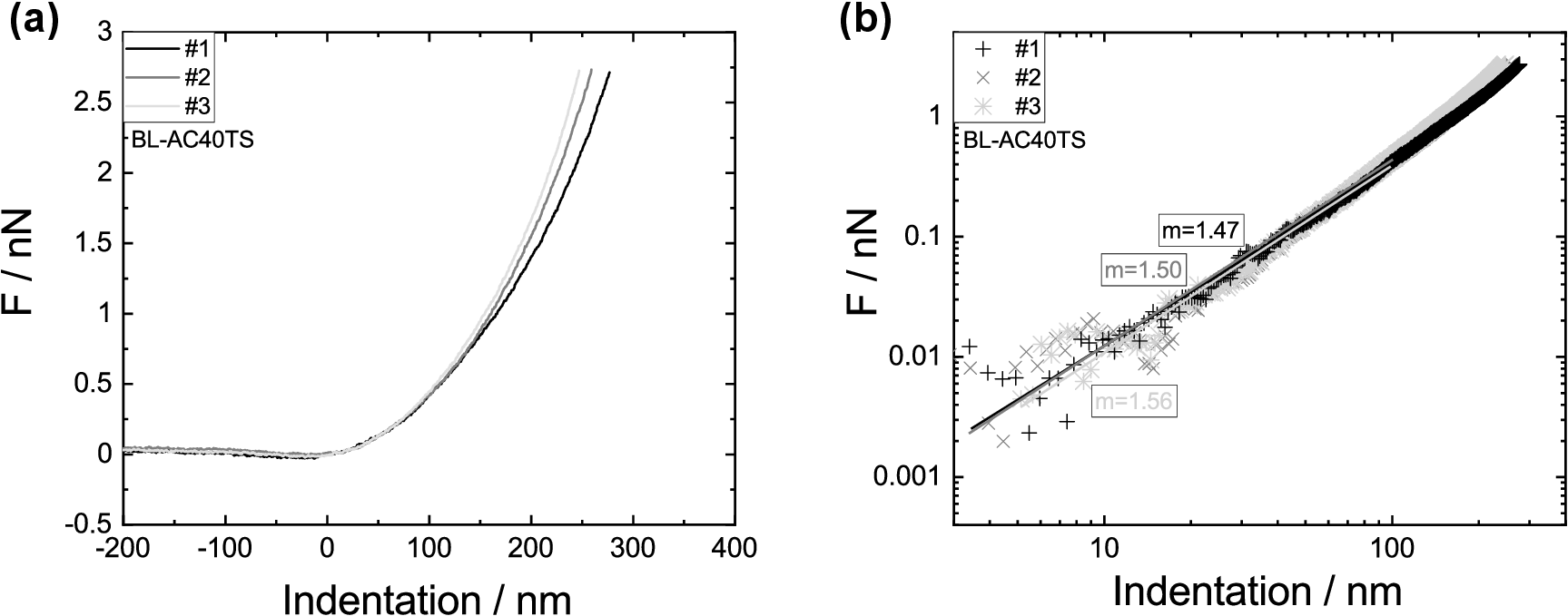}
	\caption{Exemplary force curves of MG$_\text{C2,D0}$ indented by a BL-AC40TS cantilever obtained by force spectroscopy measurements, plotted in linear (a) and logarithmic (b) scale. (b) includes linear fits with a slope of 1.47-1.56\,N\,m$^{-1}$ confirming the validity of the Hertzian model in ultrapure water up to indentation depths of 100\,nm. }
	\label{fig:Indentation_cantilevers}
\end{figure*}

\noindent In total six different trigger points from $0.5-4.9$\,nN have been studied (see SI, fig.\,S9). The trigger point must be in accordance to the sample's stiffness. While large trigger points (corresponding to large indentation depths) can destroy the sample and/or the tip, small trigger points would not fully resolve the microgel structure, i.e. by only indenting into the outer shell of the microgels, resulting in inaccurate $E$ moduli. Force curves of different trigger points are depicted in fig.\,S9\,(a) (see SI), and the resulting $E$ moduli, obtained from the centre of 3 different microgel particles, are presented in fig.\,S9\,(b) (see SI). $E$ moduli have been analysed with the Hertz model, as explained in a few paragraphs. No significant change of $E$ is observable, indicating no significant influence of the trigger point in the range from $0.5-4.9$\,nN. For the following measurements, a medium trigger point of 2.7\,nN was set. \\
The elastic modulus $E$ was determined by fitting the approach part of the force curves with the Hertz model \cite{Hertz1881}:
\begin{equation}
F=\frac{4E\sqrt[2]{R}}{3(1-\nu^2)}\delta^{\frac{3}{2}}
\label{eq:Hertz}
\end{equation}
where $F$ is the force of the indenter with radius $R$, $\nu$ the Poisson ratio of the sample, $\delta$ the indentation depth and $E$ the elastic modulus. The fit region of the Hertz model was set to $0-40$\% of the maximum indentation depth $\delta$ to exclude the impact of the underlying surface on the resulting moduli \cite{Harmon2003_2, Keenhammer2022, Witte2021}. The Poisson ratio $\nu$ for microgels varies from 0.25 to 0.5 in literature \cite{Voudouris2013, Hirotsu1991, Boon2017, Geissler1980, Backes2018}. For further analysis, a Poisson ratio of $\nu=0.4$ was used, which is close to the Poisson ratio of incompressible fluids in accordance with literature \cite{Voudouris2013} for PNIPAM microgels. Under- or overestimating the Poisson ratio by $\pm$ $0.15$ leads to errors of $11.6$\% for the elastic modulus $E$. \\
The validation of the Hertz model was studied using the reference system again (pure PNIPAM microgel MG$_\text{C2,D0}$). The BL-AC40 TS cantilever is much stiffer than the sample and the area of contact is assumed to be small \cite{Schulte2022, Popov2019} to fulfill the requirements of the Hertz model. Force curves were evaluated in the centre of 10 different particles at 20$^\circ$C. Three representative force curves are given in fig.\,\ref{fig:Indentation_cantilevers}\,(a) (for all force curves, see SI; fig.\,S10), and plotted in log-log scale in fig.\,\ref{fig:Indentation_cantilevers}\,(b). A linear fit to the logarithmic force curves provides a slope of 1.47 to 1.56\,N\,m$^{-1}$ up to indentation depths of 100\,nm, confirming the Hertzian model (see eq.\,\ref{eq:Hertz}, slope\,$=\frac{3}{2}$). \\
For the characterisation of the remaining microgels, at least 10 force curves were evaluated in the centre of 10 microgel particles at 20$^\circ$C (see SI, fig.\,S11). $E$ moduli over the particle's cross-section of the pure PNIPAM MG$_\text{C2,D0}$ and the P(NIPAM-co-DMA) microgel MG$_\text{C2,D10}$ were analysed for 5 individual microgel particles at 20$^\circ$C and 50$^\circ$C.

\section{Results} \label{Results}
\subsection{Swelling behaviour of microgels in bulk solutions}
\begin{figure}[h!]
	\centering
	\includegraphics[width=9cm, height=20cm, keepaspectratio]{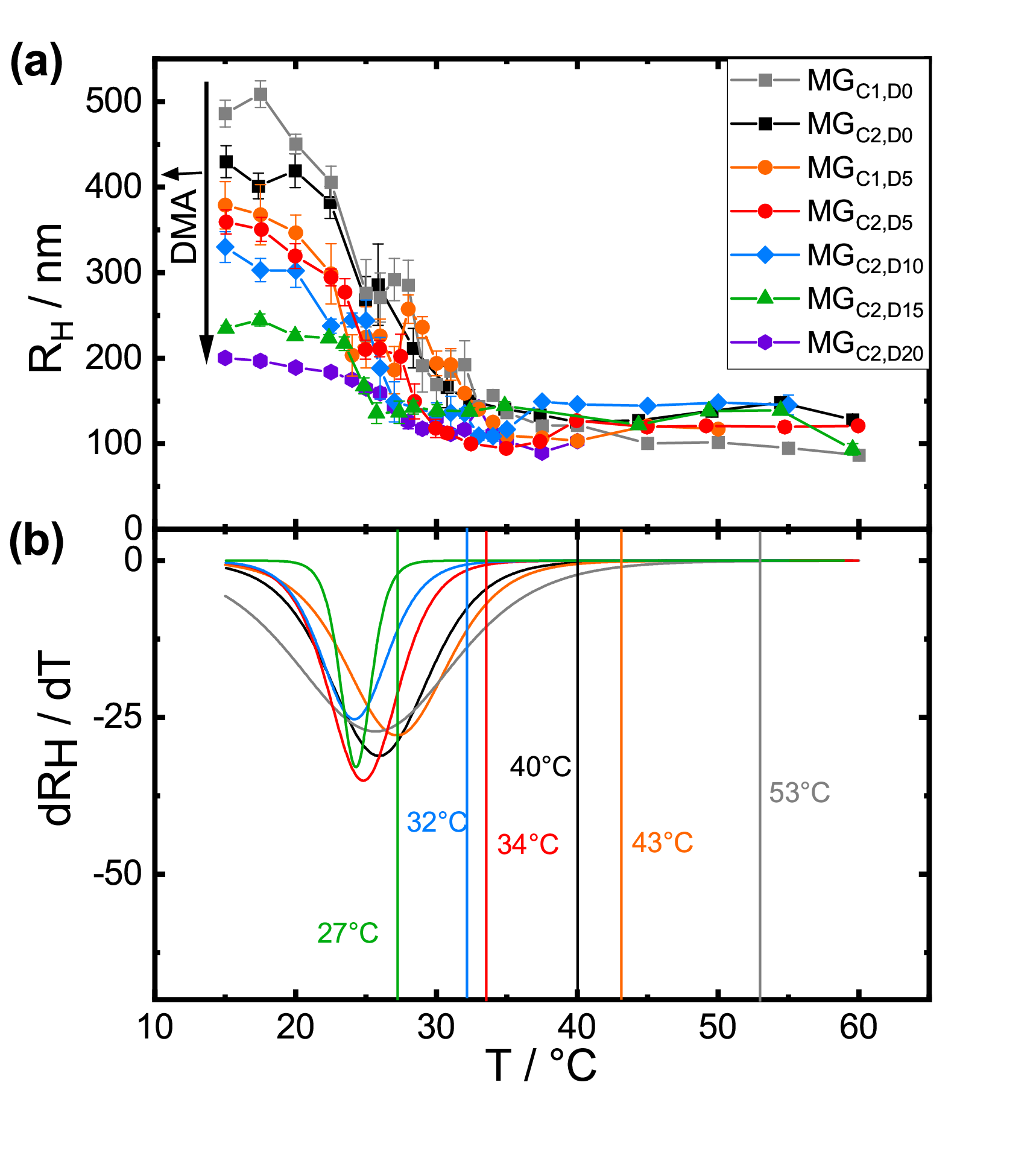}
	\caption{(a) $R_\text{H}$ in dependence of temperature $T$ for the different microgels. (b) Derivative $\text{d}R_\text{H}/\text{d}T$ of corresponding Boltzmann fits. Vertical lines show the VPT CT for the different microgels. For MG$_\text{C2,D20}$ no VPT CT could be defined, because the Boltzman fit couldn't describe the VPT (flocculation of sample at higher temperatures). }
	\label{fig:VPTT}
\end{figure}

Microgels in bulk solutions have been characterised with dynamic light scattering (DLS) to study their swelling behaviour as described in sec.\,\ref{subsec:charactTechniques}\,\ref{subsubseq:DLS}. Their swelling curves, namely their hydrodynamic radii $R_\text{H}$ in dependence of temperature $T$, are shown in fig.\,\ref{fig:VPTT}\,(a).  \\
The hydrodynamic radius $R_\text{H}(20^\circ \text{C})$ of the swollen microgels is similar for the microgels with 1\,mol\% and 2\,mol\% of cross-linker BIS (MG$_\text{C1,D0}$ compared to MG$_\text{C2,D0}$ and MG$_\text{C1,D5}$ compared to MG$_\text{C2,D5}$). It decreases continuously from 419.1 $\pm$ 19.8\,nm (MG$_\text{C2,D0}$) to 188.9 $\pm$ 5.6\,nm (MG$_\text{C2,D20}$) for an increasing amount of DMA. \\
The hydrodynamic radius $R_\text{H}(50^\circ \text{C})$ of the shrunken microgel increases around 38\,nm from the reference microgels MG$_\text{C1,D0}$ to MG$_\text{C2,D0}$, while $R_\text{H}(50^\circ \text{C})$ is similar for MG$_\text{C1,D5}$ and MG$_\text{C2,D5}$ containing 5\,mol\% DMA. $R_\text{H}(50^\circ \text{C})$ stays constant at around 138\,nm for an increasing amount of DMA (from MG$_\text{C2,D0}$ to MG$_\text{C2,D15}$). For MG$_\text{C2,D20}$, swelling curves could only be conducted for temperatures up to $40^\circ \text{C}$, as the dispersion flocculated at higher temperatures. \\
Temperature-dependent swelling curves were fitted by a Boltzmann fit (see SI, fig.\,S6). The derivatives of the Boltzmann fits  $\text{d}R_\text{H}/\text{d}T$ are plotted in fig.\,\ref{fig:VPTT}\,(b). The volume phase transition (VPT) completion temperature (CT) according to \cite{Garcia2019, Pasparakis2020, Forg2022} and as explained in sec.\,\ref{subsec:charactTechniques}\,\ref{subsubseq:DLS} is illustrated by vertical lines and is given in table\,\ref{tab:microgels_characterization_bulk}. The minimum of the derivative $\text{d}R_\text{H}/\text{d}T$ corresponds to the volume phase transition temperature (VPTT).  \\
The VPT CT significantly decreases for an increasing amount of cross-linker (MG$_\text{C1,D0}$ compared to MG$_\text{C2,D0}$ and MG$_\text{C1,D5}$ compared to MG$_\text{C2,D5}$), leading to sharper transitions. The VPTT is similar for an increasing amount of cross-linker from MG$_\text{C1,D0}$ to MG$_\text{C2,D0}$ and decreases from MG$_\text{C1,D5}$ to MG$_\text{C2,D5}$. For MG$_\text{C1,D5}$, data fluctuate largely around the VPT. For an increasing amount of incorporated DMA, the VPT CT decreases from 40$^\circ$C to 27$^\circ$C from MG$_\text{C2,D0}$ to MG$_\text{C2,D15}$, leading to sharper transitions. The incorporation of DMA reduces the VPTT. For MG$_\text{C2,D20}$, the Boltzmann fit could not describe the VPT completion adequately, since values at higher temperatures were not available due to the flocculation of the sample. Therefore, no VPT CT was determined. \\
From the swelling curves, the swelling ratio $\alpha$ and the mesh size $\xi$ were evaluated by eq.\,\ref{eq:swellingratio} and eq.\,\ref{eq:mesh size} (see table\,\ref{tab:microgels_characterization_bulk} and SI; fig.\,S7). The swelling ratio $\alpha$ is the highest for the reference microgel MG$_\text{C1,D0}$ with 1\,mol\% cross-linker BIS due to the smallest $R_\text{H}(50^\circ \text{C})$.  In comparison, the swelling ratio $\alpha$ of the reference microgel MG$_\text{C2,D0}$ (2\,mol\% cross-linker BIS) is only 1/3 of the swelling ratio of MG$_\text{C1,D0}$. However, $\alpha$ is calculated from the two hydrodynamic radii at $20^\circ \text{C}$ and $50^\circ \text{C}$, which can be prone to error. The swelling ratio decreases from the lower cross-linked MG$_\text{C1,D5}$ to the higher cross-linked MG$_\text{C2,D5}$. With increasing amount of DMA (from MG$_\text{C2,D0}$ to MG$_\text{C2,D15}$), the swelling ratio decreases from 27.9 $\pm$ 4.1 to 4.4 $\pm$ 0.3. The swelling ratio of MG$_\text{C2,D20}$ was calculated with the hydrodynamic radius of the shrunken microgel at $40^\circ \text{C}$ $R_\text{H}(40^\circ \text{C})$ due to the flocculation of the sample at higher temperatures. Therefore, the swelling ratio is prone to smaller errors and is slightly higher than the swelling ratio of MG$_\text{C2,D15}$. 

\noindent Mesh sizes decrease for an increasing amount of cross-linker BIS (MG$_\text{C1,D0}$ compared to MG$_\text{C2,D0}$ and MG$_\text{C1,D5}$ compared to MG$_\text{C2,D5}$). They further decrease for an increasing amount of DMA (from MG$_\text{C2,D0}$ to MG$_\text{C2,D15}$), from 12.2\,nm to 6.5\,nm. Again, MG$_\text{C2,D20}$ possesses a slightly higher value than MG$_\text{C2,D15}$, since it had to be calculated with the value at $40^\circ \text{C}$.\\
The electrophoretic mobility $\mu$ was measured as described in sec.\,\ref{subsec:charactTechniques}\,\ref{subsec:elecMob} at 20$^\circ$C and 50$^\circ$C to gain information about the charge of the microgels. Values are given in table\,\ref{tab:microgels_characterization_bulk}. $\mu$ is positive and in the same order for every microgel, corresponding to a positive surface charge of the microgels. It increases with increasing temperature.

\begin{table*}[h!]
\caption{Characterisation of microgels in bulk solution: Swelling ratio $\alpha$ and mesh size $\xi$ of microgels calculated from results obtained by DLS by eq.\,\ref{eq:swellingratio} and eq.\,\ref{eq:mesh size}. VPT CT determined according to \cite{Garcia2019, Pasparakis2020, Forg2022} as described in sec.\,\ref{subsec:charactTechniques}\,\ref{subsubseq:DLS}. Electrophoretic mobility $\mu$  measured as described in sec.\,\ref{subsec:charactTechniques}\,\ref{subsec:elecMob}. }
\centering
\begin{tabular}{llllll}
\hline
Microgel & swelling ratio $\alpha$ & mesh size $\xi$ & VPT CP  & $\mu$(20$^\circ$C) /  & $\mu$(50$^\circ$C) / \\
& &  / nm  & / $^\circ$C & \textmu m\,cm\,V$^{-1}$\,s$^{-1}$ & \textmu m\,cm\,V$^{-1}$\,s$^{-1}$ \\
\hline
MG$_\text{C1,D0}$ & 87.5 $\pm$ 9.0 & 24.2 & 53 &  0.55$\pm$0.02 & 3.66$\pm$0.22\\
MG$_\text{C2,D0}$ & 27.9 $\pm$ 4.1 & 12.2 &  40&  1.01$\pm$0.02 & 5.57$\pm$0.14\\
MG$_\text{C1,D5}$  & 26.0 $\pm$ 4.8 & 16.4  & 43&  0.49$\pm$0.03 & 4.71$\pm$0.13\\
MG$_\text{C2,D5}$  & 18.4 $\pm$ 2.9 & 10.3 &  34&  0.28$\pm$0.02 & 5.45$\pm$0.06\\
MG$_\text{C2,D10}$  &  8.5 $\pm$1.7  &  7.9 &  32&  0.12$\pm$0.03& 4.30$\pm$0.07\\
MG$_\text{C2,D15}$ & 4.4 $\pm$ 0.3 &  6.5 & 27&  0.65$\pm$0.02 & 4.30$\pm$0.10\\
MG$_\text{C2,D20}$  & 6.6 $\pm$ 1.1\textsuperscript{a} & 7.3 \textsuperscript{a} & - \textsuperscript{b} &  1.44$\pm$0.02& 4.84$\pm$0.04 \\
\hline
\multicolumn{6}{l}{\footnotesize{\textsuperscript{a} Swelling ratio and mesh size were calculated with the hydrodynamic radius of the shrunken microgel at}} \\
\multicolumn{6}{l}{\footnotesize{$40^\circ \text{C}$ $R_\text{H}(40^\circ \text{C})$ due to the flocculation of the sample at higher temperatures. \textsuperscript{b} not available due to}} \\
\multicolumn{6}{l}{\footnotesize{flocculation of the sample. }}
\end{tabular}
\label{tab:microgels_characterization_bulk}
\end{table*}

\subsection{Mechanical properties of microgels at the solid surface}
\subsubsection{Height profiles of adsorbed microgel particles}
\begin{figure}[h!]
	\centering
	\includegraphics[width=8.4cm, height=7cm, keepaspectratio]{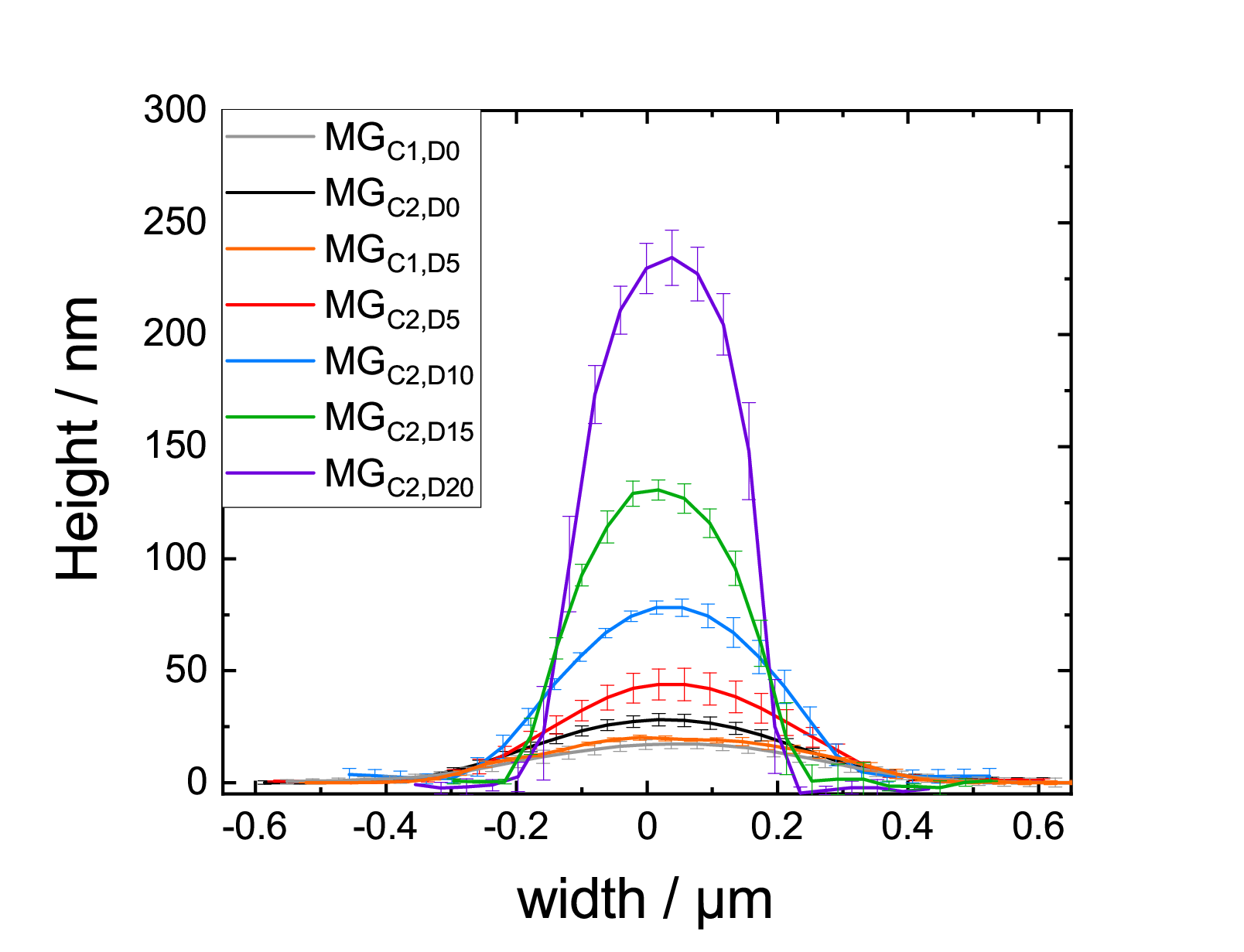}
	\caption{Height profile of microgels scanned by AFM in ambient conditions. Profiles are an average of 5 different microgel particles per sample.}
	\label{fig:dry height microgels}
\end{figure}

In fig.\,\ref{fig:dry height microgels} the height profiles of all the microgels adsorbed on silicon wafers in ambient conditions are shown, determined by AFM scanning as explained in sec.\,\ref{subsec:charactTechniques}\,\ref{subsubseq:Scanning}. The corresponding AFM scans can be found in SI, fig.\,S8. The value of the height-to-width ratio $h_\text{amb,AFM}$/$w_\text{amb,AFM}$ is given in table\,\ref{tab:microgels_characterization_surface}, which is calculated from the height $h_\text{amb,AFM}$ and width $w_\text{amb,AFM}$ of 5 different microgels per sample (values are given in the SI, table\,S1.)  \\
The $h_\text{amb,AFM}$/$w_\text{amb,AFM}$ ratio of the microgels increases with increasing amount of cross-linker (MG$_\text{C1,D0}$ compared to MG$_\text{C2,D0}$ and MG$_\text{C1,D5}$ compared to MG$_\text{C2,D5}$) and with an increasing amount of DMA from MG$_\text{C2,D0}$ to MG$_\text{C2,D20}$ at 20$^\circ$C. 

\begin{table}[h!]
\caption{Characterisation of microgels at the surface by AFM scanning in ambient conditions: height-to-width ratio ($h_\text{amb,AFM}$/$w_\text{amb,AFM}$) calculated from the height $h_\text{amb,AFM}$ and width $w_\text{amb,AFM}$ of 5 different microgels per sample (values are given in the SI, table\,S1.) }
\centering
\begin{tabular}{lc}
\hline
Microgel &  $h_\text{amb,AFM}$/$w_\text{amb,AFM}$ ratio\\
\hline
MG$_\text{C1,D0}$ & 0.014 $\pm$ 0.002\\
MG$_\text{C2,D0}$ & 0.026 $\pm$ 0.004\\
MG$_\text{C1,D5}$ & 0.020 $\pm$ 0.002\\
MG$_\text{C2,D5}$ & 0.049 $\pm$ 0.009\\
MG$_\text{C2,D10}$ & 0.110 $\pm$ 0.011\\
MG$_\text{C2,D15}$  & 0.278 $\pm$ 0.029\\
MG$_\text{C2,D20}$  & 0.540 $\pm$ 0.060\\
\hline
\end{tabular}
\label{tab:microgels_characterization_surface}
\end{table}

\subsubsection{Mechanical properties of adsorbed microgel particles} \label{subsubsec:MechanPropertiesResults}

\begin{figure*}[h!]
	\centering
	\includegraphics[width=17cm, height=17cm, keepaspectratio]{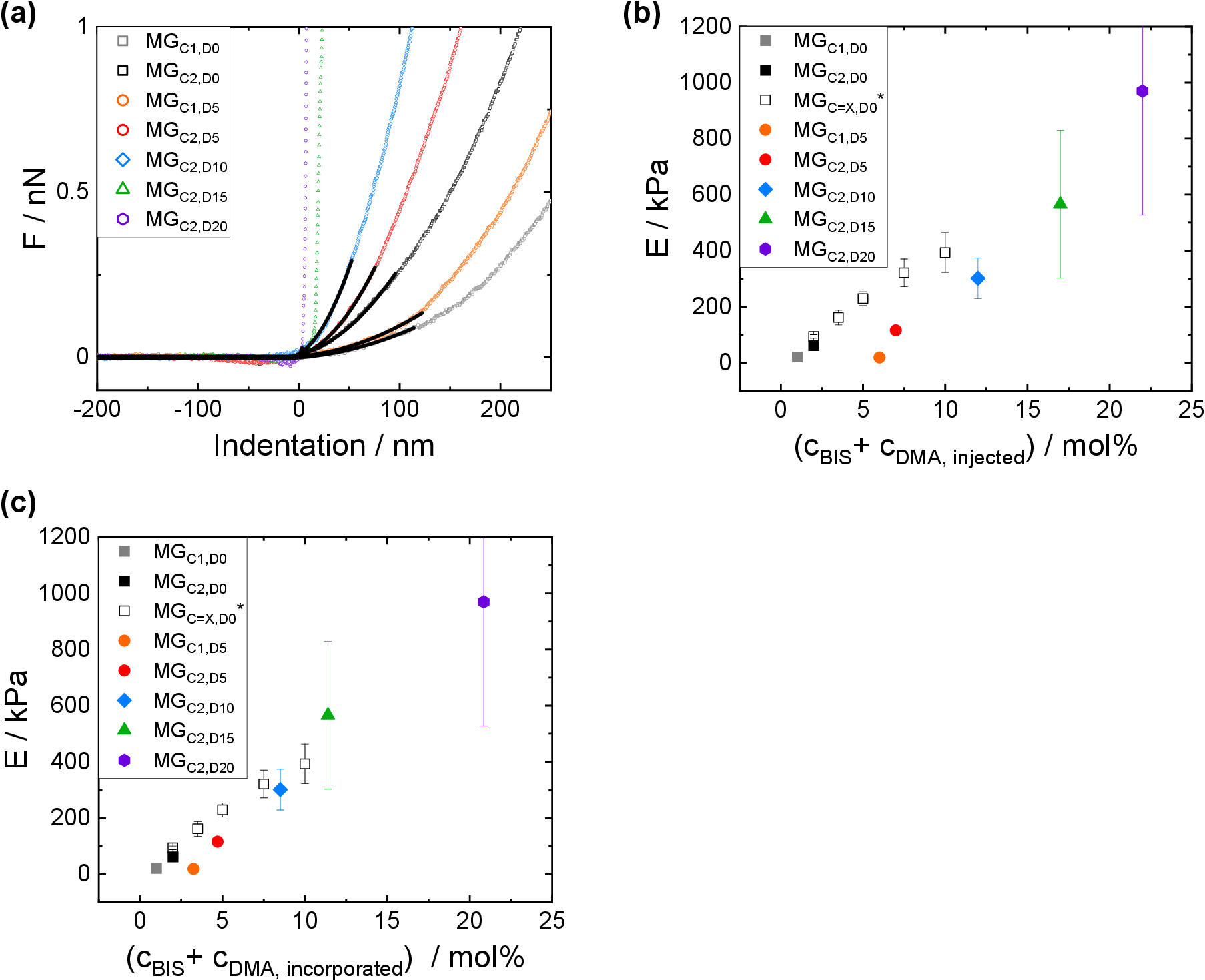}
	\caption{(a) Exemplary force curve for each type of microgel (open symbols) conducted at the centre of one particle together with the corresponding Hertzian fit (solid black line). (b) Averaged $E$ moduli from force curves conducted in the middle of 10 microgel particles of each type of microgel plotted over the sum of concentration of the cross-linker BIS $c_\text{BIS}$ and injected co-monomer DMA $c_\text{DMA, injected}$. (c) Averaged $E$ moduli from force curves conducted in the middle of 10 microgel particles of each type of microgel plotted over the sum of concentration of the cross-linker BIS $c_\text{BIS}$ and incorporated amount of DMA $c_\text{DMA, incorporated}$ determined by UV-vis and NMR spectroscopy \cite{Forg2022}. *Open symbols in (b) and (c) correspond to results obtained by Kühnhammer \textit{et al.} \cite{Keenhammer2022} for pure PNIPAM microgels without DMA with varying ($x$) amount of cross-linker BIS.}
	\label{fig:mechanicalIndentation}
\end{figure*}

Mechanical properties of the adsorbed microgels swollen in water have been studied with AFM force spectroscopy as explained in sec.\,\ref{subsec:charactTechniques}\,\ref{subsec:Indentation}. An exemplary force curve of each type of microgel measured in the centre of the respective microgel plotted together with the corresponding Hertzian fit is displayed in fig.\,\ref{fig:mechanicalIndentation}\,(a). All force curves used for the analysis of $E$ can be found in the SI, fig.\,S11. \\
The slope of the force curves increases with an increasing amount of cross-linker BIS and an increasing amount of co-monomer DMA. At high indentations ($>$\,300\,nm), the microgels cross-linked by 1\,mol\% BIS (MG$_\text{C1,D0}$ and MG$_\text{C1,D5}$) exhibit a vertical increase of force over indentation, indicating a complete penetration of the cantilever through the microgel down to the substrate (see SI; fig.\,S11\,(a)). For some force curves, negative forces are observed before reaching the actual contact point (see SI; fig.\,S11\,(b)). They are caused by attractive interactions between cantilever and tip just before the cantilever indents into the sample. The effect was most pronounced for the microgel with the highest amount of DMA MG$_\text{C2,D20}$ (see SI; fig.\,S11\,(b)). \\
The resulting $E$ moduli, from force curves conducted in the centre of 10 microgel particles of each type of microgel, over the sum of the concentrations of cross-linker BIS $c_\text{BIS}$ and injected co-monomer DMA $c_\text{DMA, injected}$ are plotted in fig.\,\ref{fig:mechanicalIndentation}\,(b). Resulting $E$ moduli are also shown in dependence of the sum of the concentrations of cross-linker BIS $c_\text{BIS}$ and incorporated co-monomer DMA $c_\text{DMA, incorporated}$ in fig.\,\ref{fig:mechanicalIndentation}\,(c), calculated as explained in sec.\,\ref{subsec:charactTechniques}\,\ref{subsec:UV-Vis } and \ref{subsubseq:NMR} by UV-vis and NMR spectroscopy. The results are shown together with results obtained by Kühnhammer \textit{et al.} \cite{Keenhammer2022} and show an increase of $E$ with an increase in cross-linker BIS amount for pure PNIPAM microgels. These PNIPAM microgels were synthesised with different cross-linker BIS amount in our group with a similar recipe. The excellent agreement between our data and the data from literature underlines that the $E$ modulus determination is valid. \\
The $E$ modulus also increases for an increasing amount of DMA equivalently to an increasing amount of cross-linker BIS. When $E$ is plotted over the sum of the concentration of BIS $c_\text{BIS}$ and the injected amount of DMA $c_\text{DMA, injected}$ in fig.\,\ref{fig:mechanicalIndentation}\,(b), the increase in $E$ moduli is shifted to lower $E$ moduli in comparison to the $E$ moduli of pure PNIPAM microgels. When $E$ is plotted over the sum of the concentration of BIS $c_\text{BIS}$ and the incorporated amount of DMA $c_\text{DMA, incorporated}$ in fig.\,\ref{fig:mechanicalIndentation}\,(c), the values of $E$ are shifted towards the values obtained for the pure PNIPAM microgels, but are still lower. \\
The error bars of $E$ enlarge for an increasing amount of cross-linker and DMA. The increase in error for increasing $E$ moduli is a well-known effect. Microgels of the same sample have been found to show some variation in heterogeneity in the centre of the particle, getting more pronounced for microgels with higher cross-linker contents \cite{Burmistrova2011, Burmistrova2011_2}. Moreover, the trigger point is set at a constant force. So for steeper force curves the number of data points is lower in a smaller indentation range \cite{Aufderhorst2018} as seen in fig.\ref{fig:mechanicalIndentation}\,(a). For the microgels, with the two highest concentrations of DMA (MG$_\text{C2,D15}$ and MG$_\text{C2,D20}$), $E$ moduli exhibit the largest errors. Therefore, a different approach was tried to fit the force curves of MG$_\text{C2,D15}$ and MG$_\text{C2,D20}$ based on the study from Aufderhorst-Roberts \textit{et al.} \cite{Aufderhorst2018} for force curves with steeper increases. The force curves were fitted with a constant force range instead of a constant indentation range (constant $y$-range instead of $x$-range). However, the maximum fit range, which could describe the force curve sufficiently was 100\,pN. 100\,pN is well above the noise of the force curves, but doesn't include significantly more data points than fitting with a constant indentation range of $0-40$\% for MG$_\text{C2,D15}$ and MG$_\text{C2,D20}$. Setting a larger fit range of 150 or 200\,pN resulted in deviances from fit and force curve, especially in the lower part of the force curve. Exemplary force curve of MG$_\text{C2,D20}$ with different fitting procedures are given in the SI, fig.\,S12. The resulting $E$ moduli are shown in the SI in fig.\,S13\,(b), but did not lead to a significant improvement of the data. For MG$_\text{C2,D15}$, both results are comparable, but the second approach resulted in slightly smaller errors. $E$ moduli for MG$_\text{C2,D20}$ deviate significantly for both methods. Fitting with a constant force range led to even larger errors. The trend that $E$ moduli increase for an increasing amount of DMA remains for both methods. 
 
\begin{figure*}[h!]
	\centering
	\includegraphics[width=17cm, height=7cm, keepaspectratio]{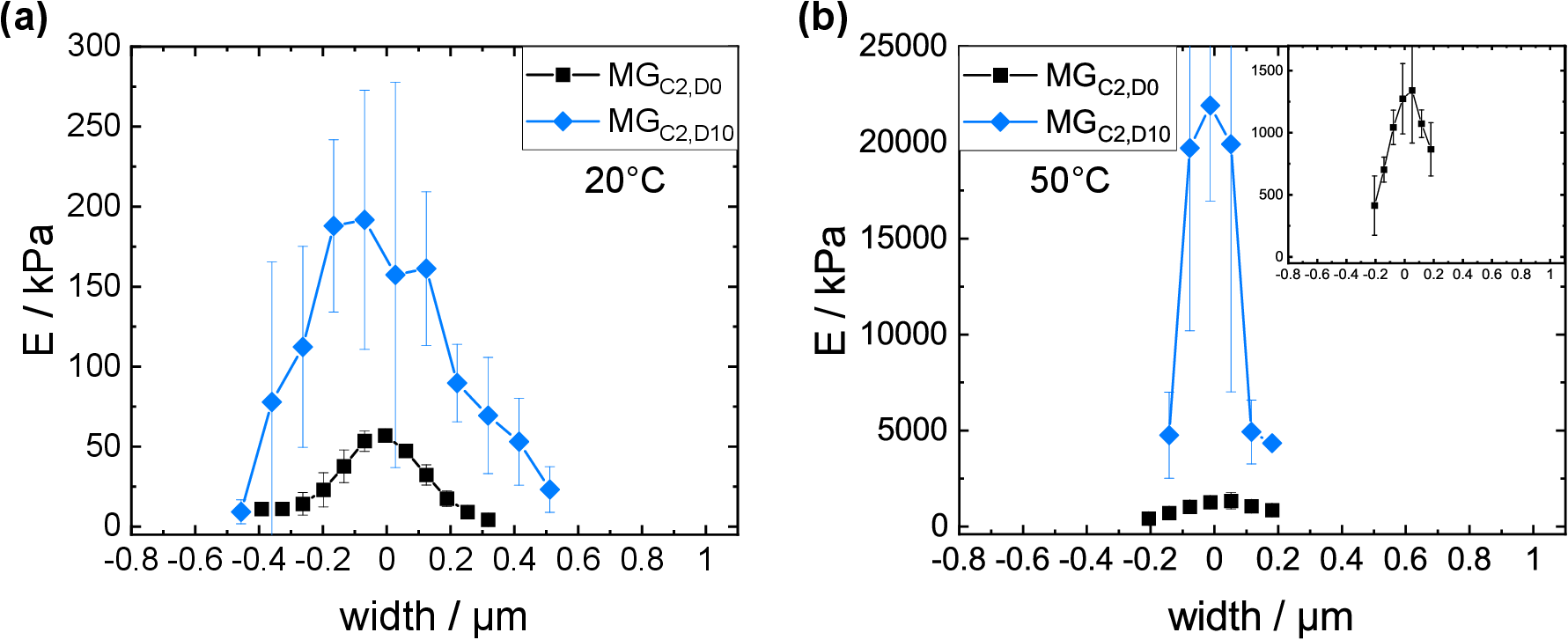}
	\caption{$E$ modulus distribution over particle cross-section for the two microgels MG$_\text{C2,D0}$ and MG$_\text{C2,D10}$ calculated as an average from 5 different microgel particles at (a) 20$^\circ$C and (b) 50$^\circ$C. }
	\label{fig:E_distribution}
\end{figure*}

\begin{figure*}[h!]
\centering
\includegraphics[width=15cm, height=15cm, keepaspectratio]{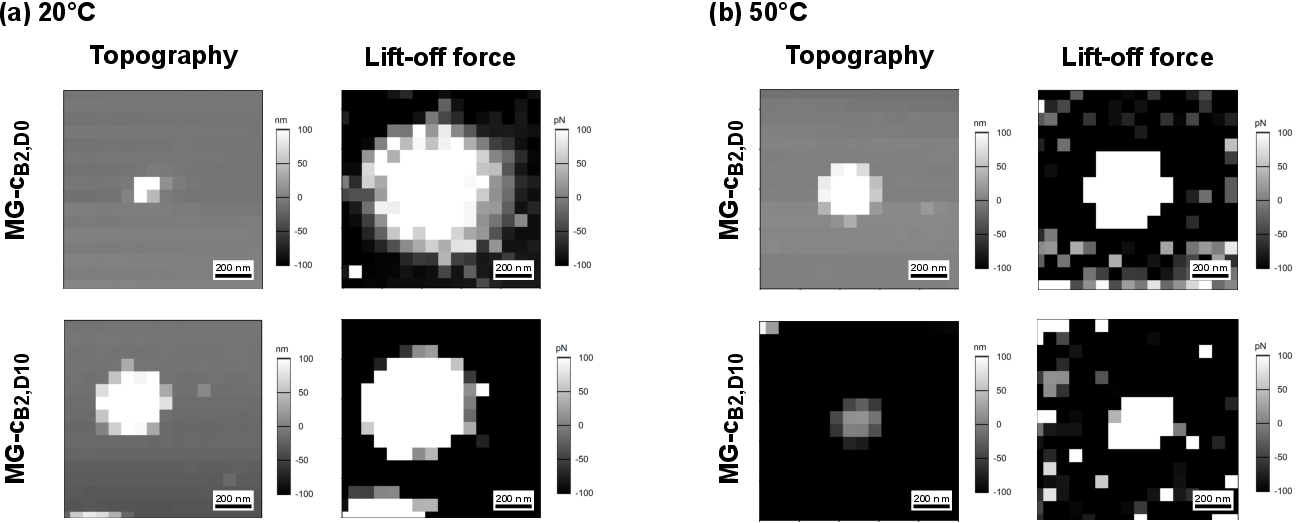}
\caption{Exemplary force maps providing information about the topography and lift-off force at (a) 20$^\circ$C and (b) 50$^\circ$C of MG$_\text{C2,D0}$ (top) and MG$_\text{C2,D10}$ (bottom).}
\label{fig:ForceMaps_example}
\end{figure*}

\noindent The $E$ modulus profile for two exemplary microgels MG$_\text{C2,D0}$ and MG$_\text{C2,D10}$ is shown in fig.\,\ref{fig:E_distribution} at (a) $20^\circ$C and (b) 50$^\circ$C. The errors of MG$_\text{C2,D10}$ can be explained as before. Comparing the pure PNIPAM microgel MG$_\text{C2,D0}$ and the P(NIPAM-co-DMA) microgel MG$_\text{C2,D10}$, both show the highest $E$ modulus in the centre, decreasing towards the outer shell. Overall, the P(NIPAM-co-DMA) microgel exhibits higher $E$ moduli than the pure PNIPAM microgel with the same BIS amount. When heating from 20$^\circ$C to 50$^\circ$C above the VPT(T), $E$ moduli increase significantly for both types of microgels. While MG$_\text{C2,D0}$ possesses a stiffening factor $E(50^\circ\text{C})/E(20^\circ\text{C})$ of around 20 in the centre of the microgel particle upon heating, MG$_\text{C2,D10}$ stiffens by more than 2 orders of magnitude. $E$ decreases again from the centre towards the outer regions of the microgels. The particle cross-section width decreases from 20$^\circ$C to 50$^\circ$C for both microgels in the same order similarly to the microgels scanned in ambient condition. \\
Fig.\,\ref{fig:ForceMaps_example} depicts the topography and lift-off map obtained from the force mapping tool (see sec.\,\ref{subsec:charactTechniques}\,\ref{subsec:Indentation}) for the two microgels MG$_\text{C2,D0}$ and MG$_\text{C2,D10}$ at (a) 20$^\circ$C and (b) 50$^\circ$C. \\
At 20$^\circ$C interestingly, the pure PNIPAM microgel particles (MG$_\text{C2,D0}$) are barely visible in the topography map while clearly identifiable in the corresponding lift-off map. In contrast to that, the P(NIPAM-co-DMA) microgel particles (MG$_\text{C2,D10}$) are visible in both maps, although exhibiting slightly larger radii in the lift-off map. \\
At 50$^\circ$C both microgels are visible in the topography map and lift-off map. Both microgels possess slightly larger radii in the lift-off map compared to the topography map, while the difference is larger for the pure PNIPAM microgel MG$_\text{C2,D0}$.

\section{Discussion}\label{Discussion Neu}
\subsection*{Swelling behaviour of microgels in bulk solutions}
The hydrodynamic radius $R_\text{H}(20^\circ \text{C})$ of the swollen microgel is known to decrease for an increase in cross-linker BIS amount \cite{Karg2013,Senff2000,Schlattmann2022, Kratz1998, Thiele2021}, since a higher cross-linked microgel structure is less capable of swelling (reflected by smaller mesh sizes and smaller swelling ratios). For the hydrodynamic radius $R_\text{H}(50^\circ \text{C})$ of the shrunken microgel, an increase in size (larger rigid core) is expected for an increasing BIS content \cite{Schlattmann2022, Friesen2021} - BIS has a higher reaction kinetics than NIPAM, therefore mainly accumulating in the microgel's centre \cite{Wu1994, Acciaro2011, Forg2022}. Nevertheless, in a lot of studies, no (systematic) effect was observed \cite{Karg2013, Thiele2021}. For the lower cross-linked pure PNIPAM microgels MG$_\text{C1,D0}$ a lower shrunken hydrodynamic $R_\text{H}(50^\circ \text{C})$ is obtained than for the higher cross-linked MG$_\text{C2,D0}$ (around $38$\,nm difference), indicating the expected smaller core and leading to significantly larger mesh sizes. However, there was no difference found for the differently cross-linked P(NIPAM-co-DMA) microgels. The values from literature \cite{Thiele2021, Lehmann2019, Witt2019} for an increase from 1 to 2\,mol\% cross-linker amount are in agreement with our results of hydrodynamic radii, swelling ratios and mesh sizes. \\
DMA is known to cross-link structures as well \cite{Forg2022, Xue2017, Yang2015}, promoting the cohesive strength of catechol-based materials. This results in a decrease of the hydrodynamic radius $R_\text{H}(20^\circ \text{C})$ of the swollen microgel in our study, same as for an increasing cross-linker amount \cite{Karg2013,Senff2000,Schlattmann2022, Kratz1998, Thiele2021}. This systematic effect, to the best of our knowledge, has not been observed for DMA cross-linked microgels so far. $R_\text{H}(50^\circ \text{C})$ of the shrunken microgel stays constant for different amounts of DMA, indicating that the core size stays unaffected by the addition of DMA. Consequently, microgels with increasing DMA are less capable to swell, underlined by the decreasing mesh size $\xi$ and swelling ratio $\alpha$. This is in good agreement with the results from the group of Xue \textit{et al.} \cite{Xue2017}. \\
It should be noted that the given values of mesh sizes $\xi$ in table\,\ref{tab:microgels_characterization_bulk} for P(NIPAM-co-DMA) microgels are maximum values, as we assume that only BIS acts as cross-linker (see eq.\,\ref{eq:mesh size}). Since DMA acts (at least partially) as an additional cross-linker, mesh sizes should be smaller. Since it is unclear how much of DMA acts as a cross-linker due to the assumption of free catechol moieties, and we are only interested in the trend of mesh sizes, eq.\,\ref{eq:mesh size} was not adjusted.\\
BIS is more hydrophilic than NIPAM and therefore increases the microgel's hydrophilicity in general, theoretically increasing the VPTT. When heating a microgel above its VPTT, hydrophobic interactions get dominant, which are now compensated by hydrophilic interactions caused by the cross-linker BIS. However, in literature, none or only slight effects are observed \cite{Karg2013, Thiele2021, Senff2000} and the VPTT remains constant at about 32$^\circ$C irrespective of the BIS content. This is in agreement with our data, where the VPTT is similar for the pure PNIPAM microgels with 1\,mol\% and 2\,mol\% cross-linker BIS (MG$_\text{C1,D0}$ compared to MG$_\text{C2,D0}$). However, from MG$_\text{C1,D5}$ to MG$_\text{C2,D5}$ the VPTT decreases, which is in contrast to the expectation. This could be a result of the fluctuating hydrodynamic radii of MG$_\text{C1,D5}$ around the VPTT. Due to conformational changes of the microgel particles around the VPTT, fluctuations in radii could be possible. That supports that the VPT CT is more reliable in our case. \\
In contrast to BIS, DMA is more hydrophobic than NIPAM, and hydrophobic groups are known to decrease the VPTT \cite{Ryo1995, Garcia2019, Vatankhah2014, Feil1993, Marcisz2017} due to an enhancement of hydrophobic interactions. DMA has the tendency to press water out of the microgel network. Especially the VPT onset was shown to decrease significantly \cite{Garcia2019}. The present study shows that the incorporation of DMA leads to a slight decrease of the VPTT (point of deflection), which is in good agreement with literature \cite{Garcia2019, Vatankhah2014, Marcisz2017}. A significant decrease of the VPT CT is observed, which leads to a sharpening of the transition in contrast to literature \cite{Garcia2019, Vatankhah2014}. In literature, a broadening of the transition is described, explained by the statistical distribution of the hydrophobic groups generating local hydrophobic inhomogeneities. We assume that the observed sharp VPT in our study is a consequence of our synthesis protocol, differing from other literature procedures \cite{Garcia2019, Vatankhah2014}. In the present study, DMA is added later during the reaction \cite{Forg2022}, when BIS is already fully consumed. This might lead to a different distribution of the more hydrophobic DMA within the microgel network. The sharp VPT is favourable for biomedical applications \cite{Li2022}, because it results in a faster and more controlled release of i.e. incorporated drugs. \\
All microgels, both at 20$^\circ$ and 50$^\circ$C, possess a positive electrophoretic mobility and therefore a positive surface charge since they were synthesised with the positively charged initiator AAPH. Assuming a constant number of charges, the collapsed microgels at 50$^\circ$C with a smaller particle surface have a higher surface charge density than the swollen microgels \cite{Burmistrova2010}, which is confirmed by our data (table\,\ref{tab:microgels_characterization_bulk}). Values are similar for all the microgels, so DMA has no effect on the surface charge. This leads to the conclusion, that the flocculation of the microgel dispersion of MG$_\text{C2,D20}$ with the highest DMA amount is not caused by a lack of electrostatic stabilization at temperatures above the VPT \cite{Senff2000, Schmidt2008}, but probably caused by the increased hydrophobicity due to the incorporation of DMA itself.

\begin{figure}[H]
	\centering
	\includegraphics[width=8.4cm, height=10cm, keepaspectratio]{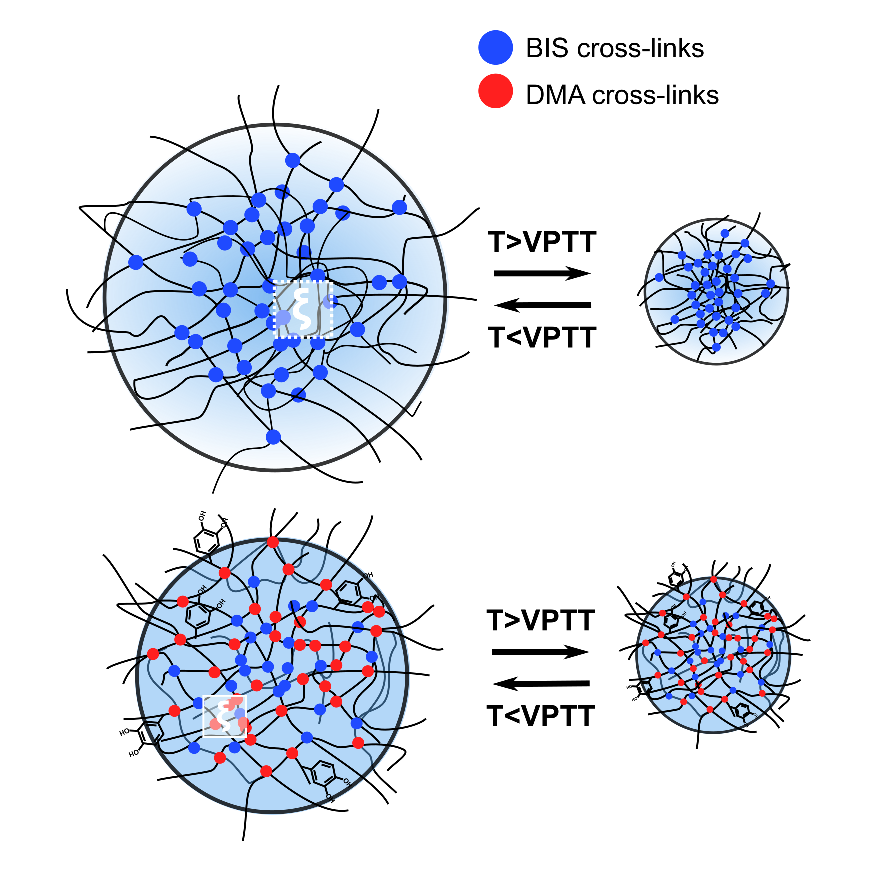}
	\caption{Summarising scheme of swelling behaviour of pure PNIPAM and P(NIPAM-co-DMA) microgels in bulk solution. For a more detailed description, see text.}
	\label{fig:scheme_bulk}
\end{figure}

\noindent To \textit{summarise}, P(NIPAM-co-DMA) microgels exhibit reduced sizes in the swollen state when compared to pure PNIPAM microgels. The size difference arises from DMA's role as an additional cross-linker, reducing the swelling capacity of P(NIPAM-co-DMA) microgels, as illustrated in fig.\,\ref{fig:scheme_bulk}. The smaller mesh sizes and swelling ratios for P(NIPAM-co-DMA) further underline this behaviour. Both types of microgels collapse to similar sizes above the VPTT, which can be attributed to similar core sizes. However, due to the incorporation of DMA with an increased hydrophobicity in comparison to NIPAM, the VPT(T) of P(NIPAM-co-DMA) microgels is shifted to lower temperatures and results in sharper phase transitions. 

\subsection*{Mechanical properties of microgels at the solid surface}
Microgels adsorb to silicon wafers due to two reasons: the positively charged microgels are electrostatically attracted by the negative charged silicon wafers. However, the adhesive properties of DMA might also contribute to the adsorption of P(NIPAM-co-DMA) microgels, which should be kept in mind.

\subsubsection{Height profiles of adsorbed microgel particles}
The $h_\text{amb,AFM}$/$w_\text{amb,AFM}$ ratio increases with increasing amount of both, cross-linker BIS and co-monomer DMA. This arises from an augmentation in the microgel's height $h_\text{amb,AFM}$ coupled with a reduction in its width $w_\text{amb,AFM}$, consistent with findings in literature \cite{Aufderhorst2018, Keenhammer2022, Burmistrova2011}. As the microgels become stiffer, due to a higher cross-linker content induced by both BIS and DMA, their decreased flexibility results in a reduced deformability when adsorbed at the silicon wafer - they protrude more from the surface. Additionally, differences in the water uptake of the microgel particles after spin coating might further enhance this effect. Here, the higher hydrophobicity of DMA might lower the water content in the microgel network structure.

\subsubsection{Mechanical properties of adsorbed microgel particles}
The sole force curves of the different microgels already show the increasing stiffness of microgels with increasing amount of cross-linker BIS and co-monomer DMA. For some force curves, attractive forces prior to the contact point are observed, which become more pronounced for higher cross-linker amounts \cite{Aufderhorst2018, Keenhammer2022}. They are likely attributed to the higher polymer concentration at the surface for higher cross-linker densities and, therefore, higher surface forces. They appear due to attractive forces between the cantilever tip and the microgel, overcoming the cantilever's spring constant (out of equilibrium) \cite{Butt2005}. The effect is most pronounced for the microgel with the highest DMA amount MG$_\text{C2,D20}$, possessing the highest stiffness. Nevertheless, not only it's high polymer density, but also the adhesive abilities of DMA itself might contribute to this phenomenon. \\
The obtained values for the $E$ modulus are in good agreement with existing literature, where $E$ moduli from 17 to 600\,kPa \cite{Burmistrova2011, Witte2021, Keenhammer2022, Aufderhorst2018, Backes2015, Schmidt2010, Witte2019} were reported for PNIPAM microgels with cross-linker BIS. For microgels containing around 1-2\,mol\% cross-linker BIS, values between 17-100\,kPa \cite{Burmistrova2011, Aufderhorst2018, Keenhammer2022} were shown. Higher cross-linked PNIPAM microgels with around 10\,mol\% cross-linker possess $E$ moduli between 312-600\,kPa \cite{Burmistrova2011, Witte2021, Keenhammer2022, Witte2019}. In the case of P(NIPAM-co-DMA) microgels, the $E$ modulus increases in consequence of an increasing DMA content, indicating DMA's role as cross-linker similar to BIS. This is in good agreement with the work of Yang \textit{et al.} \cite{Yang2015}, who found an increase in stiffening for DMA cross-linked polymers. \\
Plotting the $E$ moduli of P(NIPAM-co-DMA) microgels against the sum of concentration of BIS $c_\text{BIS}$ and \textit{incorporated} DMA $c_\text{DMA, incorporated}$, (calculated by UV-vis standard addition and NMR spectroscopy), $E$ moduli are shifted towards the $E$ moduli of pure PNIPAM microgels. However, the $E$ moduli for microgels with DMA remain lower than for pure ones. This suggests that some free dopamine moieties are still present in the microgel network structure, not serving as cross-linker, as supported by literature. The group of Xue \textit{et al.}\cite{Xue2017} found out that the catechol group of DMA forms interchains with another polymer chain - either a C-O or a C-C bond - but some free catechol moieties are still present. These free dopamine moieties can potentially support the adhesiveness of these materials in the future.\\
As for pure PNIPAM microgels, the $E$ modulus decreases from the centre to the outer periphery of the microgel particles exemplary shown for the P(NIPAM-co-DMA) microgel MG$_\text{C2,D10}$, confirming a heterogeneous microgel structure with a stiffer core and a fluffier shell / dangling ends. $E$ moduli increase significantly when heating upon the VPTT, consistent with prior research \cite{Burmistrova2011, Harmon2003_2, Junk2010, Schmidt2010, Matzelle2003}. Additionally, the cross-section of the particles undergoes changes, with a decrease in width and an increase in height upon heating \cite{Burmistrova2011, Burmistrova2011_2, Schmidt2010}, equivalently as for an increase of cross-linker amount. The stiffening factor $E(50^\circ\text{C})/E(20^\circ\text{C})$ of 20 in the centre of the pure PNIPAM microgel and of 2 orders of magnitude in the centre of the P(NIPAM-co-DMA) microgel is in the same order as found in literature \cite{Matzelle2003, Harmon2003_2}, where values of 42-110 have been reported, increasing from low to high cross-linked PNIPAM. It is attributed to the higher cross-linker density \cite{Matzelle2003} and a decrease in elasticity of the polymer chains in between two cross-links upon collapsing \cite{Junk2010}. This highlights the likeness of P(NIPAM-co-DMA) microgels with microgels of high cross-linker content, but without DMA. \\
The studied topography and lift-off maps of MG$_\text{C2,D0}$ and MG$_\text{C2,D10}$ support the results received by AFM scanning of the microgels in ambient condition and the results of the $E$ modulus distribution: MG$_\text{C2,D0}$ particles are barely detectable in the topography map at 20$^\circ$C due to their strong deformation when adsorbed at silicon wafers. Specifically, adsorbed PNIPAM microgels are known to form the so-called "fried egg" structure \cite{Aufderhorst2018, Mourran2016}, where the dangling ends spread at the surface. Therefore, they are not detectable in the topography map due to their small dimensions. MG$_\text{C2,D10}$ is however clearly detectable, since it protrudes more from the surface due to a decreased flexibility, and due to an overall stiffening of the microgel upon the incorporation of DMA (see $E$ modulus distribution). Both types of particles can be clearly identified in the lift-off map, because the cantilever adheres to the particles, including their dangling ends, during the retraction. The microgel particles exhibit in both cases a larger diameter in the lift-off map than in the topography map, due to the heterogeneous nature of both microgel types, in accordance with the results obtained for the $E$ modulus distribution.\\
Upon heating to 50$^\circ$C, both particles increase in stiffness, therefore the pure PNIPAM microgel MG$_\text{C2,D0}$ is now visible in the topography map as well. Both microgels possess slightly larger radii in the lift-off map compared to the topography map, while the difference is larger for the pure PNIPAM microgel MG$_\text{C2,D0}$, which agrees well with the data of $E$ moduli across the microgel's cross-section: Even if the pure PNIPAM microgel MG$_\text{C2,D0}$ stiffens from 20$^\circ$C to 50$^\circ$C, the P(NIPAM-co-DMA) microgel MG$_\text{C2,D10}$ stiffens much more. \\

\begin{figure}[h!]
	\centering
	\includegraphics[width=8.4cm, height=7cm, keepaspectratio]{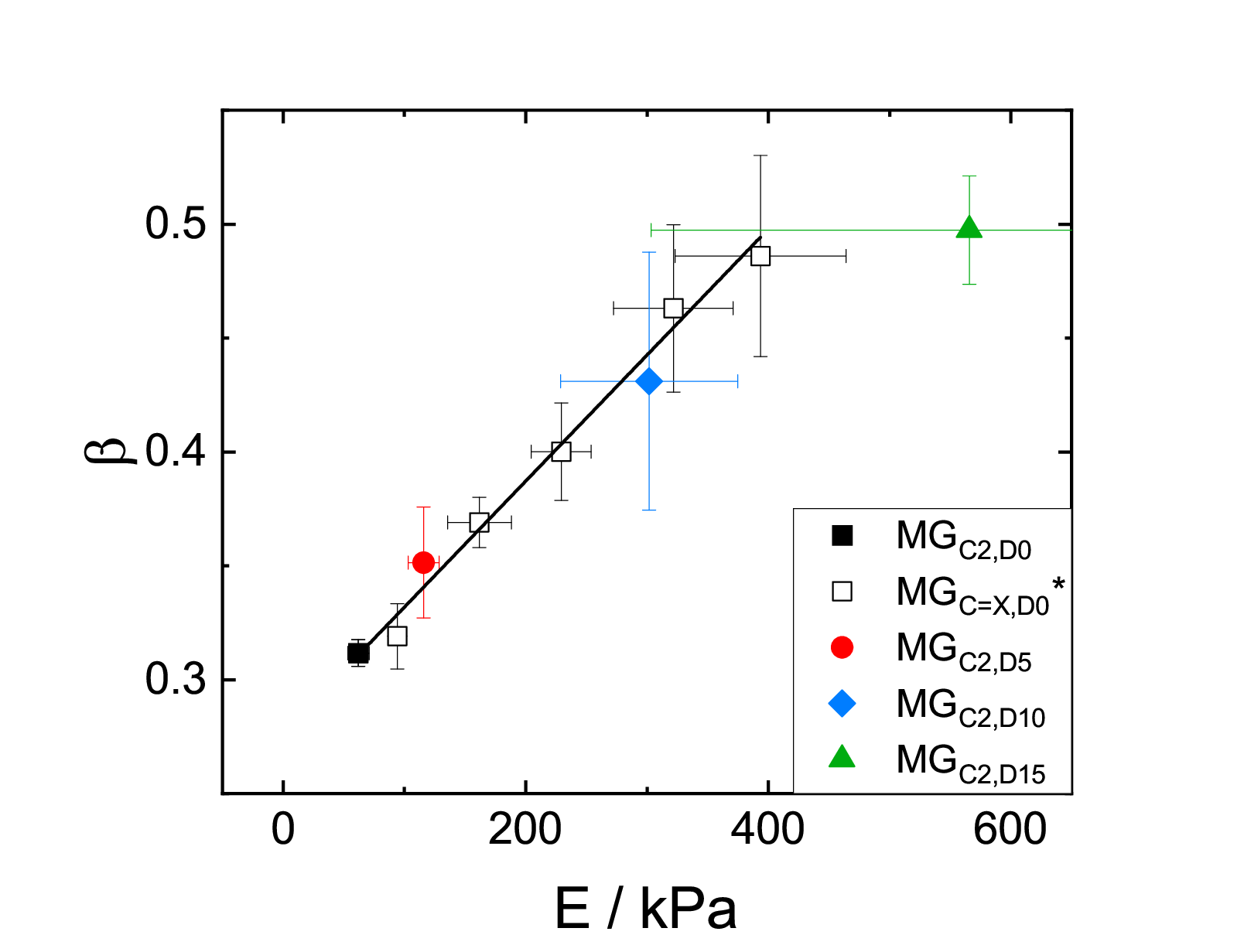}
	\caption{Affine network factor $\beta$ (calculated as described in sec.\,\ref{subsubseq:DLS} by eq.\,\ref{eq:affine}) and $E$ moduli for an increasing amount of sole BIS cross-links and increasing amount of DMA cross-links (from MG$_\text{C2,D0}$ compared to MG$_\text{C2,D15}$). *Open symbols correspond to results obtained by Kühnhammer \textit{et al.} \cite{Keenhammer2022} for pure PNIPAM microgels without DMA with varying ($x$) amount of cross-linker BIS, which can be fitted by a linear function.}
	\label{fig:affine}
\end{figure}

\begin{figure}[h!]
	\centering
	\includegraphics[width=8.4cm, height=10cm, keepaspectratio]{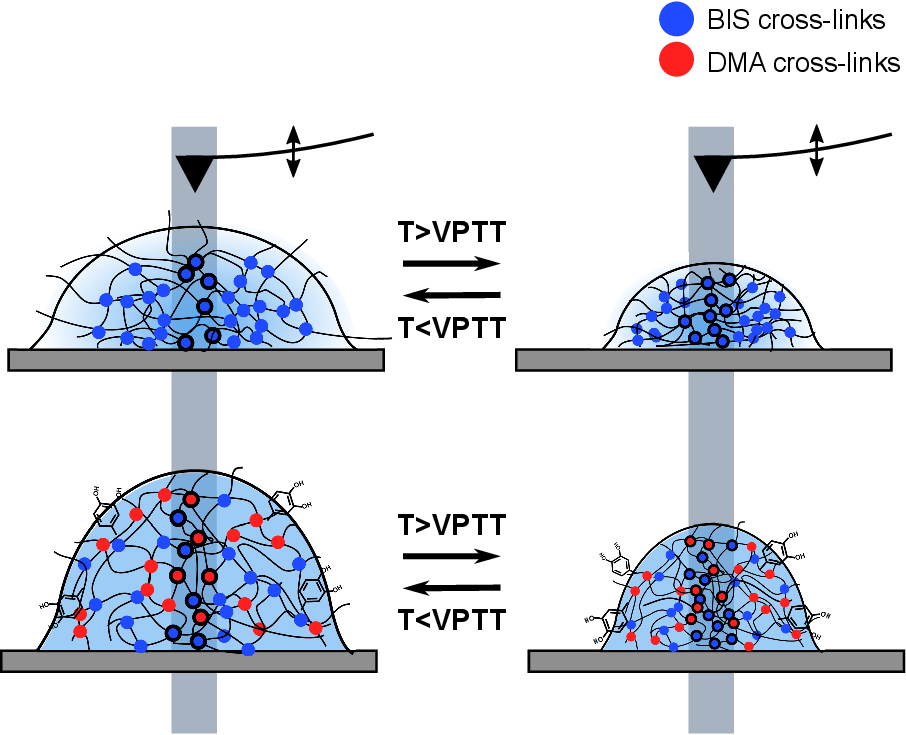}
	\caption{Summarising scheme of pure PNIPAM and P(NIPAM-co-DMA) microgels at the surface, measured by AFM force spectroscopy: the cantilever indents into the adsorbed microgel particle at a defined position, compressing the polymer including its cross-links (illustrated by grey area) and then retracts. For a more detailed description, see text.}
	\label{fig:scheme_surface}
\end{figure}

\noindent The affine network factor $\beta$, calculated as described in sec.\,\ref{subsec:charactTechniques}\,\ref{subsubseq:DLS} by eq.\,\ref{eq:affine}, is shown in dependency of the microgel's $E$ moduli in fig.\,\ref{fig:affine}. According to the Flory-Rehner theory the $E$ modulus of the microgels is linearly correlated with $\beta$, which was experimentally proven \cite{Keenhammer2022}. \\
For pure PNIPAM microgels a linear dependency is confirmed with increasing cross-linker BIS amount, including data obtained from Kühnhammer \textit{et al.} \cite{Keenhammer2022} for pure PNIPAM microgels without DMA.\footnote{Results of MG$_\text{C1,D0}$ with 1\,mol\% cross-linker where excluded due to its significantly larger swelling ratio.} P(NIPAM-co-DMA) microgels with an increasing amount of DMA (from MG$_\text{C2,D0}$ compared to MG$_\text{C2,D10}$) show the same linear trend in agreement with data for pure PNIPAM microgels and fall on the same master curve.  MG$_\text{C2,D15}$, however, deviates from the linear dependency. Even if its $E$ modulus shows larger error bars, due to the smaller number of data points (i.e. small indentations), the deviation is distinct.\footnote{MG$_\text{C2,D20}$ was excluded due to the flocculation of the sample at higher temperatures.} DMA surprisingly decreases the affine network factor $\beta$ more than the cross-linker BIS, even when possessing a higher hydrophobicity than NIPAM. Catechols can interact with each other by a variety of mechanism, such as hydrogen bonding, covalent cross-linking, metal coordination or $\pi-\pi$ stacking \cite{Li2019, Saiz-Poseu2019, Zhang2020, Yang2014}. Due to the small mesh sizes of the microgel, catechol groups of DMA might interact with each other and consequently suppress the collapse of the microgels, while increasing the stiffness.\\
\\
To \textit{summarise:} P(NIPAM-co-DMA) microgels have a higher effective cross-linker density than pure PNIPAM microgels, due to the additional cross-linking of DMA, also in the shell of the microgel particles. Therefore, the cantilever tip compresses more cross-links while indenting as illustrated in fig.\,\ref{fig:scheme_surface}. As a consequence, the microgels protrude more from the surface and $E$ moduli increase. This enables tuning the $E$ modulus by incorporating DMA, making it more applicable for biomedical applications, while further enabling more advantages arising from the versatility of free catechol groups. Upon shrinking above the VPTT, both types of microgels collapse in size (lateral and horizontal), which leads to a significant stiffening of the microgels. P(NIPAM-co-DMA) microgels with an increasing DMA amount show the same linear trend of the affine network factor $\beta$ and $E$ moduli than pure PNIPAM microgels for an increasing BIS amount, but only until a DMA percentage of 10\,mol\%. For a DMA amount of 15\,mol\%, DMA surprisingly hinders the shrinking of the microgel in a greater tendency as expected for a cross-linker like BIS, while still exhibiting a high stiffness, possibly attributed to catechol interactions within the microgel network. 

\section{Conclusion}
This present study addresses the influence of dopamine methacrylamide (DMA) as a co-monomer on the behaviour of PNIPAM microgels, particularly focusing on their swelling behaviour in bulk solution and nanomechanical properties after adsorption at the solid surface. Temperature-dependent dynamic light scattering (DLS) measurements show a decrease in the swelling ability when incorporating DMA, since DMA acts as an additional cross-linker. The volume phase transition (temperature) (VPT(T)) is shifted to smaller temperatures as expected for the introduction of a co-monomer with a higher hydrophobicity than NIPAM like DMA, while showing a sharpening of the VPT. Atomic force microscopy (AFM) measurements performed on adsorbed P(NIPAM-co-DMA) microgels show an increase in $E$ moduli, and therefore in stiffness, as the DMA content increases. This is similar as shown for pure PNIPAM microgels with increasing BIS amount, confirming the role of DMA as a cross-linking agent. $E$ moduli increase over the whole microgel's cross-section when comparing P(NIPAM-co-DMA) microgels to pure PNIPAM microgels without DMA. For both microgels, the smallest $E$ moduli are found at the outer regions of the microgel particles, verifying the microgels' heterogeneous character. Nevertheless, a stiffening of the shell is obtained by the incorporation of DMA. Temperature-dependent AFM measurements below and above the VPTT show a significant stiffening of the microgels, attributed to a higher cross-linker and polymer density in the collapsed state. The affine network parameter $\beta$ is proportional to the $E$ modulus of the microgels and falls on the same master curve for both types of microgels. Nonetheless, a large DMA amount of 15\,mol\% suppresses the shrinkage of the microgel, while maintaining large $E$ moduli, potentially owing to catechol interactions within the network. The results enable the possibility to design catechol-containing PNIPAM microgel systems customised for specific applications, especially by controlling their cohesive strength. Free catechol groups incorporated in the microgel network might further promote the applicability of these systems in the future by increasing the adhesion of these systems.

\section*{Author Information}\label{Author Info}
\textbf{Author Information} \\
* E-Mail: klitzing@smi.tu-darmstadt.de \\
\textbf{Author Contributions}
All authors have given approval to the final version of the manuscript.  \\
\textbf{Notes} \\
The authors declare no competing financial interest.

\section*{Acknowledgement}\label{Acknowledgement}
The authors thank Cassia Lux from the group Soft Matter at Interfaces, Physics department (TU Darmstadt) for help with the manuscript. The authors greatly thank the support by the mass spectrometry core facility team of the Chemistry Department (TU Darmstadt) for measurements of the ESI/APCI spectra and the German Research Foundation (DFG) through grant no INST 163/444-1 FUGG (QTOF MS). We greatfully acknowledge the NMR team of the Chemistry Department (TU Darmstadt) for providing the measurements and for the corresponding discussions. We gratefully acknowledge the funding of the Deutsche Forschungsgemeinschaft via a Sino-German research project (DFG, German Research Foundation; Grant number 392298985) and of the federal state of Hesse (Landes-Offensive zur Entwicklung wissenschaftlich-ökonomischer Exzellenz, shortly LOEWE) within the project "Flow for Life". \\

\section*{Associated Content}\label{Associated Content}
\textbf{Supporting Information} \\
The following files are available free of charge. $^1$H and $^{13}$C NMR spectra of exemplary DMA batch in DMSO-d$_6$; Relative monomer consumptions $cc_0^{-1}$ of reactants NIPAM, BIS and DMA obtained by MS; UV-vis standard addition calibration curves of P(NIPAM-co-DMA) microgels; $^1$H NMR spectra of P(NIPAM-co-DMA) in DMSO-d$_6$ and methanol-d$_4$; swelling curves of microgels fitted by Boltzmann function; swelling ratio $\alpha$ and mesh size $\xi$ plotted as a function of the sum of BIS and injected DMA concentration;  AFM scans of microgels at ambient conditions at 20$^\circ$C; height $h_\text{amb,AFM}$ and width $w_\text{amb,AFM}$ values of microgels scanned in ambient condition at 20$^\circ$C, derived from the AFM scans; Force curves received for BL-AC40 TS cantilever on a pure PNIPAM microgel sample in dependence of different trigger points and corresponding $E$ moduli; Force curves used for the analysis of $E$, measured with a BL-AC40 TS cantilever for pure PNIPAM microgel MG$_\text{C2,D0}$; force curves used for the analysis of $E$ for the different microgels; Exemplary force curve of MG$_\text{C2,D20}$ fitted by different methods; $E$ moduli fitted with constant force range in comparison to moduli calculated with constant indentation range for MG$_\text{C2,D15}$ and MG$_\text{C2,D20}$ \\

\bibliography{Bib_finalversion.bib}

\providecommand{\latin}[1]{#1}
\makeatletter
\providecommand{\doi}
  {\begingroup\let\do\@makeother\dospecials
  \catcode`\{=1 \catcode`\}=2 \doi@aux}
\providecommand{\doi@aux}[1]{\endgroup\texttt{#1}}
\makeatother
\providecommand*\mcitethebibliography{\thebibliography}
\csname @ifundefined\endcsname{endmcitethebibliography}
  {\let\endmcitethebibliography\endthebibliography}{}
\begin{mcitethebibliography}{90}
\providecommand*\natexlab[1]{#1}
\providecommand*\mciteSetBstSublistMode[1]{}
\providecommand*\mciteSetBstMaxWidthForm[2]{}
\providecommand*\mciteBstWouldAddEndPuncttrue
  {\def\EndOfBibitem{\unskip.}}
\providecommand*\mciteBstWouldAddEndPunctfalse
  {\let\EndOfBibitem\relax}
\providecommand*\mciteSetBstMidEndSepPunct[3]{}
\providecommand*\mciteSetBstSublistLabelBeginEnd[3]{}
\providecommand*\EndOfBibitem{}
\mciteSetBstSublistMode{f}
\mciteSetBstMaxWidthForm{subitem}{(\alph{mcitesubitemcount})}
\mciteSetBstSublistLabelBeginEnd
  {\mcitemaxwidthsubitemform\space}
  {\relax}
  {\relax}

\bibitem[McNaught and Wilkinson(1997)McNaught, and Wilkinson]{IUPAC1997}
McNaught,~A.; Wilkinson,~A. \emph{IUPAC. Compendium of Chemical Terminology,
  2nd ed. (the Gold Book)}; Blackwell Scientific Publications, Oxford,
  1997\relax
\mciteBstWouldAddEndPuncttrue
\mciteSetBstMidEndSepPunct{\mcitedefaultmidpunct}
{\mcitedefaultendpunct}{\mcitedefaultseppunct}\relax
\EndOfBibitem
\bibitem[Liu \latin{et~al.}(2015)Liu, Zheng, Poh, Machens, and
  Schilling]{Liu2015}
Liu,~J.; Zheng,~H.; Poh,~P. S.~P.; Machens,~H.-G.; Schilling,~A.~F. Hydrogels
  for Engineering of Perfusable Vascular Networks. \emph{International Journal
  of Molecular Sciences} \textbf{2015}, \emph{16}, 15997--16016\relax
\mciteBstWouldAddEndPuncttrue
\mciteSetBstMidEndSepPunct{\mcitedefaultmidpunct}
{\mcitedefaultendpunct}{\mcitedefaultseppunct}\relax
\EndOfBibitem
\bibitem[Chai \latin{et~al.}(2017)Chai, Jiao, and Yu]{Chai2017}
Chai,~Q.; Jiao,~Y.; Yu,~X. Hydrogels for Biomedical Applications: Their
  Characteristics and the Mechanisms behind Them. \emph{Gels} \textbf{2017},
  \emph{3}, 2310--2861\relax
\mciteBstWouldAddEndPuncttrue
\mciteSetBstMidEndSepPunct{\mcitedefaultmidpunct}
{\mcitedefaultendpunct}{\mcitedefaultseppunct}\relax
\EndOfBibitem
\bibitem[Li and Mooney(2016)Li, and Mooney]{Li2016}
Li,~J.; Mooney,~D. Designing hydrogels for controlled drug delivery. \emph{Nat
  Rev Mater} \textbf{2016}, \emph{1}\relax
\mciteBstWouldAddEndPuncttrue
\mciteSetBstMidEndSepPunct{\mcitedefaultmidpunct}
{\mcitedefaultendpunct}{\mcitedefaultseppunct}\relax
\EndOfBibitem
\bibitem[Lee and Mooney(2001)Lee, and Mooney]{Lee2001}
Lee,~K.~Y.; Mooney,~D.~J. Hydrogels for Tissue Engineering. \emph{Chemical
  Reviews} \textbf{2001}, \emph{101}, 1869--1880\relax
\mciteBstWouldAddEndPuncttrue
\mciteSetBstMidEndSepPunct{\mcitedefaultmidpunct}
{\mcitedefaultendpunct}{\mcitedefaultseppunct}\relax
\EndOfBibitem
\bibitem[Mi \latin{et~al.}(2011)Mi, Xue, Li, and Jiang]{Mi2011}
Mi,~L.; Xue,~H.; Li,~Y.; Jiang,~S. A Thermoresponsive Antimicrobial Wound
  Dressing Hydrogel Based on a Cationic Betaine Ester. \emph{Advanced
  Functional Materials} \textbf{2011}, \emph{21}, 4028--4034\relax
\mciteBstWouldAddEndPuncttrue
\mciteSetBstMidEndSepPunct{\mcitedefaultmidpunct}
{\mcitedefaultendpunct}{\mcitedefaultseppunct}\relax
\EndOfBibitem
\bibitem[Zhao \latin{et~al.}(2022)Zhao, Song, Ren, Zhang, Lin, and
  Zhao]{Zhao2022}
Zhao,~Y.; Song,~S.; Ren,~X.; Zhang,~J.; Lin,~Q.; Zhao,~Y. Supramolecular
  Adhesive Hydrogels for Tissue Engineering Applications. \emph{Chemical
  Reviews} \textbf{2022}, \emph{122}, 5604--5640\relax
\mciteBstWouldAddEndPuncttrue
\mciteSetBstMidEndSepPunct{\mcitedefaultmidpunct}
{\mcitedefaultendpunct}{\mcitedefaultseppunct}\relax
\EndOfBibitem
\bibitem[Gupta \latin{et~al.}(2014)Gupta, Martin, Werfel, Shen, Page, and
  Duvall]{Gupta2014}
Gupta,~M.~K.; Martin,~J.~R.; Werfel,~T.~A.; Shen,~T.; Page,~J.~M.;
  Duvall,~C.~L. Cell Protective, ABC Triblock Polymer-Based Thermoresponsive
  Hydrogels with ROS-Triggered Degradation and Drug Release. \emph{Journal of
  the American Chemical Society} \textbf{2014}, \emph{136}, 14896--14902\relax
\mciteBstWouldAddEndPuncttrue
\mciteSetBstMidEndSepPunct{\mcitedefaultmidpunct}
{\mcitedefaultendpunct}{\mcitedefaultseppunct}\relax
\EndOfBibitem
\bibitem[Satarkar and Hilt(2008)Satarkar, and Hilt]{Satarkar2008}
Satarkar,~N.~S.; Hilt,~J.~Z. Magnetic hydrogel nanocomposites for remote
  controlled pulsatile drug release. \emph{Journal of Controlled Release}
  \textbf{2008}, \emph{130}, 246--251\relax
\mciteBstWouldAddEndPuncttrue
\mciteSetBstMidEndSepPunct{\mcitedefaultmidpunct}
{\mcitedefaultendpunct}{\mcitedefaultseppunct}\relax
\EndOfBibitem
\bibitem[Vatankhah-Varnoosfaderani
  \latin{et~al.}(2014)Vatankhah-Varnoosfaderani, Hashmi, GhavamiNejad, and
  Stadler]{Vatankhah2014}
Vatankhah-Varnoosfaderani,~M.; Hashmi,~S.; GhavamiNejad,~A.; Stadler,~F.~J.
  {Rapid self-healing and triple stimuli responsiveness of a supramolecular
  polymer gel based on boron–catechol interactions in a novel water-soluble
  mussel-inspired copolymer}. \emph{Polym. Chem.} \textbf{2014}, \emph{5},
  512--523\relax
\mciteBstWouldAddEndPuncttrue
\mciteSetBstMidEndSepPunct{\mcitedefaultmidpunct}
{\mcitedefaultendpunct}{\mcitedefaultseppunct}\relax
\EndOfBibitem
\bibitem[Krogsgaard \latin{et~al.}(2013)Krogsgaard, Behrens, Pedersen, and
  Birkedal]{Krogsgaard2013}
Krogsgaard,~M.; Behrens,~M.~A.; Pedersen,~J.~S.; Birkedal,~H. {Self-Healing
  Mussel-Inspired Multi-pH-Responsive Hydrogels}. \emph{Biomacromolecules}
  \textbf{2013}, \emph{14}, 297--301\relax
\mciteBstWouldAddEndPuncttrue
\mciteSetBstMidEndSepPunct{\mcitedefaultmidpunct}
{\mcitedefaultendpunct}{\mcitedefaultseppunct}\relax
\EndOfBibitem
\bibitem[Wang \latin{et~al.}(2015)Wang, Xu, Li, Li, Liu, von Klitzing, Ober,
  Kayitmazer, Li, and Guo]{Wang2015}
Wang,~W.; Xu,~Y.; Li,~A.; Li,~T.; Liu,~M.; von Klitzing,~R.; Ober,~C.~K.;
  Kayitmazer,~A.~B.; Li,~L.; Guo,~X. Zinc induced polyelectrolyte coacervate
  bioadhesive and its transition to a self-healing hydrogel. \emph{RSC Adv.}
  \textbf{2015}, \emph{5}, 66871--66878\relax
\mciteBstWouldAddEndPuncttrue
\mciteSetBstMidEndSepPunct{\mcitedefaultmidpunct}
{\mcitedefaultendpunct}{\mcitedefaultseppunct}\relax
\EndOfBibitem
\bibitem[Daly \latin{et~al.}(2020)Daly, Riley, Segura, and Burdick]{Daly2020}
Daly,~A.~C.; Riley,~L.; Segura,~T.; Burdick,~J.~A. Hydrogel microparticles for
  biomedical applications. \emph{Nature Reviews Materials} \textbf{2020},
  \emph{5}, 20--34\relax
\mciteBstWouldAddEndPuncttrue
\mciteSetBstMidEndSepPunct{\mcitedefaultmidpunct}
{\mcitedefaultendpunct}{\mcitedefaultseppunct}\relax
\EndOfBibitem
\bibitem[Alzanbaki \latin{et~al.}(2021)Alzanbaki, Moretti, and
  Hauser]{Alzanbaki2021}
Alzanbaki,~H.; Moretti,~M.; Hauser,~C.~A. Microgels-Their Manufacturing and
  Biomedical Applications. \emph{Micromachines (Basel)} \textbf{2021},
  \emph{12}\relax
\mciteBstWouldAddEndPuncttrue
\mciteSetBstMidEndSepPunct{\mcitedefaultmidpunct}
{\mcitedefaultendpunct}{\mcitedefaultseppunct}\relax
\EndOfBibitem
\bibitem[Wellert \latin{et~al.}(2015)Wellert, Richter, Hellweg, von Klitzing,
  and Hertle]{Wellert2015}
Wellert,~S.; Richter,~M.; Hellweg,~T.; von Klitzing,~R.; Hertle,~Y. Responsive
  Microgels at Surfaces and Interfaces. \emph{Zeitschrift für Physikalische
  Chemie} \textbf{2015}, \emph{229}, 1225--1250\relax
\mciteBstWouldAddEndPuncttrue
\mciteSetBstMidEndSepPunct{\mcitedefaultmidpunct}
{\mcitedefaultendpunct}{\mcitedefaultseppunct}\relax
\EndOfBibitem
\bibitem[Uhlig \latin{et~al.}(2018)Uhlig, Wegener, Hertle, Bookhold, Jaeger,
  Hellweg, Fery, and Duschl]{Uhlig2018}
Uhlig,~K.; Wegener,~T.; Hertle,~Y.; Bookhold,~J.; Jaeger,~M.; Hellweg,~T.;
  Fery,~A.; Duschl,~C. Thermoresponsive Microgel Coatings as Versatile
  Functional Compounds for Novel Cell Manipulation Tools. \emph{Polymers}
  \textbf{2018}, \emph{10}\relax
\mciteBstWouldAddEndPuncttrue
\mciteSetBstMidEndSepPunct{\mcitedefaultmidpunct}
{\mcitedefaultendpunct}{\mcitedefaultseppunct}\relax
\EndOfBibitem
\bibitem[Metawea \latin{et~al.}(2021)Metawea, Abdelmoneem, Haiba, Khalil,
  Teleb, Elzoghby, Khafaga, Noreldin, Albericio, and Khattab]{Metawea2021}
Metawea,~O.~R.; Abdelmoneem,~M.~A.; Haiba,~N.~S.; Khalil,~H.~H.; Teleb,~M.;
  Elzoghby,~A.~O.; Khafaga,~A.~F.; Noreldin,~A.~E.; Albericio,~F.;
  Khattab,~S.~N. A novel 'smart' PNIPAM-based copolymer for breast cancer
  targeted therapy: Synthesis, and characterization of dual
  pH/temperature-responsive lactoferrin-targeted PNIPAM-co-AA. \emph{Colloids
  Surf B Biointerfaces} \textbf{2021}, \emph{202}\relax
\mciteBstWouldAddEndPuncttrue
\mciteSetBstMidEndSepPunct{\mcitedefaultmidpunct}
{\mcitedefaultendpunct}{\mcitedefaultseppunct}\relax
\EndOfBibitem
\bibitem[Lima \latin{et~al.}(2016)Lima, Morales, and Cabral]{Lima2016}
Lima,~L.~H.; Morales,~Y.; Cabral,~T. Ocular Biocompatibility of
  Poly-N-Isopropylacrylamide (pNIPAM). \emph{J Ophthalmol.} \textbf{2016},
  \emph{2016}\relax
\mciteBstWouldAddEndPuncttrue
\mciteSetBstMidEndSepPunct{\mcitedefaultmidpunct}
{\mcitedefaultendpunct}{\mcitedefaultseppunct}\relax
\EndOfBibitem
\bibitem[Wang \latin{et~al.}(2001)Wang, Gan, Lyon, and El-Sayed]{Wang2001}
Wang,~J.; Gan,~D.; Lyon,~L.~A.; El-Sayed,~M.~A. {Temperature-Jump
  Investigations of the Kinetics of Hydrogel Nanoparticle Volume Phase
  Transitions}. \emph{Journal of the American Chemical Society} \textbf{2001},
  \emph{123}, 11284--11289\relax
\mciteBstWouldAddEndPuncttrue
\mciteSetBstMidEndSepPunct{\mcitedefaultmidpunct}
{\mcitedefaultendpunct}{\mcitedefaultseppunct}\relax
\EndOfBibitem
\bibitem[Li and Liang(2022)Li, and Liang]{Li2022}
Li,~Z.; Liang,~B. Modulation of phase transition of
  poly(N-isopropylacrylamide)-based microgels for pulsatile drug release.
  \emph{Polymers for Advanced Technologies} \textbf{2022}, \emph{33},
  710--722\relax
\mciteBstWouldAddEndPuncttrue
\mciteSetBstMidEndSepPunct{\mcitedefaultmidpunct}
{\mcitedefaultendpunct}{\mcitedefaultseppunct}\relax
\EndOfBibitem
\bibitem[Waite and Tanzer(1981)Waite, and Tanzer]{Waite1981}
Waite,~J.~H.; Tanzer,~M.~L. {Polyphenolic Substance of Mytilus edulis: Novel
  Adhesive Containing L-Dopa and Hydroxyproline.} \emph{Science} \textbf{1981},
  \emph{212}, 1038--40\relax
\mciteBstWouldAddEndPuncttrue
\mciteSetBstMidEndSepPunct{\mcitedefaultmidpunct}
{\mcitedefaultendpunct}{\mcitedefaultseppunct}\relax
\EndOfBibitem
\bibitem[Kaushik \latin{et~al.}(2015)Kaushik, Kaushik, Pardeshi, Sharma, Lee,
  and Choi]{Kaushik2015}
Kaushik,~N.~K.; Kaushik,~N.; Pardeshi,~S.; Sharma,~J.~G.; Lee,~S.~H.;
  Choi,~E.~H. Biomedical and Clinical Importance of Mussel-Inspired Polymers
  and Materials. \emph{Marine Drugs} \textbf{2015}, \emph{13}, 6792--6817\relax
\mciteBstWouldAddEndPuncttrue
\mciteSetBstMidEndSepPunct{\mcitedefaultmidpunct}
{\mcitedefaultendpunct}{\mcitedefaultseppunct}\relax
\EndOfBibitem
\bibitem[Balkenende \latin{et~al.}(2019)Balkenende, Winkler, and
  Messersmith]{Balkenende2019}
Balkenende,~D. W.~R.; Winkler,~S.~M.; Messersmith,~P.~B. Marine-Inspired
  Polymers in Medical Adhesion. \emph{European polymer journal} \textbf{2019},
  \emph{116}, 134–143\relax
\mciteBstWouldAddEndPuncttrue
\mciteSetBstMidEndSepPunct{\mcitedefaultmidpunct}
{\mcitedefaultendpunct}{\mcitedefaultseppunct}\relax
\EndOfBibitem
\bibitem[Zhang \latin{et~al.}(2017)Zhang, Zhang, Song, Fan, and
  Wang]{Zhang2017}
Zhang,~K.; Zhang,~F.; Song,~Y.; Fan,~J.-B.; Wang,~S. Recent Progress of
  Mussel-Inspired Underwater Adhesives. \emph{Chinese Journal of Chemistry}
  \textbf{2017}, \emph{35}, 811--820\relax
\mciteBstWouldAddEndPuncttrue
\mciteSetBstMidEndSepPunct{\mcitedefaultmidpunct}
{\mcitedefaultendpunct}{\mcitedefaultseppunct}\relax
\EndOfBibitem
\bibitem[Cha \latin{et~al.}(2008)Cha, Hwang, and Lim]{Cha2008}
Cha,~H.~J.; Hwang,~D.~S.; Lim,~S. Development of bioadhesives from marine
  mussels. \emph{Biotechnology Journal} \textbf{2008}, \emph{3}, 631--638\relax
\mciteBstWouldAddEndPuncttrue
\mciteSetBstMidEndSepPunct{\mcitedefaultmidpunct}
{\mcitedefaultendpunct}{\mcitedefaultseppunct}\relax
\EndOfBibitem
\bibitem[Morgan(1990)]{Morgan1990}
Morgan,~D. Two firms race to derive profits from mussels glue: despite gaps in
  their knowledge of how the mollusk produces the adhesive, scientists hope to
  recreate it. \emph{Scientist} \textbf{1990}, \emph{4}\relax
\mciteBstWouldAddEndPuncttrue
\mciteSetBstMidEndSepPunct{\mcitedefaultmidpunct}
{\mcitedefaultendpunct}{\mcitedefaultseppunct}\relax
\EndOfBibitem
\bibitem[Castillo \latin{et~al.}(2017)Castillo, Shanbhag, and He]{Castillo2017}
Castillo,~J.~J.; Shanbhag,~B.~K.; He,~L. In \emph{Food Bioactives: Extraction
  and Biotechnology Applications}; Puri,~M., Ed.; Springer International
  Publishing: Cham, 2017; pp 111--135\relax
\mciteBstWouldAddEndPuncttrue
\mciteSetBstMidEndSepPunct{\mcitedefaultmidpunct}
{\mcitedefaultendpunct}{\mcitedefaultseppunct}\relax
\EndOfBibitem
\bibitem[Forg \latin{et~al.}(2022)Forg, Karbacher, Ye, Guo, and von
  Klitzing]{Forg2022}
Forg,~S.; Karbacher,~A.; Ye,~Z.; Guo,~X.; von Klitzing,~R. Copolymerization
  Kinetics of Dopamine Methacrylamide during PNIPAM Microgel Synthesis for
  Increased Adhesive Properties. \emph{Langmuir} \textbf{2022}, \emph{38},
  5275--5285\relax
\mciteBstWouldAddEndPuncttrue
\mciteSetBstMidEndSepPunct{\mcitedefaultmidpunct}
{\mcitedefaultendpunct}{\mcitedefaultseppunct}\relax
\EndOfBibitem
\bibitem[Xue \latin{et~al.}(2017)Xue, Zhang, Nie, and Du]{Xue2017}
Xue,~J.; Zhang,~Z.; Nie,~J.; Du,~B. Formation of {M}icrogels by {U}tilizing the
  {R}eactivity of {C}atechols with {R}adicals. \emph{Macromolecules}
  \textbf{2017}, \emph{50}, 5285--5292\relax
\mciteBstWouldAddEndPuncttrue
\mciteSetBstMidEndSepPunct{\mcitedefaultmidpunct}
{\mcitedefaultendpunct}{\mcitedefaultseppunct}\relax
\EndOfBibitem
\bibitem[Yang \latin{et~al.}(2015)Yang, Keijsers, van Heek, Stuiver,
  Cohen~Stuart, and Kamperman]{Yang2015}
Yang,~J.; Keijsers,~J.; van Heek,~M.; Stuiver,~A.; Cohen~Stuart,~M.~A.;
  Kamperman,~M. The effect of molecular composition and crosslinking on
  adhesion of a bio-inspired adhesive. \emph{Polym. Chem.} \textbf{2015},
  \emph{6}, 3121--3130\relax
\mciteBstWouldAddEndPuncttrue
\mciteSetBstMidEndSepPunct{\mcitedefaultmidpunct}
{\mcitedefaultendpunct}{\mcitedefaultseppunct}\relax
\EndOfBibitem
\bibitem[Glass \latin{et~al.}(2009)Glass, Chung, Washburn, and
  Sitti]{Glass2009}
Glass,~P.; Chung,~H.; Washburn,~N.~R.; Sitti,~M. {Enhanced Reversible Adhesion
  of Dopamine Methacrylamide-Coated Elastomer Microfibrillar Structures under
  Wet Conditions}. \emph{Langmuir} \textbf{2009}, \emph{25}, 6607--6612\relax
\mciteBstWouldAddEndPuncttrue
\mciteSetBstMidEndSepPunct{\mcitedefaultmidpunct}
{\mcitedefaultendpunct}{\mcitedefaultseppunct}\relax
\EndOfBibitem
\bibitem[Tiu \latin{et~al.}(2019)Tiu, Delparastan, Ney, Gerst, and
  Messersmith]{Tiu2019}
Tiu,~B. D.~B.; Delparastan,~P.; Ney,~M.~R.; Gerst,~M.; Messersmith,~P.~B.
  Enhanced Adhesion and Cohesion of Bioinspired Dry/Wet Pressure-Sensitive
  Adhesives. \emph{ACS Applied Materials \& Interfaces} \textbf{2019},
  \emph{11}, 28296--28306\relax
\mciteBstWouldAddEndPuncttrue
\mciteSetBstMidEndSepPunct{\mcitedefaultmidpunct}
{\mcitedefaultendpunct}{\mcitedefaultseppunct}\relax
\EndOfBibitem
\bibitem[Xiong \latin{et~al.}(2018)Xiong, Liu, Shi, Zhang, Weng, and
  Qu]{Xiong2018}
Xiong,~X.; Liu,~Y.; Shi,~F.; Zhang,~G.; Weng,~J.; Qu,~S. Enhanced Adhesion of
  Mussel-inspired Adhesive through Manipulating Contents of Dopamine
  Methacrylamide and Molecular Weight of Polymer. \emph{Journal of Bionic
  Engineering volume} \textbf{2018}, \emph{15}, 461–470\relax
\mciteBstWouldAddEndPuncttrue
\mciteSetBstMidEndSepPunct{\mcitedefaultmidpunct}
{\mcitedefaultendpunct}{\mcitedefaultseppunct}\relax
\EndOfBibitem
\bibitem[Thavasi \latin{et~al.}(2009)Thavasi, Bettens, and Leong]{Thavasi2009}
Thavasi,~V.; Bettens,~R. P.~A.; Leong,~L.~P. {Temperature and Solvent Effects
  on Radical Scavenging Ability of Phenols}. \emph{The Journal of Physical
  Chemistry A} \textbf{2009}, \emph{113}, 3068--3077\relax
\mciteBstWouldAddEndPuncttrue
\mciteSetBstMidEndSepPunct{\mcitedefaultmidpunct}
{\mcitedefaultendpunct}{\mcitedefaultseppunct}\relax
\EndOfBibitem
\bibitem[Marcisz \latin{et~al.}(2017)Marcisz, Romanski, Stojek, and
  Karbarz]{Marcisz2017}
Marcisz,~K.; Romanski,~J.; Stojek,~Z.; Karbarz,~M. Environmentally sensitive
  hydrogel functionalized with electroactive and complexing-iron(III) catechol
  groups. \emph{Journal of Polymer Science Part A: Polymer Chemistry}
  \textbf{2017}, \emph{55}, 3236--3242\relax
\mciteBstWouldAddEndPuncttrue
\mciteSetBstMidEndSepPunct{\mcitedefaultmidpunct}
{\mcitedefaultendpunct}{\mcitedefaultseppunct}\relax
\EndOfBibitem
\bibitem[Garc\'{i}a-Pe\~{n}as \latin{et~al.}(2019)Garc\'{i}a-Pe\~{n}as, Biswas,
  Liang, Wang, Yang, and Stadler]{Garcia2019}
Garc\'{i}a-Pe\~{n}as,~A.; Biswas,~C.~S.; Liang,~W.; Wang,~Y.; Yang,~P.;
  Stadler,~F.~J. Effect of Hydrophobic Interactions on Lower Critical Solution
  Temperature for Poly(N-isopropylacrylamide-co-dopamine Methacrylamide)
  Copolymers. \emph{Polymers} \textbf{2019}, \emph{11}\relax
\mciteBstWouldAddEndPuncttrue
\mciteSetBstMidEndSepPunct{\mcitedefaultmidpunct}
{\mcitedefaultendpunct}{\mcitedefaultseppunct}\relax
\EndOfBibitem
\bibitem[Yoshida \latin{et~al.}(1995)Yoshida, Sakai, Okano, and
  Sakurai]{Ryo1995}
Yoshida,~R.; Sakai,~K.; Okano,~T.; Sakurai,~Y. Modulating the phase transition
  temperature and thermosensitivity in N-isopropylacrylamide copolymer gels.
  \emph{Journal of Biomaterials Science, Polymer Edition} \textbf{1995},
  \emph{6}, 585--598\relax
\mciteBstWouldAddEndPuncttrue
\mciteSetBstMidEndSepPunct{\mcitedefaultmidpunct}
{\mcitedefaultendpunct}{\mcitedefaultseppunct}\relax
\EndOfBibitem
\bibitem[Feil \latin{et~al.}(1993)Feil, Bae, Feijen, and Kim]{Feil1993}
Feil,~H.; Bae,~Y.~H.; Feijen,~J.; Kim,~S.~W. Effect of comonomer hydrophilicity
  and ionization on the lower critical solution temperature of
  N-isopropylacrylamide copolymers. \emph{Macromolecules} \textbf{1993},
  \emph{26}, 2496--2500\relax
\mciteBstWouldAddEndPuncttrue
\mciteSetBstMidEndSepPunct{\mcitedefaultmidpunct}
{\mcitedefaultendpunct}{\mcitedefaultseppunct}\relax
\EndOfBibitem
\bibitem[Flory and Rehner(1943)Flory, and Rehner]{Flory1943}
Flory,~P.~J.; Rehner,~J.,~John {Statistical Mechanics of Cross‐Linked Polymer
  Networks I. Rubberlike Elasticity}. \emph{The Journal of Chemical Physics}
  \textbf{1943}, \emph{11}, 512--520\relax
\mciteBstWouldAddEndPuncttrue
\mciteSetBstMidEndSepPunct{\mcitedefaultmidpunct}
{\mcitedefaultendpunct}{\mcitedefaultseppunct}\relax
\EndOfBibitem
\bibitem[Flory(1953)]{Flory1953}
Flory,~P.~J. \emph{Principles of Polymer Chemistry}; Ithaca N.Y: Cornell
  University Press, 1953\relax
\mciteBstWouldAddEndPuncttrue
\mciteSetBstMidEndSepPunct{\mcitedefaultmidpunct}
{\mcitedefaultendpunct}{\mcitedefaultseppunct}\relax
\EndOfBibitem
\bibitem[Pelton and Chibante(1986)Pelton, and Chibante]{PeltonChibante1986}
Pelton,~R.~H.; Chibante,~P. {Preparation of aqueous latices with
  N-isopropylacrylamide}. \emph{Colloids and Surfaces} \textbf{1986},
  \emph{20}, 247--256\relax
\mciteBstWouldAddEndPuncttrue
\mciteSetBstMidEndSepPunct{\mcitedefaultmidpunct}
{\mcitedefaultendpunct}{\mcitedefaultseppunct}\relax
\EndOfBibitem
\bibitem[Witt \latin{et~al.}(2019)Witt, Hinrichs, M\"oller, Backes, Fischer,
  and von Klitzing]{Witt2019}
Witt,~M.~U.; Hinrichs,~S.; M\"oller,~N.; Backes,~S.; Fischer,~B.; von
  Klitzing,~R. {Distribution of CoFe2O4 Nanoparticles Inside PNIPAM-Based
  Microgels of Different Cross-linker Distributions}. \emph{The Journal of
  Physical Chemistry B} \textbf{2019}, \emph{123}, 2405--2413\relax
\mciteBstWouldAddEndPuncttrue
\mciteSetBstMidEndSepPunct{\mcitedefaultmidpunct}
{\mcitedefaultendpunct}{\mcitedefaultseppunct}\relax
\EndOfBibitem
\bibitem[Acciaro \latin{et~al.}(2011)Acciaro, Gilányi, and Varga]{Acciaro2011}
Acciaro,~R.; Gilányi,~T.; Varga,~I. {Preparation of Monodisperse
  Poly(N-isopropylacrylamide) Microgel Particles with Homogenous Cross-Link
  Density Distribution}. \emph{Langmuir} \textbf{2011}, \emph{27},
  7917--7925\relax
\mciteBstWouldAddEndPuncttrue
\mciteSetBstMidEndSepPunct{\mcitedefaultmidpunct}
{\mcitedefaultendpunct}{\mcitedefaultseppunct}\relax
\EndOfBibitem
\bibitem[Peng \latin{et~al.}(2019)Peng, Zhang, Rong, and Zhang]{Peng2019}
Peng,~W.; Zhang,~Z.; Rong,~M.; Zhang,~M. Core-Shell Structure Design of Hollow
  Mesoporous Silica Nanospheres Based on Thermo-Sensitive PNIPAM and
  pH-Responsive Catechol-Fe3+ Complex. \emph{Polymers} \textbf{2019},
  \emph{11}\relax
\mciteBstWouldAddEndPuncttrue
\mciteSetBstMidEndSepPunct{\mcitedefaultmidpunct}
{\mcitedefaultendpunct}{\mcitedefaultseppunct}\relax
\EndOfBibitem
\bibitem[Pasparakis and Tsitsilianis(2020)Pasparakis, and
  Tsitsilianis]{Pasparakis2020}
Pasparakis,~G.; Tsitsilianis,~C. LCST polymers: Thermoresponsive nanostructured
  assemblies towards bioapplications. \emph{Polymer} \textbf{2020}, \emph{211},
  123146\relax
\mciteBstWouldAddEndPuncttrue
\mciteSetBstMidEndSepPunct{\mcitedefaultmidpunct}
{\mcitedefaultendpunct}{\mcitedefaultseppunct}\relax
\EndOfBibitem
\bibitem[Kuhn(1946)]{Kuhn1946}
Kuhn,~W. Dependence of the average transversal on the longitudinal dimensions
  of statistical coils formed by chain molecules. \emph{Journal of Polymer
  Science} \textbf{1946}, \emph{1}, 380--388\relax
\mciteBstWouldAddEndPuncttrue
\mciteSetBstMidEndSepPunct{\mcitedefaultmidpunct}
{\mcitedefaultendpunct}{\mcitedefaultseppunct}\relax
\EndOfBibitem
\bibitem[Canal and Peppas(1989)Canal, and Peppas]{Canal1989}
Canal,~T.; Peppas,~N.~A. {Correlation between mesh size and equilibrium degree
  of swelling of polymeric networks}. \emph{Journal of Biomedical Materials
  Research} \textbf{1989}, \emph{23}, 1183--1193\relax
\mciteBstWouldAddEndPuncttrue
\mciteSetBstMidEndSepPunct{\mcitedefaultmidpunct}
{\mcitedefaultendpunct}{\mcitedefaultseppunct}\relax
\EndOfBibitem
\bibitem[Peppas \latin{et~al.}(1985)Peppas, Moynihan, and Lucht]{Peppas1985}
Peppas,~N.~A.; Moynihan,~H.~J.; Lucht,~L.~M. The structure of highly
  crosslinked poly(2-hydroxyethyl methacrylate) hydrogels. \emph{Journal of
  Biomedical Materials Research} \textbf{1985}, \emph{19}, 397--411\relax
\mciteBstWouldAddEndPuncttrue
\mciteSetBstMidEndSepPunct{\mcitedefaultmidpunct}
{\mcitedefaultendpunct}{\mcitedefaultseppunct}\relax
\EndOfBibitem
\bibitem[Fänger \latin{et~al.}(2006)Fänger, Wack, and Ulbricht]{Faenger2006}
Fänger,~C.; Wack,~H.; Ulbricht,~M. Macroporous Poly(N-isopropylacrylamide)
  Hydrogels with Adjustable Size “Cut-off” for the Efficient and Reversible
  Immobilization of Biomacromolecules. \emph{Macromolecular Bioscience}
  \textbf{2006}, \emph{6}, 393--402\relax
\mciteBstWouldAddEndPuncttrue
\mciteSetBstMidEndSepPunct{\mcitedefaultmidpunct}
{\mcitedefaultendpunct}{\mcitedefaultseppunct}\relax
\EndOfBibitem
\bibitem[Schulte \latin{et~al.}(2022)Schulte, Izak-Nau, Braun, Pich,
  Richtering, and Göstl]{Schulte2022}
Schulte,~M.~F.; Izak-Nau,~E.; Braun,~S.; Pich,~A.; Richtering,~W.; Göstl,~R.
  Microgels react to force: mechanical properties{,} syntheses{,} and
  force-activated functions. \emph{Chem. Soc. Rev.} \textbf{2022}, \emph{51},
  2939--2956\relax
\mciteBstWouldAddEndPuncttrue
\mciteSetBstMidEndSepPunct{\mcitedefaultmidpunct}
{\mcitedefaultendpunct}{\mcitedefaultseppunct}\relax
\EndOfBibitem
\bibitem[Matzelle \latin{et~al.}(2003)Matzelle, Geuskens, and
  Kruse]{Matzelle2003}
Matzelle,~T.~R.; Geuskens,~G.; Kruse,~N. Elastic Properties of
  Poly(N-isopropylacrylamide) and Poly(acrylamide) Hydrogels Studied by
  Scanning Force Microscopy. \emph{Macromolecules} \textbf{2003}, \emph{36},
  2926--2931\relax
\mciteBstWouldAddEndPuncttrue
\mciteSetBstMidEndSepPunct{\mcitedefaultmidpunct}
{\mcitedefaultendpunct}{\mcitedefaultseppunct}\relax
\EndOfBibitem
\bibitem[Junk \latin{et~al.}(2010)Junk, Berger, and Jonas]{Junk2010}
Junk,~M. J.~N.; Berger,~R.; Jonas,~U. Atomic Force Spectroscopy of
  Thermoresponsive Photo-Cross-Linked Hydrogel Films. \emph{Langmuir}
  \textbf{2010}, \emph{26}, 7262--7269\relax
\mciteBstWouldAddEndPuncttrue
\mciteSetBstMidEndSepPunct{\mcitedefaultmidpunct}
{\mcitedefaultendpunct}{\mcitedefaultseppunct}\relax
\EndOfBibitem
\bibitem[Harmon \latin{et~al.}(2003)Harmon, Kuckling, and Frank]{Harmon2003_2}
Harmon,~M.~E.; Kuckling,~D.; Frank,~C.~W. Photo-Cross-Linkable PNIPAAm
  Copolymers. 5. Mechanical Properties of Hydrogel Layers. \emph{Langmuir}
  \textbf{2003}, \emph{19}, 10660--10665\relax
\mciteBstWouldAddEndPuncttrue
\mciteSetBstMidEndSepPunct{\mcitedefaultmidpunct}
{\mcitedefaultendpunct}{\mcitedefaultseppunct}\relax
\EndOfBibitem
\bibitem[Kühnhammer \latin{et~al.}(2022)Kühnhammer, Gräff, Loran, Soltwedel,
  Löhmann, Frielinghaus, and von Klitzing]{Keenhammer2022}
Kühnhammer,~M.; Gräff,~K.; Loran,~E.; Soltwedel,~O.; Löhmann,~O.;
  Frielinghaus,~H.; von Klitzing,~R. Structure formation of PNIPAM microgels in
  foams and foam films. \emph{Soft Matter} \textbf{2022}, \emph{18},
  9249--9262\relax
\mciteBstWouldAddEndPuncttrue
\mciteSetBstMidEndSepPunct{\mcitedefaultmidpunct}
{\mcitedefaultendpunct}{\mcitedefaultseppunct}\relax
\EndOfBibitem
\bibitem[Zemła \latin{et~al.}(2020)Zemła, Bobrowska, Kubiak, Zieliński,
  Pabijan, Pogoda, Bobrowski, and Lekka]{Zemla2020}
Zemła,~J.; Bobrowska,~J.; Kubiak,~A.; Zieliński,~T.; Pabijan,~J.; Pogoda,~K.;
  Bobrowski,~P.; Lekka,~M. Indenting soft samples (hydrogels and cells) with
  cantilevers possessing various shapes of probing tip. \emph{Eur Biophys J}
  \textbf{2020}, \emph{49}, 485--495\relax
\mciteBstWouldAddEndPuncttrue
\mciteSetBstMidEndSepPunct{\mcitedefaultmidpunct}
{\mcitedefaultendpunct}{\mcitedefaultseppunct}\relax
\EndOfBibitem
\bibitem[Sokolov and Dokukin(2014)Sokolov, and Dokukin]{Sokolov2014}
Sokolov,~I.; Dokukin,~M. {Mechanics of Biological Cells Studied with Atomic
  Force Microscopy}. \emph{Microscopy and Microanalysis} \textbf{2014},
  \emph{20}, 2076--2077\relax
\mciteBstWouldAddEndPuncttrue
\mciteSetBstMidEndSepPunct{\mcitedefaultmidpunct}
{\mcitedefaultendpunct}{\mcitedefaultseppunct}\relax
\EndOfBibitem
\bibitem[Alcaraz \latin{et~al.}(2018)Alcaraz, Otero, Jorba, and
  Navajas]{Alcaraz2018}
Alcaraz,~J.; Otero,~J.; Jorba,~I.; Navajas,~D. Bidirectional mechanobiology
  between cells and their local extracellular matrix probed by atomic force
  microscopy. \emph{Seminars in Cell \& Developmental Biology} \textbf{2018},
  \emph{73}, 71--81\relax
\mciteBstWouldAddEndPuncttrue
\mciteSetBstMidEndSepPunct{\mcitedefaultmidpunct}
{\mcitedefaultendpunct}{\mcitedefaultseppunct}\relax
\EndOfBibitem
\bibitem[Hutter and Bechhoefer(1993)Hutter, and Bechhoefer]{Hutter1993}
Hutter,~J.~L.; Bechhoefer,~J. Calibration of atomic‐force microscope tips.
  \emph{Review of Scientific Instruments} \textbf{1993}, \emph{64},
  1868--1873\relax
\mciteBstWouldAddEndPuncttrue
\mciteSetBstMidEndSepPunct{\mcitedefaultmidpunct}
{\mcitedefaultendpunct}{\mcitedefaultseppunct}\relax
\EndOfBibitem
\bibitem[Hashmi and Dufresne(2009)Hashmi, and Dufresne]{Hashmi2009}
Hashmi,~S.~M.; Dufresne,~E.~R. Mechanical properties of individual microgel
  particles through the deswelling transition. \emph{Soft Matter}
  \textbf{2009}, \emph{5}, 3682--3688\relax
\mciteBstWouldAddEndPuncttrue
\mciteSetBstMidEndSepPunct{\mcitedefaultmidpunct}
{\mcitedefaultendpunct}{\mcitedefaultseppunct}\relax
\EndOfBibitem
\bibitem[Hertz(1881)]{Hertz1881}
Hertz,~H. Über die Berührung fester elastischer Körper. \emph{Journal für
  die reine und angewandte Mathematik} \textbf{1881}, \emph{92}, 156--171\relax
\mciteBstWouldAddEndPuncttrue
\mciteSetBstMidEndSepPunct{\mcitedefaultmidpunct}
{\mcitedefaultendpunct}{\mcitedefaultseppunct}\relax
\EndOfBibitem
\bibitem[Witte \latin{et~al.}(2021)Witte, Kyrey, Lutzki, Dahl, Kühnhammer,
  Klitzing, Holderer, and Wellert]{Witte2021}
Witte,~J.; Kyrey,~T.; Lutzki,~J.; Dahl,~A.~M.; Kühnhammer,~M.;
  Klitzing,~R.~v.; Holderer,~O.; Wellert,~S. Looking inside
  Poly(N-isopropylacrylamide) Microgels: Nanomechanics and Dynamics at
  Solid–Liquid Interfaces. \emph{ACS Applied Polymer Materials}
  \textbf{2021}, \emph{3}, 976--985\relax
\mciteBstWouldAddEndPuncttrue
\mciteSetBstMidEndSepPunct{\mcitedefaultmidpunct}
{\mcitedefaultendpunct}{\mcitedefaultseppunct}\relax
\EndOfBibitem
\bibitem[Voudouris \latin{et~al.}(2013)Voudouris, Florea, van~der Schoot, and
  Wyss]{Voudouris2013}
Voudouris,~P.; Florea,~D.; van~der Schoot,~P.; Wyss,~H.~M. Micromechanics of
  temperature sensitive microgels: dip in the Poisson ratio near the LCST.
  \emph{Soft Matter} \textbf{2013}, \emph{9}, 7158--7166\relax
\mciteBstWouldAddEndPuncttrue
\mciteSetBstMidEndSepPunct{\mcitedefaultmidpunct}
{\mcitedefaultendpunct}{\mcitedefaultseppunct}\relax
\EndOfBibitem
\bibitem[Hirotsu(1991)]{Hirotsu1991}
Hirotsu,~S. Softening of bulk modulus and negative Poisson’s ratio near the
  volume phase transition of polymer gels. \emph{The Journal of Chemical
  Physics} \textbf{1991}, \emph{94}, 3949--3957\relax
\mciteBstWouldAddEndPuncttrue
\mciteSetBstMidEndSepPunct{\mcitedefaultmidpunct}
{\mcitedefaultendpunct}{\mcitedefaultseppunct}\relax
\EndOfBibitem
\bibitem[Boon and Schurtenberger(2017)Boon, and Schurtenberger]{Boon2017}
Boon,~N.; Schurtenberger,~P. Swelling of micro-hydrogels with a crosslinker
  gradient. \emph{Phys. Chem. Chem. Phys.} \textbf{2017}, \emph{19},
  23740--23746\relax
\mciteBstWouldAddEndPuncttrue
\mciteSetBstMidEndSepPunct{\mcitedefaultmidpunct}
{\mcitedefaultendpunct}{\mcitedefaultseppunct}\relax
\EndOfBibitem
\bibitem[Geissler and Hecht(1980)Geissler, and Hecht]{Geissler1980}
Geissler,~E.; Hecht,~A.~M. The Poisson Ratio in Polymer Gels.
  \emph{Macromolecules} \textbf{1980}, \emph{13}, 1276--1280\relax
\mciteBstWouldAddEndPuncttrue
\mciteSetBstMidEndSepPunct{\mcitedefaultmidpunct}
{\mcitedefaultendpunct}{\mcitedefaultseppunct}\relax
\EndOfBibitem
\bibitem[Backes and Von~Klitzing(2018)Backes, and Von~Klitzing]{Backes2018}
Backes,~S.; Von~Klitzing,~R. Nanomechanics and Nanorheology of Microgels at
  Interfaces. \emph{Polymers} \textbf{2018}, \emph{10}\relax
\mciteBstWouldAddEndPuncttrue
\mciteSetBstMidEndSepPunct{\mcitedefaultmidpunct}
{\mcitedefaultendpunct}{\mcitedefaultseppunct}\relax
\EndOfBibitem
\bibitem[Popov \latin{et~al.}(2019)Popov, Heß, and Willert]{Popov2019}
Popov,~V.~L.; Heß,~M.; Willert,~E. \emph{Handbook of Contact Mechanics};
  Springer Berlin, Heidelberg, 2019\relax
\mciteBstWouldAddEndPuncttrue
\mciteSetBstMidEndSepPunct{\mcitedefaultmidpunct}
{\mcitedefaultendpunct}{\mcitedefaultseppunct}\relax
\EndOfBibitem
\bibitem[Burmistrova \latin{et~al.}(2011)Burmistrova, Richter, Uzum, and
  Klitzing]{Burmistrova2011}
Burmistrova,~A.; Richter,~M.; Uzum,~C.; Klitzing,~R.~v. Effect of cross-linker
  density of P(NIPAM-co-AAc) microgels at solid surfaces on the
  swelling/shrinking behaviour and the Young’s modulus. \emph{Colloid and
  Polymer Science} \textbf{2011}, \emph{289}, 613--624\relax
\mciteBstWouldAddEndPuncttrue
\mciteSetBstMidEndSepPunct{\mcitedefaultmidpunct}
{\mcitedefaultendpunct}{\mcitedefaultseppunct}\relax
\EndOfBibitem
\bibitem[Burmistrova \latin{et~al.}(2011)Burmistrova, Richter, Eisele, Üzüm,
  and Von~Klitzing]{Burmistrova2011_2}
Burmistrova,~A.; Richter,~M.; Eisele,~M.; Üzüm,~C.; Von~Klitzing,~R. The
  Effect of Co-Monomer Content on the Swelling/Shrinking and Mechanical
  Behaviour of Individually Adsorbed PNIPAM Microgel Particles. \emph{Polymers}
  \textbf{2011}, \emph{3}, 1575--1590\relax
\mciteBstWouldAddEndPuncttrue
\mciteSetBstMidEndSepPunct{\mcitedefaultmidpunct}
{\mcitedefaultendpunct}{\mcitedefaultseppunct}\relax
\EndOfBibitem
\bibitem[Aufderhorst-Roberts \latin{et~al.}(2018)Aufderhorst-Roberts, Baker,
  Foster, Cayre, Mattsson, and Connell]{Aufderhorst2018}
Aufderhorst-Roberts,~A.; Baker,~D.; Foster,~R.~J.; Cayre,~O.; Mattsson,~J.;
  Connell,~S.~D. Nanoscale mechanics of microgel particles. \emph{Nanoscale}
  \textbf{2018}, \emph{10}, 16050--16061\relax
\mciteBstWouldAddEndPuncttrue
\mciteSetBstMidEndSepPunct{\mcitedefaultmidpunct}
{\mcitedefaultendpunct}{\mcitedefaultseppunct}\relax
\EndOfBibitem
\bibitem[Karg \latin{et~al.}(2013)Karg, Pr{\'e}vost, Brandt, Wallacher, von
  Klitzing, and Hellweg]{Karg2013}
Karg,~M.; Pr{\'e}vost,~S.; Brandt,~A.; Wallacher,~D.; von Klitzing,~R.;
  Hellweg,~T. Poly-NIPAM Microgels with Different Cross-Linker Densities.
  Intelligent Hydrogels. Cham, 2013; pp 63--76\relax
\mciteBstWouldAddEndPuncttrue
\mciteSetBstMidEndSepPunct{\mcitedefaultmidpunct}
{\mcitedefaultendpunct}{\mcitedefaultseppunct}\relax
\EndOfBibitem
\bibitem[Senff and Richtering(2000)Senff, and Richtering]{Senff2000}
Senff,~H.; Richtering,~W. Influence of cross-link density on rheological
  properties of temperature-sensitive microgel suspensions. \emph{Colloid and
  Polymer Science} \textbf{2000}, \emph{278}, 830–840\relax
\mciteBstWouldAddEndPuncttrue
\mciteSetBstMidEndSepPunct{\mcitedefaultmidpunct}
{\mcitedefaultendpunct}{\mcitedefaultseppunct}\relax
\EndOfBibitem
\bibitem[Schlattmann and Schönhoff(2022)Schlattmann, and
  Schönhoff]{Schlattmann2022}
Schlattmann,~D.; Schönhoff,~M. Interplay of the Influence of Crosslinker
  Content and Model Drugs on the Phase Transition of Thermoresponsive
  PNiPAM-BIS Microgels. \emph{Gels} \textbf{2022}, \emph{8}\relax
\mciteBstWouldAddEndPuncttrue
\mciteSetBstMidEndSepPunct{\mcitedefaultmidpunct}
{\mcitedefaultendpunct}{\mcitedefaultseppunct}\relax
\EndOfBibitem
\bibitem[Kratz \latin{et~al.}(1998)Kratz, Hellweg, and Eimer]{Kratz1998}
Kratz,~K.; Hellweg,~T.; Eimer,~W. Effect of connectivity and charge density on
  the swelling and local structural and dynamic properties of colloidal PNIPAM
  microgels. \emph{Berichte der Bunsengesellschaft für physikalische Chemie}
  \textbf{1998}, \emph{102}, 1603--1608\relax
\mciteBstWouldAddEndPuncttrue
\mciteSetBstMidEndSepPunct{\mcitedefaultmidpunct}
{\mcitedefaultendpunct}{\mcitedefaultseppunct}\relax
\EndOfBibitem
\bibitem[Thiele \latin{et~al.}(2021)Thiele, Andersson, Dahlin, and
  Hailes]{Thiele2021}
Thiele,~S.; Andersson,~J.; Dahlin,~A.; Hailes,~R. L.~N. Tuning the
  Thermoresponsive Behavior of Surface-Attached PNIPAM Networks: Varying the
  Crosslinker Content in SI-ATRP. \emph{Langmuir} \textbf{2021}, \emph{37},
  3391--3398\relax
\mciteBstWouldAddEndPuncttrue
\mciteSetBstMidEndSepPunct{\mcitedefaultmidpunct}
{\mcitedefaultendpunct}{\mcitedefaultseppunct}\relax
\EndOfBibitem
\bibitem[Friesen \latin{et~al.}(2021)Friesen, Hannappel, Kakorin, and
  Hellweg]{Friesen2021}
Friesen,~S.; Hannappel,~Y.; Kakorin,~S.; Hellweg,~T. Accounting for
  Cooperativity in the Thermotropic Volume Phase Transition of Smart Microgels.
  \emph{Gels} \textbf{2021}, \emph{7}\relax
\mciteBstWouldAddEndPuncttrue
\mciteSetBstMidEndSepPunct{\mcitedefaultmidpunct}
{\mcitedefaultendpunct}{\mcitedefaultseppunct}\relax
\EndOfBibitem
\bibitem[Wu \latin{et~al.}(1994)Wu, Pelton, Hamielec, Woods, and
  McPhee]{Wu1994}
Wu,~X.; Pelton,~R.~H.; Hamielec,~A.~E.; Woods,~D.~R.; McPhee,~W. {The kinetics
  of poly(N-isopropylacrylamide) microgel latex formation}. \emph{Colloid and
  Polymer Science} \textbf{1994}, \emph{272}, 467--477\relax
\mciteBstWouldAddEndPuncttrue
\mciteSetBstMidEndSepPunct{\mcitedefaultmidpunct}
{\mcitedefaultendpunct}{\mcitedefaultseppunct}\relax
\EndOfBibitem
\bibitem[Lehmann \latin{et~al.}(2019)Lehmann, Krause, Miruchna, and von
  Klitzing]{Lehmann2019}
Lehmann,~M.; Krause,~P.; Miruchna,~V.; von Klitzing,~R. Tailoring PNIPAM
  hydrogels for large temperature-triggered changes in mechanical properties.
  \emph{Colloid and Polymer Science} \textbf{2019}, \emph{297}, 633--640\relax
\mciteBstWouldAddEndPuncttrue
\mciteSetBstMidEndSepPunct{\mcitedefaultmidpunct}
{\mcitedefaultendpunct}{\mcitedefaultseppunct}\relax
\EndOfBibitem
\bibitem[Burmistrova and von Klitzing(2010)Burmistrova, and von
  Klitzing]{Burmistrova2010}
Burmistrova,~A.; von Klitzing,~R. Control of number density and
  swelling/shrinking behavior of P(NIPAM–AAc) particles at solid surfaces.
  \emph{J. Mater. Chem.} \textbf{2010}, \emph{20}, 3502--3507\relax
\mciteBstWouldAddEndPuncttrue
\mciteSetBstMidEndSepPunct{\mcitedefaultmidpunct}
{\mcitedefaultendpunct}{\mcitedefaultseppunct}\relax
\EndOfBibitem
\bibitem[Schmidt \latin{et~al.}(2008)Schmidt, Motschmann, Hellweg, and {von
  Klitzing}]{Schmidt2008}
Schmidt,~S.; Motschmann,~H.; Hellweg,~T.; {von Klitzing},~R. Thermoresponsive
  surfaces by spin-coating of PNIPAM-co-PAA microgels: A combined AFM and
  ellipsometry study. \emph{Polymer} \textbf{2008}, \emph{49}, 749--756\relax
\mciteBstWouldAddEndPuncttrue
\mciteSetBstMidEndSepPunct{\mcitedefaultmidpunct}
{\mcitedefaultendpunct}{\mcitedefaultseppunct}\relax
\EndOfBibitem
\bibitem[Butt \latin{et~al.}(2005)Butt, Cappella, and Kappl]{Butt2005}
Butt,~H.-J.; Cappella,~B.; Kappl,~M. Force measurements with the atomic force
  microscope: Technique, interpretation and applications. \emph{Surface Science
  Reports} \textbf{2005}, \emph{59}, 1--152\relax
\mciteBstWouldAddEndPuncttrue
\mciteSetBstMidEndSepPunct{\mcitedefaultmidpunct}
{\mcitedefaultendpunct}{\mcitedefaultseppunct}\relax
\EndOfBibitem
\bibitem[Backes \latin{et~al.}(2015)Backes, Witt, Roeben, Kuhrts, Aleed,
  Schmidt, and von Klitzing]{Backes2015}
Backes,~S.; Witt,~M.~U.; Roeben,~E.; Kuhrts,~L.; Aleed,~S.; Schmidt,~A.~M.; von
  Klitzing,~R. Loading of PNIPAM Based Microgels with CoFe2O4 Nanoparticles and
  Their Magnetic Response in Bulk and at Surfaces. \emph{The Journal of
  Physical Chemistry B} \textbf{2015}, \emph{119}, 12129--12137\relax
\mciteBstWouldAddEndPuncttrue
\mciteSetBstMidEndSepPunct{\mcitedefaultmidpunct}
{\mcitedefaultendpunct}{\mcitedefaultseppunct}\relax
\EndOfBibitem
\bibitem[Schmidt \latin{et~al.}(2010)Schmidt, Zeiser, Hellweg, Duschl, Fery,
  and Möhwald]{Schmidt2010}
Schmidt,~S.; Zeiser,~M.; Hellweg,~T.; Duschl,~C.; Fery,~A.; Möhwald,~H.
  Adhesion and Mechanical Properties of PNIPAM Microgel Films and Their
  Potential Use as Switchable Cell Culture Substrates. \emph{Advanced
  Functional Materials} \textbf{2010}, \emph{20}, 3235--3243\relax
\mciteBstWouldAddEndPuncttrue
\mciteSetBstMidEndSepPunct{\mcitedefaultmidpunct}
{\mcitedefaultendpunct}{\mcitedefaultseppunct}\relax
\EndOfBibitem
\bibitem[Witte \latin{et~al.}(2019)Witte, Kyrey, Lutzki, Dahl, Houston,
  Radulescu, Pipich, Stingaciu, Kühnhammer, Witt, von Klitzing, Holderer, and
  Wellert]{Witte2019}
Witte,~J.; Kyrey,~T.; Lutzki,~J.; Dahl,~A.~M.; Houston,~J.; Radulescu,~A.;
  Pipich,~V.; Stingaciu,~L.; Kühnhammer,~M.; Witt,~M.~U.; von Klitzing,~R.;
  Holderer,~O.; Wellert,~S. A comparison of the network structure and inner
  dynamics of homogeneously and heterogeneously crosslinked PNIPAM microgels
  with high crosslinker content. \emph{Soft Matter} \textbf{2019}, \emph{15},
  1053--1064\relax
\mciteBstWouldAddEndPuncttrue
\mciteSetBstMidEndSepPunct{\mcitedefaultmidpunct}
{\mcitedefaultendpunct}{\mcitedefaultseppunct}\relax
\EndOfBibitem
\bibitem[Mourran \latin{et~al.}(2016)Mourran, Wu, Gumerov, Rudov, Potemkin,
  Pich, and Möller]{Mourran2016}
Mourran,~A.; Wu,~Y.; Gumerov,~R.~A.; Rudov,~A.~A.; Potemkin,~I.~I.; Pich,~A.;
  Möller,~M. When Colloidal Particles Become Polymer Coils. \emph{Langmuir}
  \textbf{2016}, \emph{32}, 723--730\relax
\mciteBstWouldAddEndPuncttrue
\mciteSetBstMidEndSepPunct{\mcitedefaultmidpunct}
{\mcitedefaultendpunct}{\mcitedefaultseppunct}\relax
\EndOfBibitem
\bibitem[Li and Cao(2019)Li, and Cao]{Li2019}
Li,~Y.; Cao,~Y. The molecular mechanisms underlying mussel adhesion.
  \emph{Nanoscale Adv.} \textbf{2019}, \emph{1}, 4246--4257\relax
\mciteBstWouldAddEndPuncttrue
\mciteSetBstMidEndSepPunct{\mcitedefaultmidpunct}
{\mcitedefaultendpunct}{\mcitedefaultseppunct}\relax
\EndOfBibitem
\bibitem[Saiz-Poseu \latin{et~al.}(2019)Saiz-Poseu, Mancebo-Aracil, Nador,
  Busqué, and Ruiz-Molina]{Saiz-Poseu2019}
Saiz-Poseu,~J.; Mancebo-Aracil,~J.; Nador,~F.; Busqué,~F.; Ruiz-Molina,~D.
  {The Chemistry behind Catechol-Based Adhesion}. \emph{Angewandte Chemie
  International Edition} \textbf{2019}, \emph{58}, 696--714\relax
\mciteBstWouldAddEndPuncttrue
\mciteSetBstMidEndSepPunct{\mcitedefaultmidpunct}
{\mcitedefaultendpunct}{\mcitedefaultseppunct}\relax
\EndOfBibitem
\bibitem[Zhang \latin{et~al.}(2020)Zhang, Wu, Zhou, Zhou, Liu, and
  Wang]{Zhang2020}
Zhang,~C.; Wu,~B.; Zhou,~Y.; Zhou,~F.; Liu,~W.; Wang,~Z. Mussel-inspired
  hydrogels: from design principles to promising applications. \emph{Chem. Soc.
  Rev.} \textbf{2020}, \emph{49}, 3605--3637\relax
\mciteBstWouldAddEndPuncttrue
\mciteSetBstMidEndSepPunct{\mcitedefaultmidpunct}
{\mcitedefaultendpunct}{\mcitedefaultseppunct}\relax
\EndOfBibitem
\bibitem[Yang \latin{et~al.}(2014)Yang, Cohen~Stuart, and Kamperman]{Yang2014}
Yang,~J.; Cohen~Stuart,~M.~A.; Kamperman,~M. Jack of all trades: versatile
  catechol crosslinking mechanisms. \emph{Chem. Soc. Rev.} \textbf{2014},
  \emph{43}, 8271--8298\relax
\mciteBstWouldAddEndPuncttrue
\mciteSetBstMidEndSepPunct{\mcitedefaultmidpunct}
{\mcitedefaultendpunct}{\mcitedefaultseppunct}\relax
\EndOfBibitem
\end{mcitethebibliography}
\newpage

\end{document}


\beginsupplement
\maketitle

\newpage

\begin{figure}[H]
\centering
\includegraphics[scale=0.6]{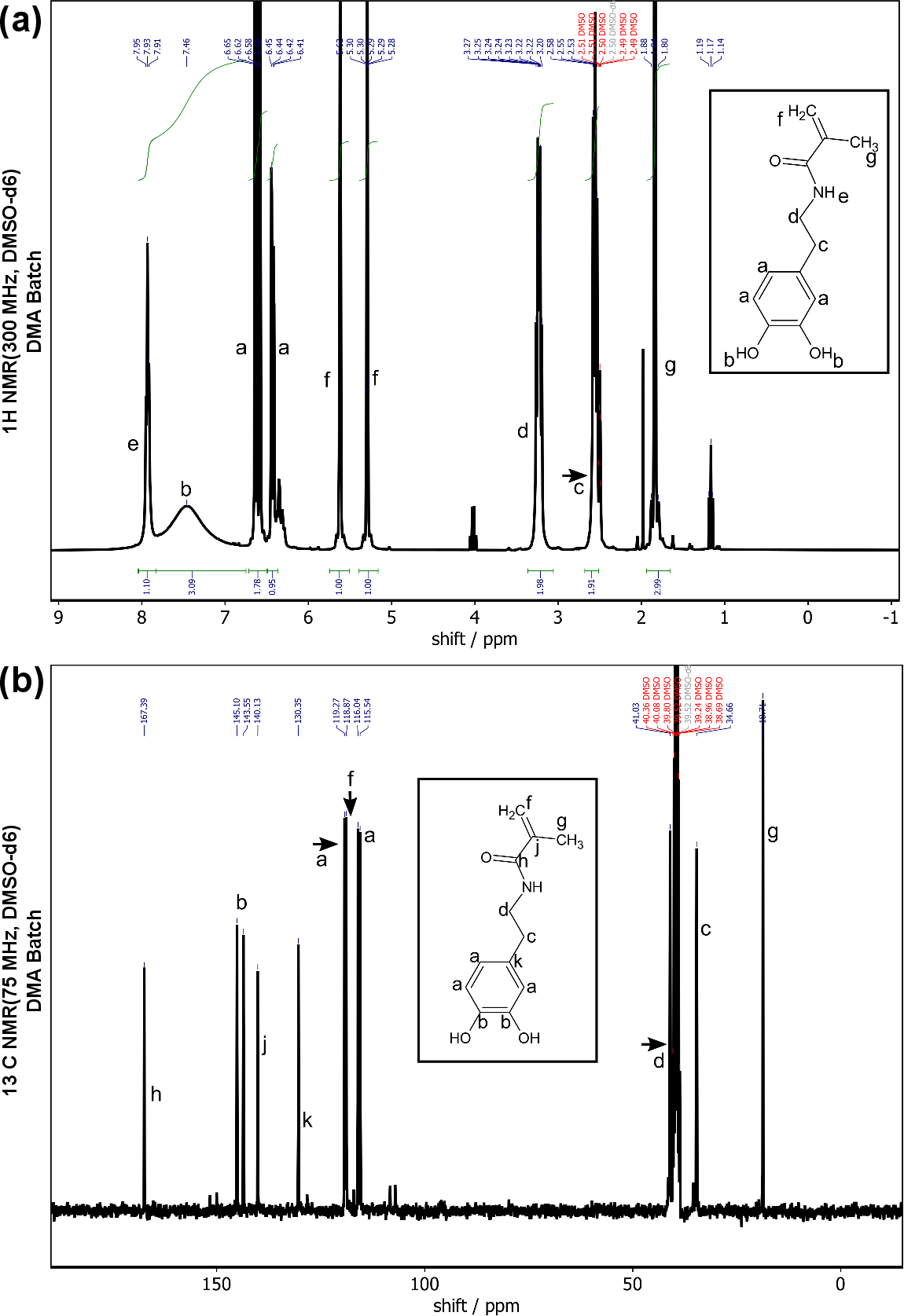}
\caption{(a) $^1$H NMR and (b) $^{13}$C NMR spectra of exemplary DMA Batch in \mbox{DMSO-d$_6$}.}
\label{SI-fig:NMR_DMA_Batch}
\end{figure}

\begin{figure}[H]
\centering
\includegraphics[scale=0.6]{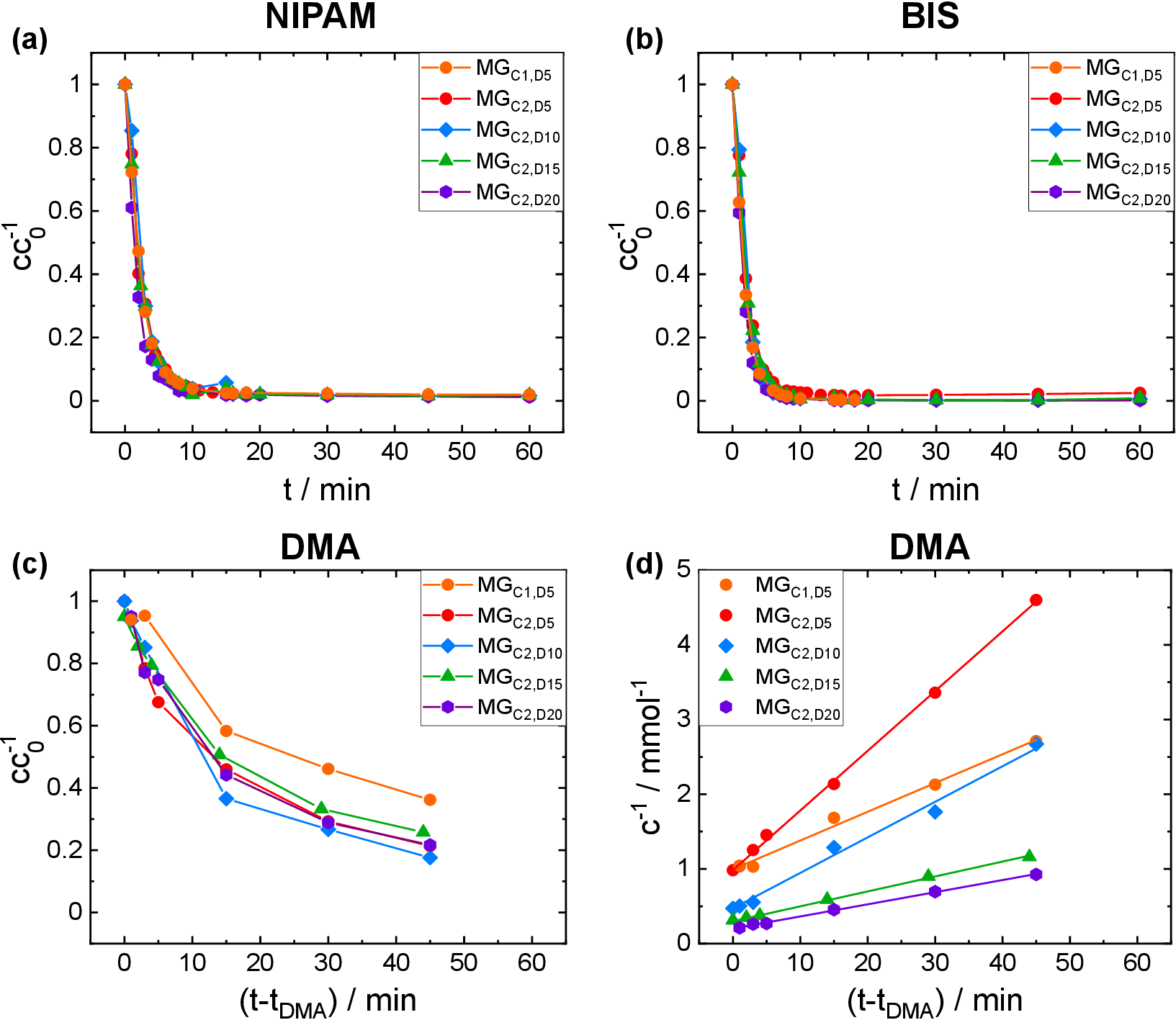}
\caption{Relative monomer consumptions $cc_0^{-1}$ of reactants NIPAM (a), BIS (b) and DMA (c) as a function of the reaction time $t$ for (a) and (b) and as a function of $t-t_\text{DMA}$ for (c) obtained by MS. (d) shows the inverse monomer consumption $c^{-1}$  of DMA. A linear fit can be applied, which underlines that DMA possesses a second-order reaction kinetics.}
\label{SI-fig:MS_Mikrogele}
\end{figure}

\begin{figure}[H]
\centering
\includegraphics[scale=0.8]{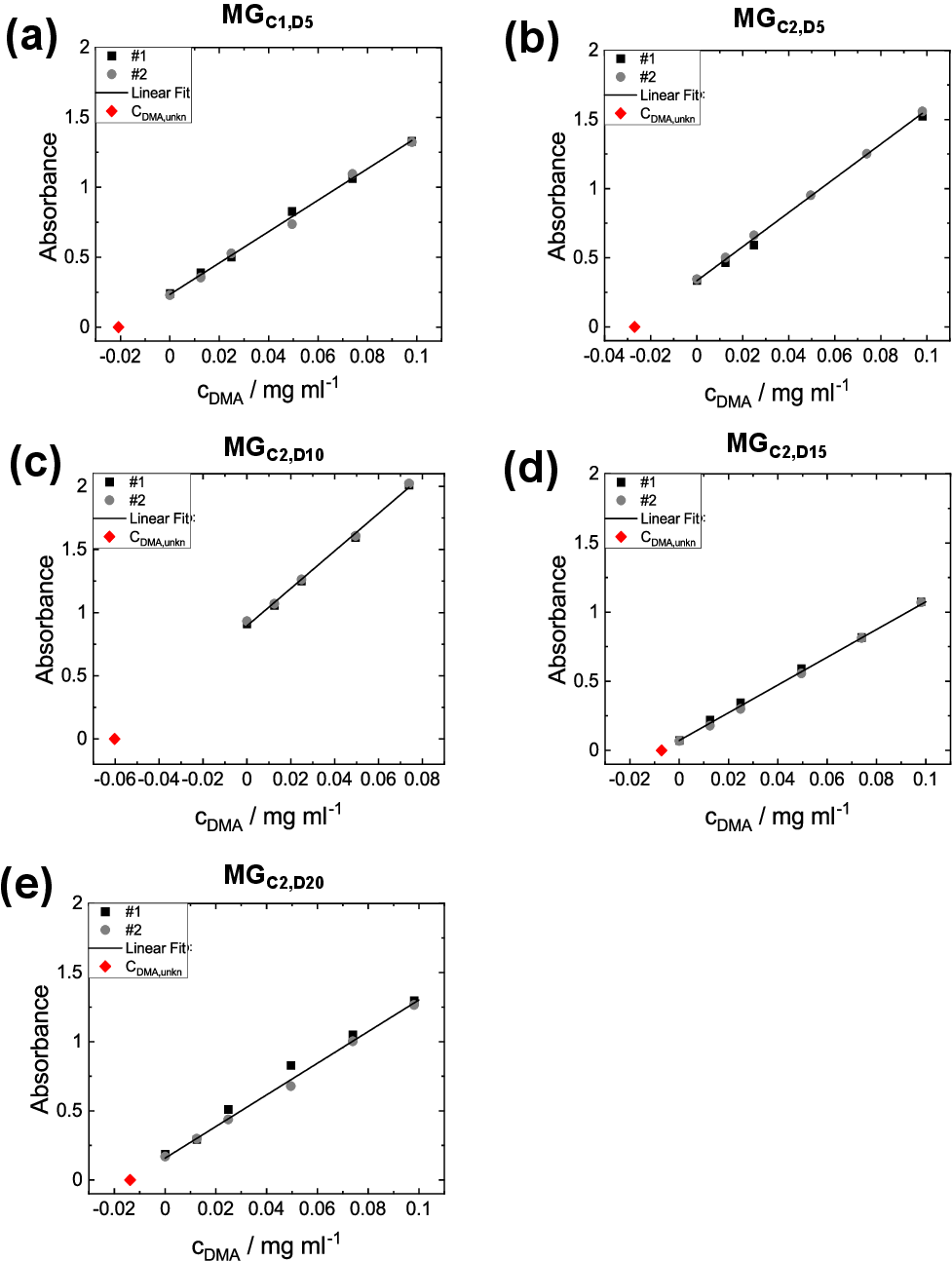}
\caption{Calibration curves obtained by UV-Vis standard addition for (a) MG$_\text{C1,D5}$, (b) MG$_\text{C2,D5}$, (c) MG$_\text{C2,D10}$, (d) MG$_\text{C2,D15}$ and (e) MG$_\text{C2,D20}$. The microgel sample concentration was 0.05\,wt\% for (a)-(c) and 0.005\,wt\% for (d)-(e). The absorbance at $\lambda_\text{max}=282$\,nm was plotted over the added concentration of DMA solution. Calibration curves were repeated 2\,times. A linear function was fitted to the medium of the two curves to obtain the unknown DMA concentration $c_\text{DMA,unkn}$.}
\label{SI-fig:UV-Vis_Calibration}
\end{figure}

\begin{figure}[H]
\centering
\includegraphics[scale=0.6]{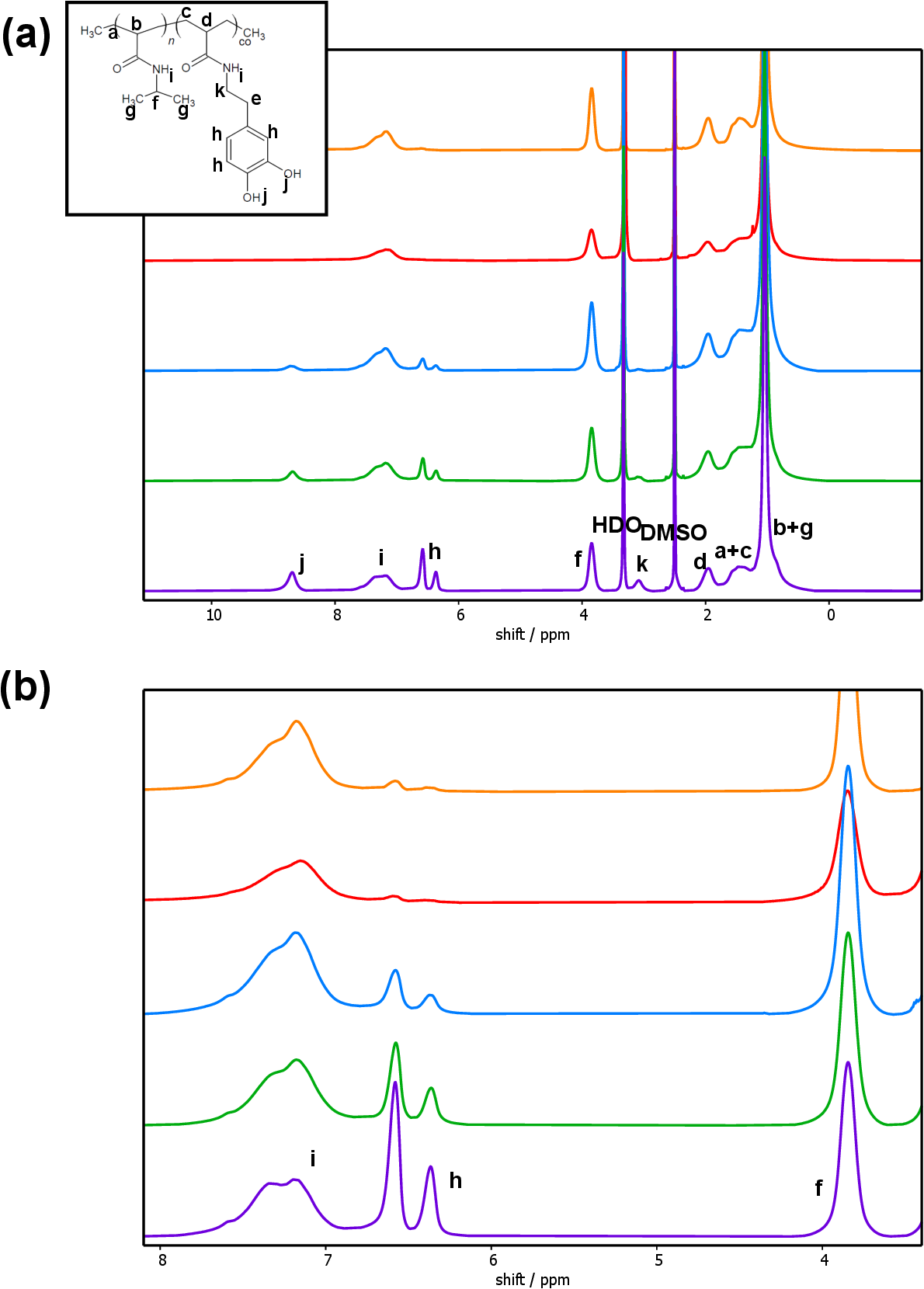}
\caption{(a)$^1$H NMR spectra of different microgels MG$_\text{C1,D5}$ (orange), MG$_\text{C2,D5}$ (red),  MG$_\text{C2,D10}$ (blue), MG$_\text{C2,D15}$ (green) and MG$_\text{C2,D20}$ (purple) in DMSO-d$_6$. An enlargement of the important shift region is shown in (b).}
\label{SI-fig:NMR_DMSO}
\end{figure}

\begin{figure}[H]
\centering
\includegraphics[scale=0.6]{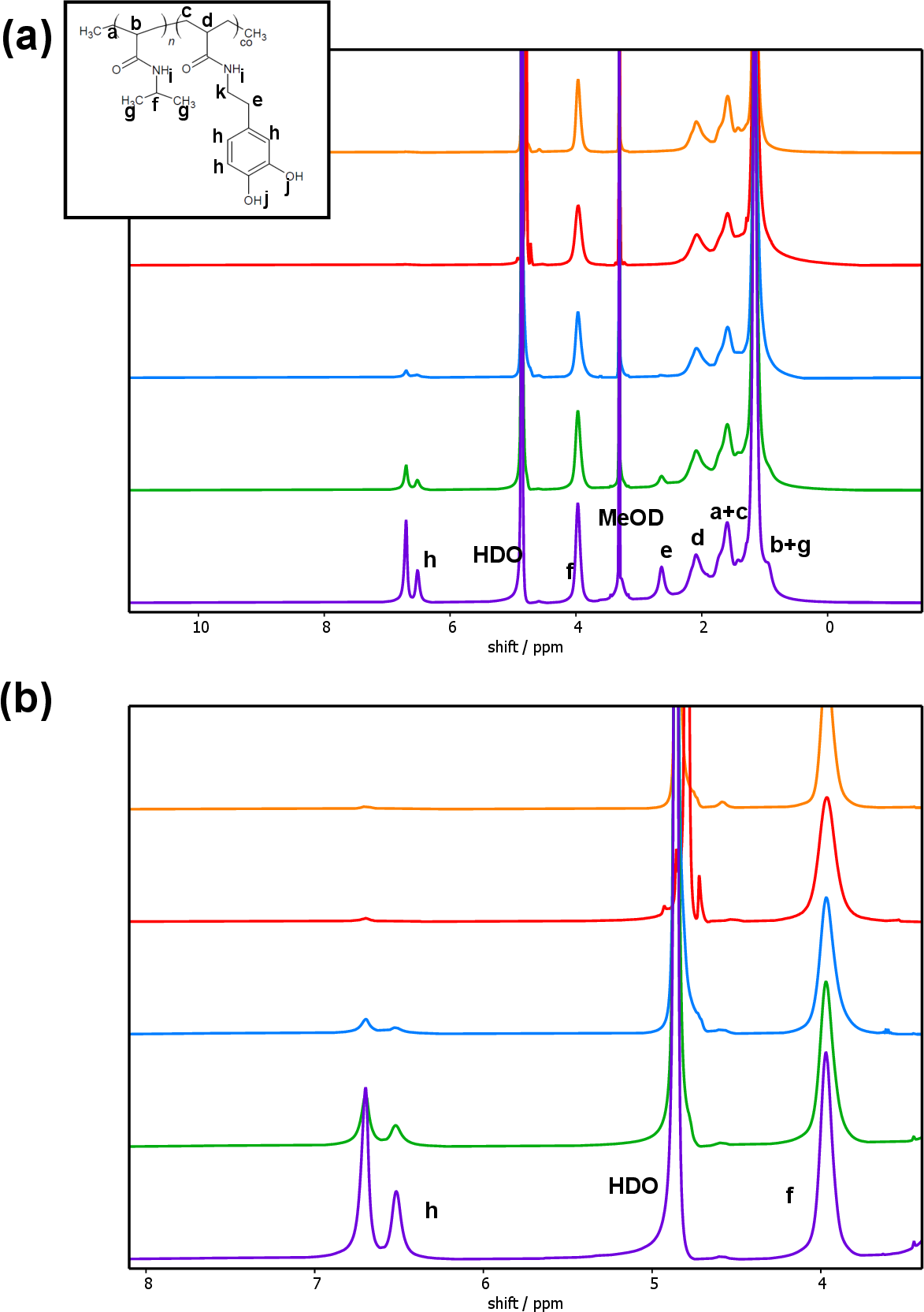}
\caption{(a)$^1$H NMR spectra of different microgels MG$_\text{C1,D5}$ (orange), MG$_\text{C2,D5}$ (red), MG$_\text{C2,D10}$ (blue), MG$_\text{C2,D15}$ (green) and MG$_\text{C2,D20}$ (purple) in methanol-d$_4$. An enlargement of the important shift region is shown in (b).}
\label{SI-fig:NMR_CD3OD}
\end{figure}

\begin{figure}[H]
\centering
\includegraphics[scale=0.35]{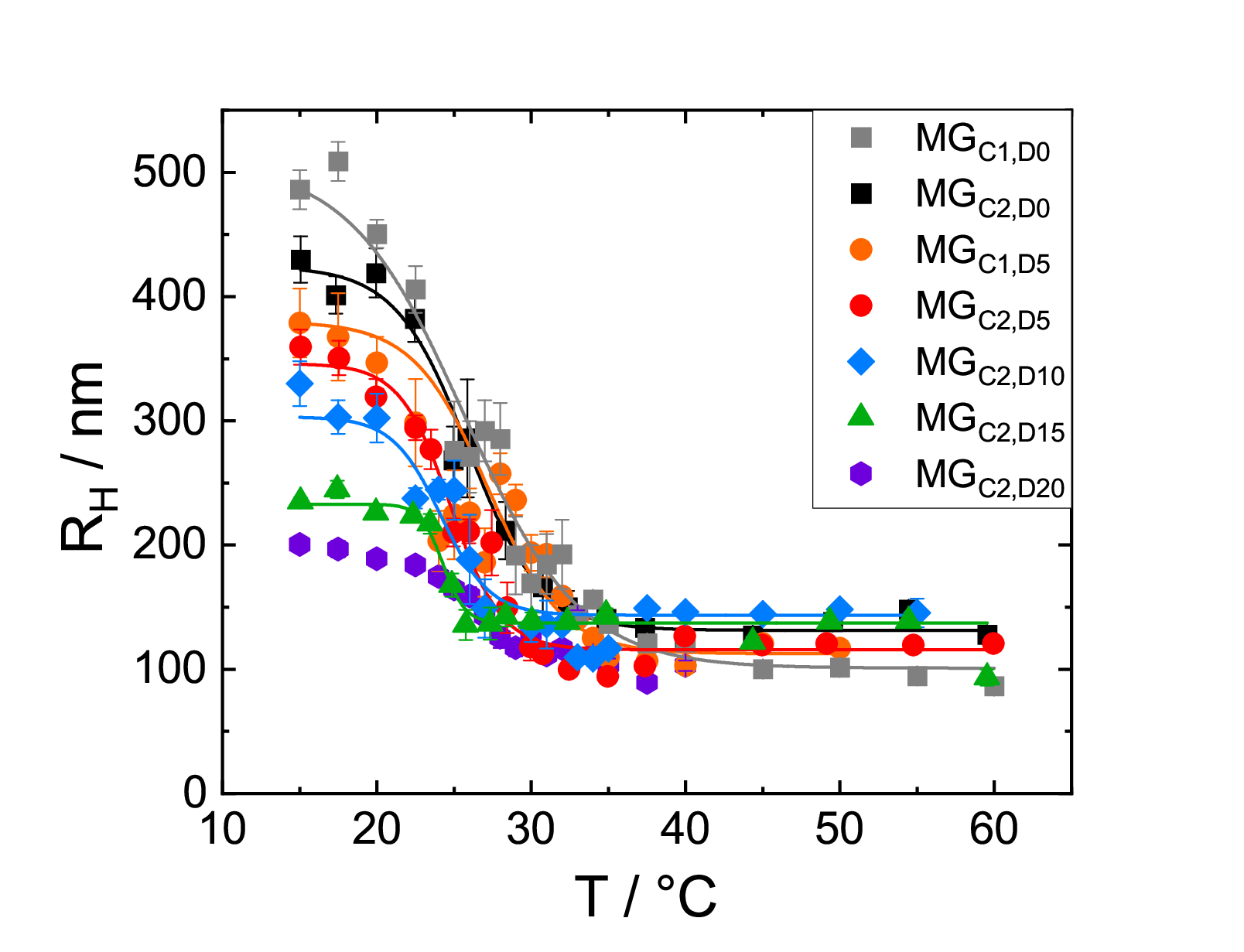}
\caption{Swelling curves of microgels fitted by a Boltzman function.}
\label{SI-fig:DLS_Boltzmann}
\end{figure}

\begin{figure}[H]
\centering
\includegraphics[scale=0.35]{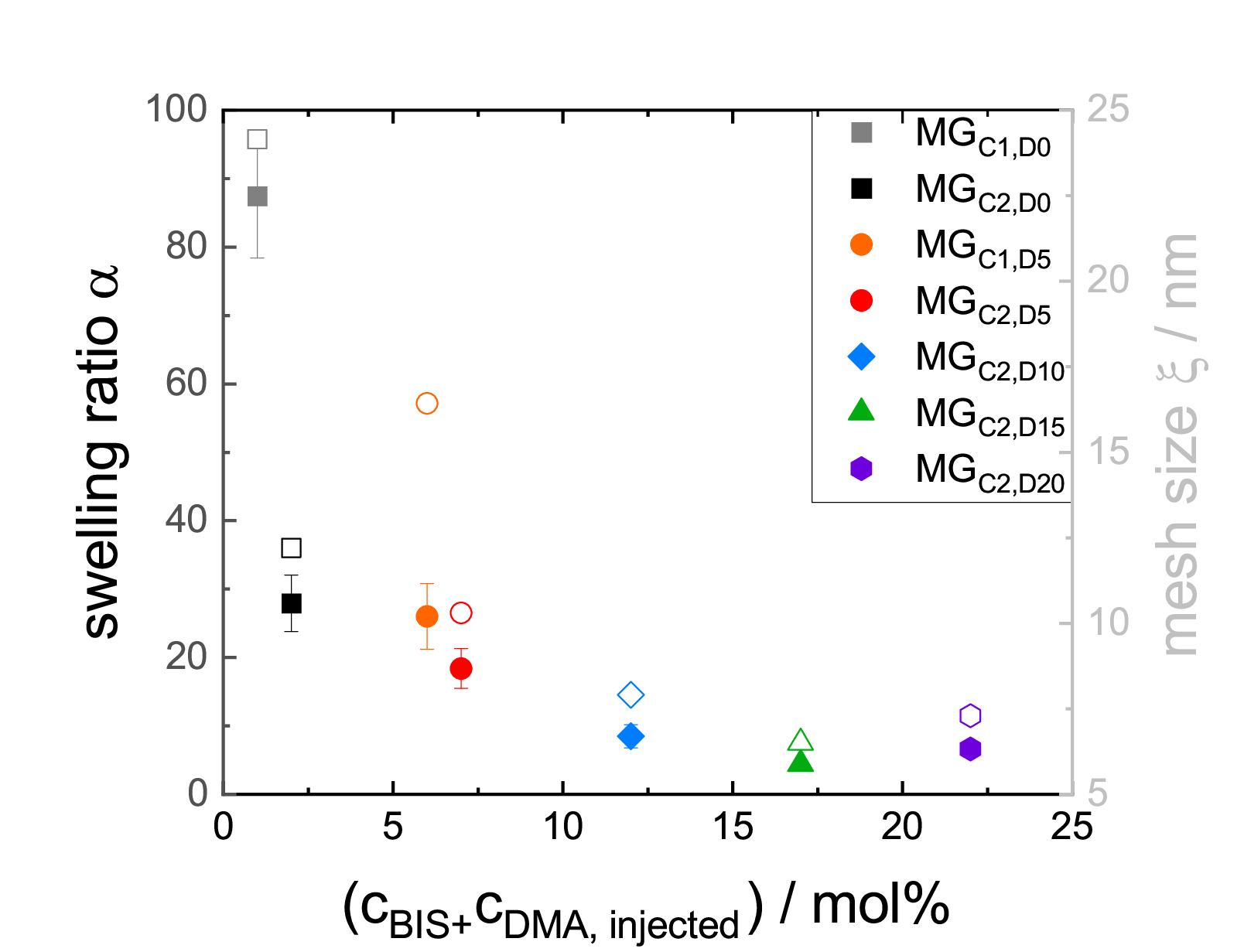}
\caption{Calculated swelling ratio $\alpha$ (filled symbols) and mesh size $\xi$ plotted as a function of the concentration of BIS and injected DMA.}
\label{SI-fig:mesh size}
\end{figure}

\begin{figure}[H]
\centering
\includegraphics[scale=0.6]{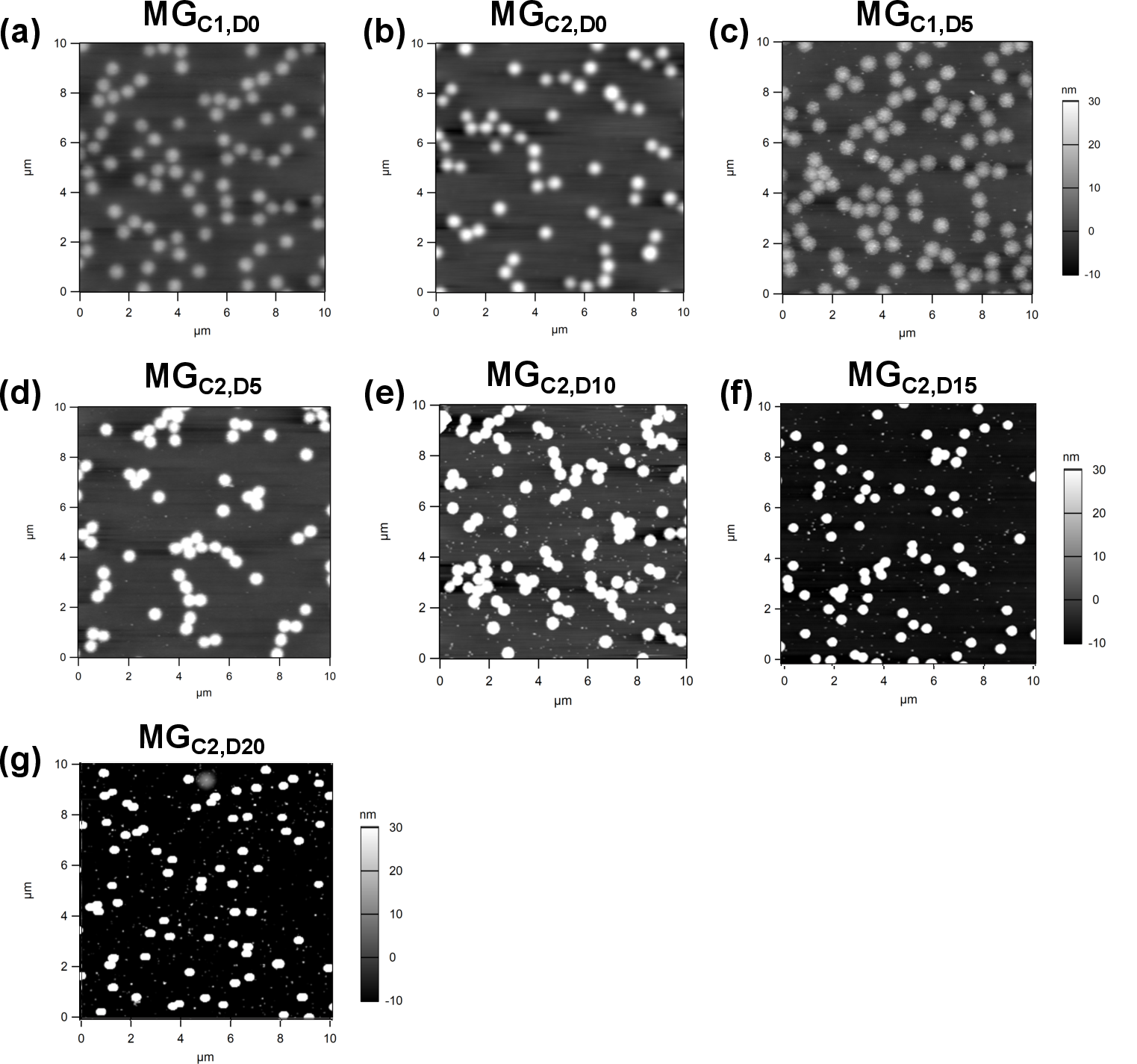}
\caption{AFM scans taken in air at a temperature of $T=20^\circ$C for (a) MG$_\text{C1,D0}$, (b) MG$_\text{C2,D0}$, (c) MG$_\text{C1,D5}$, (d) MG$_\text{C2,D5}$, (e) MG$_\text{C2,D10}$, (f) MG$_\text{C2,D15}$ and (g) MG$_\text{C2,D20}$ used to calculate the dry height $h_\text{dry,AFM}(20^\circ\,\text{C})$ and the width $w_\text{dry,AFM}(20^\circ\,\text{C})$. }
\label{SI-fig:AFM_dryheight}
\end{figure}

\begin{table}[h!]
\caption{Characterisation of microgels at the surface by AFM scanning in ambient conditions. Values are derived from an average of 5 different microgels per sample. While the error of the height $h_\text{amb,AFM}$ was obtained directly by standard deviation of the 5 particles, the error of the width $w_\text{amb,AFM}$ was estimated to be 10\%, estimated by eye for the microgel with the highest width MG$_\text{C2,D0}$.}
\centering
\begin{tabular}{lll}
\hline
Microgel &  $h_\text{amb,AFM}$ / nm & $w_\text{amb,AFM}$ / nm  \\
\hline
MG$_\text{C1,D0}$ &  17 $\pm$ 2  & 1177 $\pm$ 118 \\
MG$_\text{C2,D0}$ &  28 $\pm$ 3 & 1061 $\pm$ 106 \\
MG$_\text{C1,D5}$ &  20 $\pm$ 1 & 984 $\pm$ 98 \\
MG$_\text{C2,D5}$ &  44 $\pm$ 7& 904 $\pm$ 90 \\
MG$_\text{C2,D10}$ &   78 $\pm$ 3 &706 $\pm$ 71 \\
MG$_\text{C2,D15}$  &  131 $\pm$ 4 &471 $\pm$ 47 \\
MG$_\text{C2,D20}$  &  234 $\pm$ 12 &433 $\pm$ 43\\
\hline
\end{tabular}
\label{SI-tab:microgels_characterization_surface_height_width}
\end{table}

\begin{figure}[H]
\centering
\includegraphics[scale=0.55]{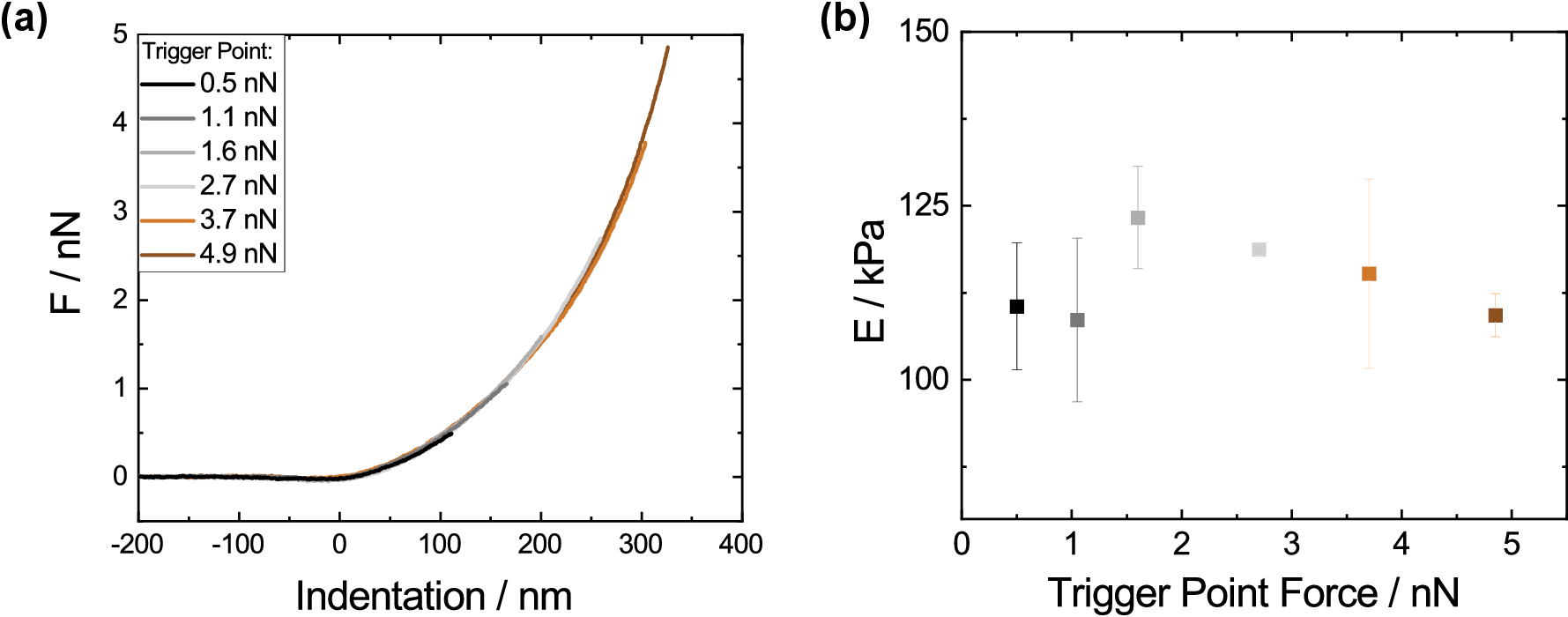}
\caption{(a) Force curves received for the cantilever BL-AC40 TS for a reference PNIPAM microgel MG$_\text{C2,D0}$ with different trigger points, defining the indentation depth of the cantilever. (b) The effect of the trigger point on the $E$ modulus of a pure PNIPAM microgel MG$_\text{C2,D0}$. A trigger point of 2.7\,nN was defined for further measurements. }
\label{SI-fig:Cantilever Vergleich_Trigger Points_pure Force Curves}
\end{figure}

\begin{figure}[H]
\centering
\includegraphics[scale=0.35]{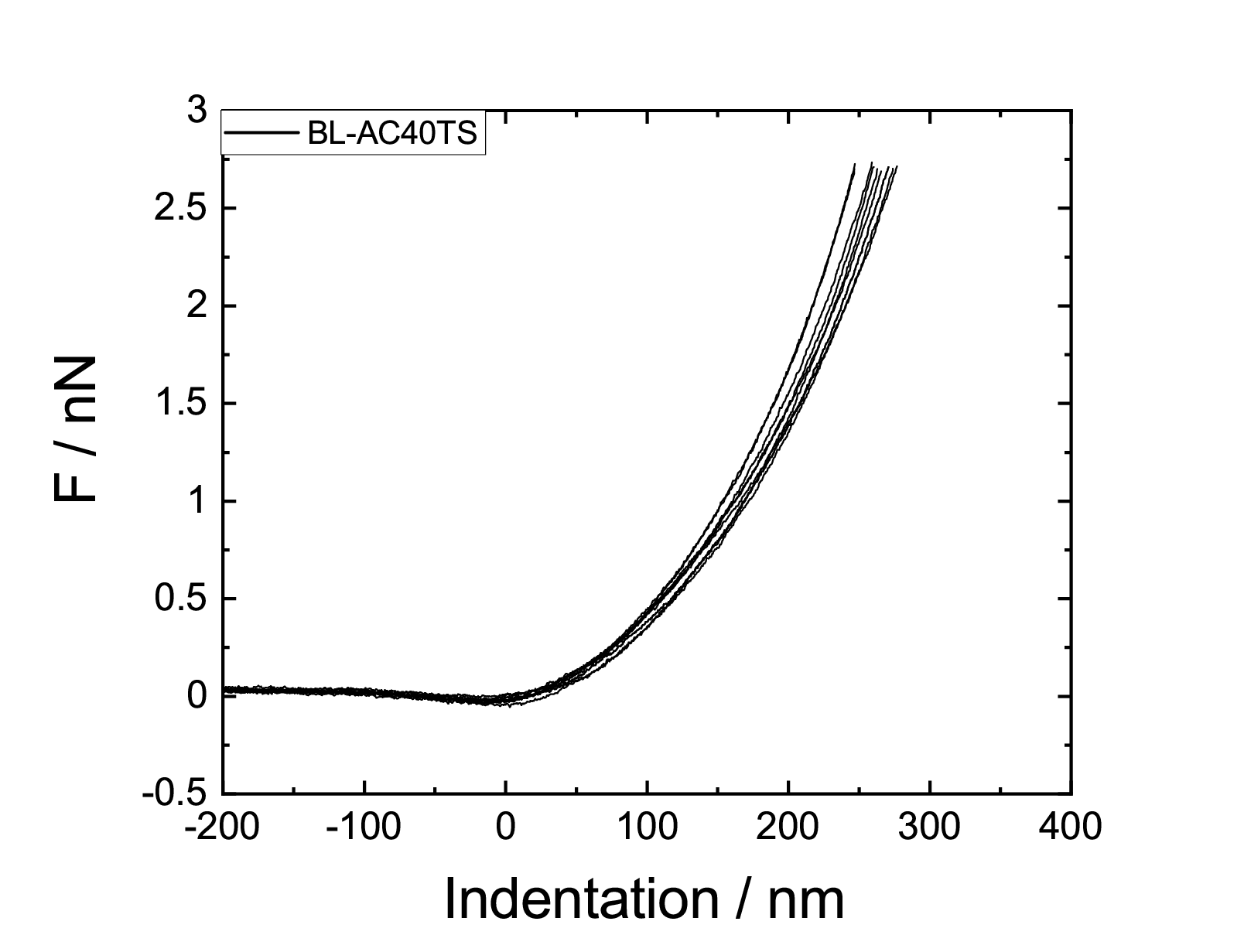}
\caption{Pure force curves received for a BL-AC40TS cantilever for a reference PNIPAM microgel MG$_\text{C2,D0}$ used for the calculation of $E$ moduli. }
\label{SI-fig:Cantilever Vergleich_pure Force Curves}
\end{figure}

\begin{figure}[H]
\centering
\includegraphics[scale=0.55]{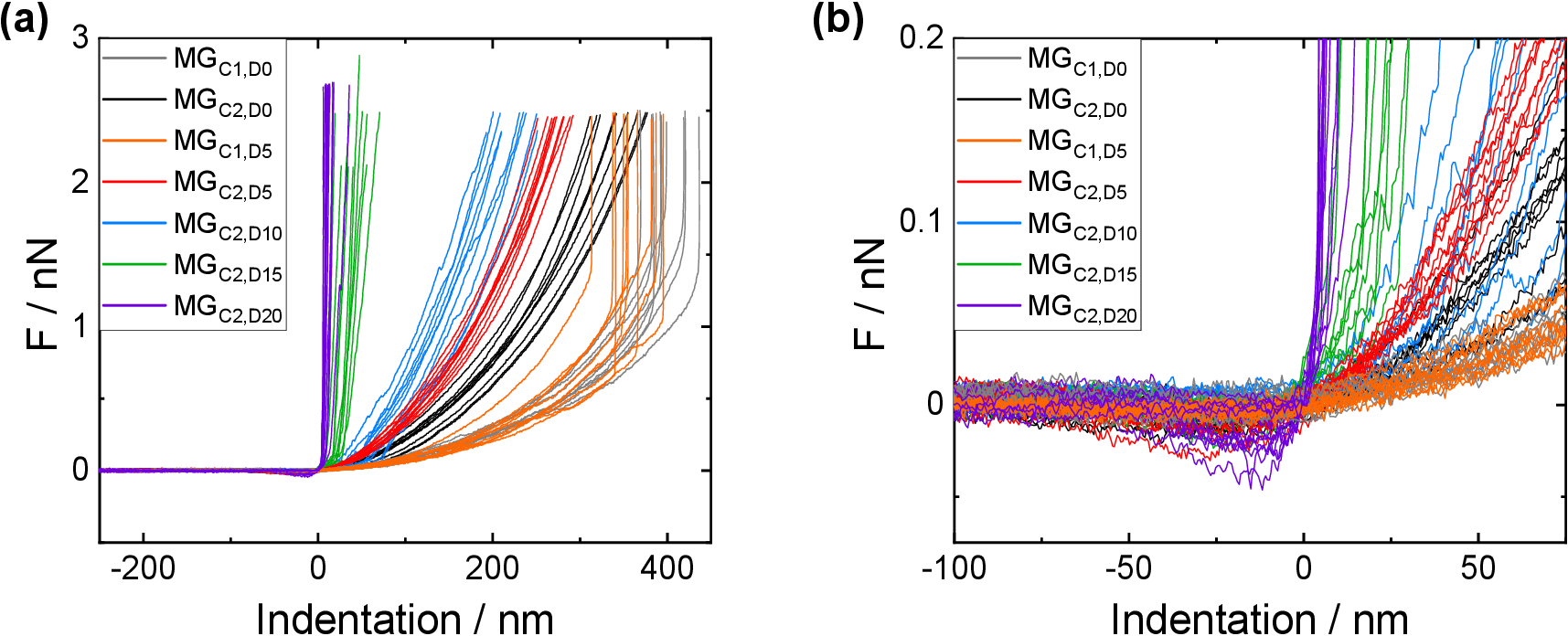}
\caption{(a) Pure force curves received for the different microgels with a BL-AC40 TS cantilever, used for the calculation of $E$ moduli. (b) Inset of (a) highlighting the attraction before the actual indentation. For the microgels with lower cross-linker amount (MG$_\text{C1,D0}$ and MG$_\text{C1,D5}$), the microgel's height is so small that the cantilever already touches the substrate's surface before reaching the trigger point. This could damage the tip when forces get too high. However, the measurement of 10 different microgels with the same tip showed no large deviations in force curves and $E$ moduli, so we conclude that the tip stays intact.}
\label{SI-fig:all pure force curves}
\end{figure}

\begin{figure}[H]
\centering
\includegraphics[scale=0.55]{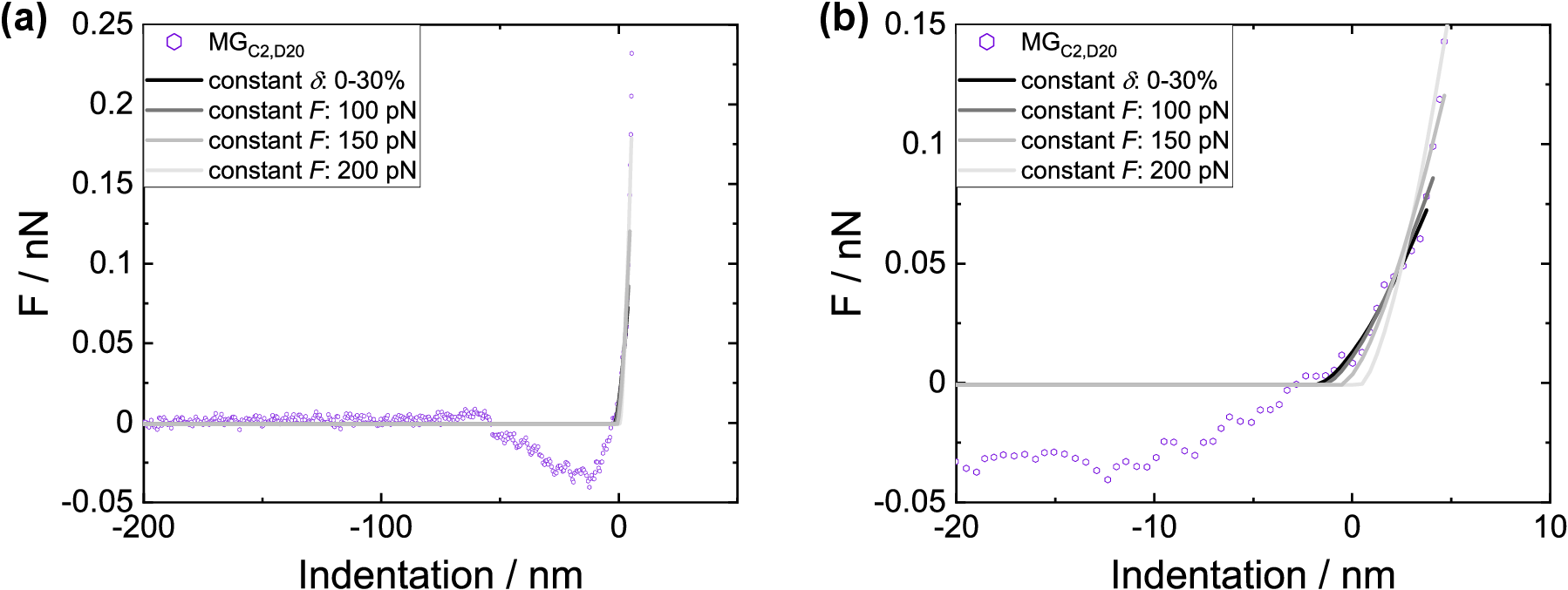}
\caption{(a) Exemplary force curve of MG$_\text{C2,D20}$ microgel fitted by different methods: either a constant indentation range of force range  (b) Inset of (a) visualising the choice of values for fitting. Constant force above 100\,pN cannot describe the force curve well anymore. }
\label{SI-fig:constant_fit_forcecurve}
\end{figure}

\begin{figure}[H]
\centering
\includegraphics[scale=0.4]{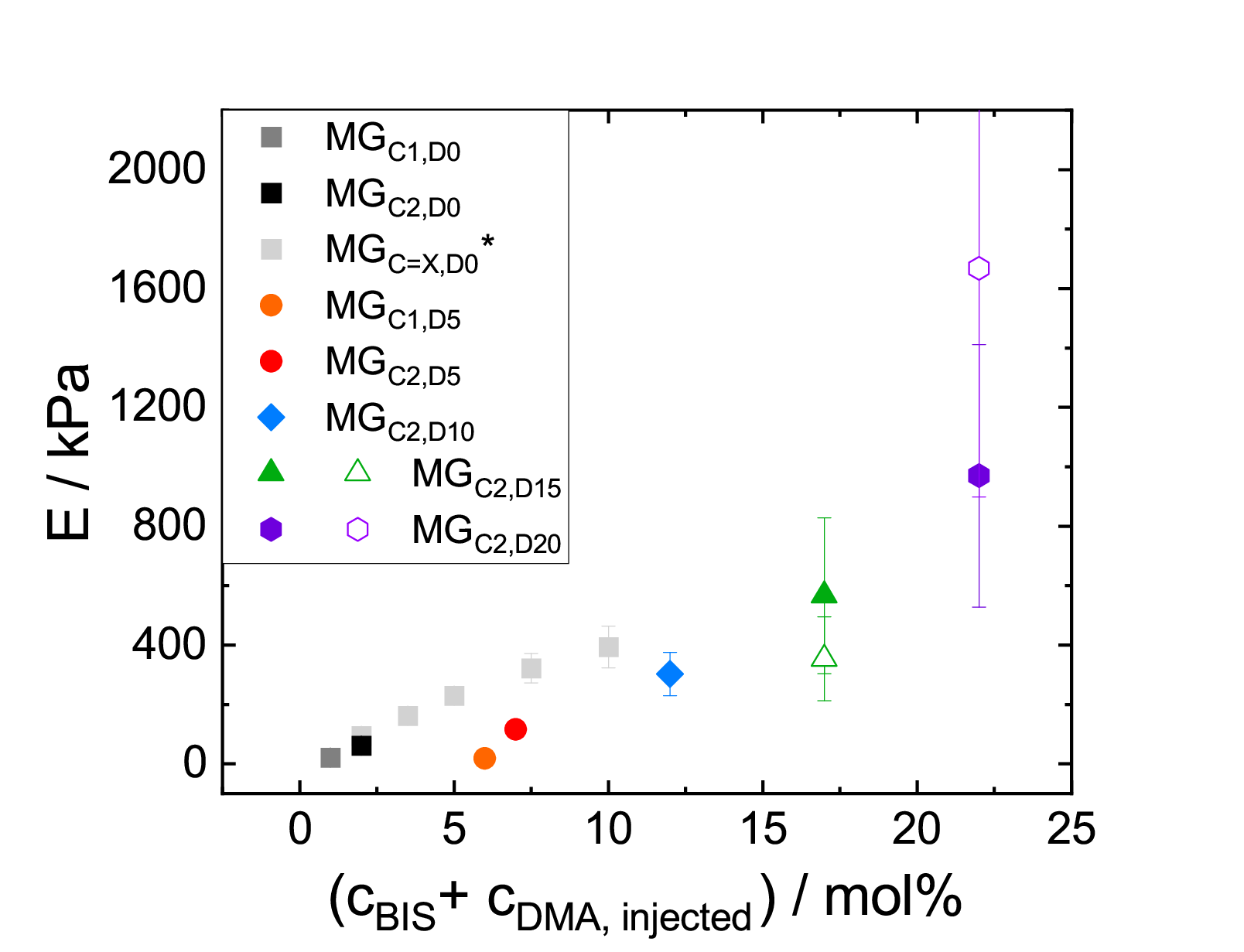}
\caption{(a) $E$ moduli obtained from 10 force curves of each microgel particle plotted over the concentration of BIS and feeded DMA, calculated by fitting the approach force curves with a constant indentation range of $0-40\%$. Light grey symbols correspond to results obtained by Kühnhammer \textit{et al.}$^{50}$. For MG$_\text{C2,D15}$ and MG$_\text{C2,D20}$, the approach force curves were fitted additionally with a constant force fit range of 100\,pN (open symbols).}
\label{SI-fig:E_microgels}
\end{figure}